\documentclass[fontsize=11pt, twoside]{scrbook}
\pdfoutput=1

\setcapindent{0pt}
\setkomafont{captionlabel}{\sffamily\bfseries}

\KOMAoptions{toc=bib}
\RedeclareSectionCommand[beforeskip=1.25ex plus 0.5ex minus .2ex]{paragraph}

\usepackage[T1]{fontenc}
\usepackage[a4paper]{geometry}

\usepackage[english]{babel}
\usepackage{amsmath}
\usepackage{amssymb}
\usepackage{graphicx}

\usepackage[usenames,dvipsnames]{xcolor}

\usepackage{tikz}
\usetikzlibrary{shapes,arrows,calc,through}%

\usepackage[sc,osf]{mathpazo}
\usepackage{fixltx2e}

\usepackage{xspace}

\usepackage{setspace}
\usepackage[bottom,perpage,symbol*]{footmisc}

\usepackage{float}
\usepackage{wrapfig}

\usepackage{caption}
\usepackage[labelformat=simple]{subcaption}

\usepackage{booktabs}
\usepackage{multirow}

\usepackage{fancyhdr}

\usepackage{microtype}

\usepackage{placeins}

\usepackage{rotating}
\usepackage{url}
\usepackage{cite}

\usepackage{mathtools}

\usepackage{siunitx}
\DeclareSIUnit\kcal{kcal}
\DeclareSIUnit\day{day}
\DeclareSIUnit\M{M}
\numberwithin{equation}{section}

\newcommand*{\mycleardouble}{\clearpage{\thispagestyle{empty}\cleardoublepage\thispagestyle{plain}}}

\newcolumntype{v}[1]{S[
table-format = #1,
parse-numbers = false,
input-symbols=.,
input-decimal-markers = x
]}

\newcommand{\specialcell}[2][c]{%
  \begin{tabular}{@{}#1@{}}#2\end{tabular}}

\newcommand{\centercell}[1]{%
  \multicolumn{1}{c}{#1}}

\newcommand{\textcell}[1]{\centercell{\specialcell{#1}}}

\usepackage{framed}
\usepackage{enumerate}

\usepackage{braket}

\usepackage{bm}

\usepackage{chemfig}

\usepackage[bookmarks=true, final=true]{hyperref}
\usepackage[capitalise]{cleveref}

\newcommand\inputpgf[2]{{
\let\pgfimageWithoutPath\pgfimage
\renewcommand{\pgfimage}[2][]{\pgfimageWithoutPath[##1]{#1/##2}}
\input{#1/#2}
}}

\newcommand*{\etal}{\textit{et al.\ }}
\newcommand*{\ie}{\textit{i.e.},\ }
\newcommand*{\eg}{\textit{e.g.},\ }
\newcommand*{\cf}{\textit{cf.}\ }
\newcommand*{\vs}{\textit{vs.}\ }

\newcommand*{\nth}{\textsuperscript{th}\xspace}

\def\vec#1{{\bm{#1}}}
\newcommand*{\mat}[1]{\ensuremath{#1}}

\newcommand*{\matSigma}{\bm{\Sigma}}

\newcommand*{\transposed}[1]{{#1}^\intercal}

\newcommand*{\covariance}{\operatorname{cov}}

\newcommand*{\half}{\tfrac{1}{2}}

\providecommand*{\abs}[1]{|#1|}

\providecommand*{\order}[1]{\ensuremath{\mathcal{O}(#1)}}
\providecommand*{\normaldist}[1]{\ensuremath{\mathcal{N}(#1)}}

\newcommand*{\identity}{\mathbb{1}}

\newcommand*{\kB}{\ensuremath{{k_{\operatorname{B}}}}}
\newcommand*{\kT}{\ensuremath{\mathit{{\kB}T}}}

\providecommand*{\ee}{\ensuremath{\operatorname{e}}}
\providecommand*{\dd}{\ensuremath{\operatorname{d}}}

\providecommand*{\myfrac}[2]{\mathchoice{\frac{#1}{#2}}{#1/#2}{#1/#2}{#1/#2}}
\providecommand*{\derivform}[4][]{\ensuremath{\myfrac{{#2}^{#1}#3}{#2 {#4}^{#1}}}}
\providecommand*{\deriv}[3][]{\derivform[#1]{\dd}{#2}{\! #3}}
\providecommand*{\pderiv}[3][]{\derivform[#1]{\partial}{#2}{#3}}
\providecommand*{\pcrossderiv}[3]{\ensuremath{\myfrac{\partial^2 #1}{\partial #2 \partial #3}}}
\providecommand*{\fderiv}[3][]{\derivform[#1]{\delta}{#2}{#3}}
\providecommand*{\inlinederiv}[2]{\dd\! #1/\!\dd\! #2}

\providecommand*{\expp}[1]{\ensuremath{\ee^{#1}}}

\newcommand*{\vecr}{\vec{r}}
\newcommand{\ints}{\int\!}
\newcommand*{\dr}[1][\vecr]{\dd\!#1\,}

\newcommand{\intdd}[1]{\ints\dr[#1]}

\newcommand*{\chem}[1]{\ensuremath{\mathrm{#1}}}

\newcommand*{\ceil}[1]{\ensuremath{\lceil #1 \rceil}}
\newcommand*{\average}[2][]{\ensuremath{#1\langle #2 #1\rangle}}
\newcommand*{\Average}[1]{\ensuremath{\left\langle #1 \right\rangle}}

\newcommand*{\norm}[1]{\ensuremath{\|#1\|}}

\newcommand{\expectval}{\mathbb{E}}
\newcommand{\variance}{\mathbb{V}}

\newcommand*{\myvspace}{\vspace{1.5ex}}

\newcommand*{\properacronym}[1]{\textsc{#1}\xspace}
\newcommand*{\AMBER}{\properacronym{amber}}
\newcommand*{\GAUSSIAN}{\properacronym{gaussian}}
\newcommand*{\VOTCA}{\properacronym{votca}}
\newcommand*{\LAMMPS}{\properacronym{lammps}}
\newcommand*{\PMFlib}{\textsc{PMF}lib\xspace}
\newcommand*{\QUIP}{\properacronym{quip}}

\newcommand*{\captionheading}[1]{\textbf{\textsf{#1}}}
\newcommand{\mycaption}[2]{\caption[#1]{\captionheading{#1} #2}}
\newcommand*{\textcite}[1]{Ref.~\citen{#1}}
\newcommand*{\mypageref}[1]{page~\pageref{#1}}

\begin{document}

\title{Many-Body Coarse-Grained Interactions using Gaussian Approximation Potentials}
\author{S.T. John}
\newgeometry{tmargin=30mm,bmargin=35mm,lmargin=35mm,rmargin=34mm}
\include{titlepage}

\section*{Summary}
Coarse-graining (CG) is an approach to molecular simulations in which molecules are described at a lower resolution than individual atoms, thereby reducing the number of interaction sites and speeding up the simulation. Considering the formal connection between all-atom and CG models given by statistical mechanics, a CG model is deemed to be \emph{consistent} with the all-atom model when the CG interactions are described by the many-body potential of mean force (PMF).

In practice, all current CG methods only consider a limited set of interaction terms between a very small number of CG sites; non-bonded interactions are typically described by pairwise radial interactions between sites. Effective interactions can be obtained by matching certain properties of the all-atom model, such as pairwise distance distributions or forces. However, this results in a lack of \emph{representability}: CG models that only include these limited site-based interactions cannot represent both structural and thermodynamic properties of the underlying all-atom system.

In the field of atomistic simulations, many-body interactions have been successfully described using the Gaussian Approximation Potential (GAP), a machine-learning approach based on Gaussian process regression for obtaining flexible, high-dimensional interaction potentials.
The present work transfers GAP to coarse-graining (GAP-CG), in which the CG interactions are derived from observations of forces in the all-atom model. This includes the development of solutions to specific challenges such as inhomogeneous length scales and insufficient sampling.
GAP-CG allows us to approximate the PMF using \emph{molecular} terms that correspond to pairs and triplets of entire molecules.

In applying GAP-CG to methanol and benzene clusters and bulk systems, this thesis demonstrates that this approach can recover the PMF, allowing a representation of both structural (including orientational) distributions \emph{and} forces. Furthermore, this thesis studies how the accuracy of force predictions depends on the hyperparameters.

Many-body interactions are computationally expensive. At the example of water-solvated benzene, this thesis shows that a solvent-free CG model using GAP interactions can be faster than the corresponding all-atom model while being significantly more accurate than current approaches to coarse-graining.

\cleardoublepage

\vspace*{5cm}
\begin{flushright}
  \itshape
  Dedicated to my father\\
  who would have liked to read this
\end{flushright}
\cleardoublepage

\restoregeometry

\pagestyle{fancyplain}
\fancyhf{} 
\fancyfoot[EL]{\thepage}
\fancyhead[EL]{\textsl{\textsf{\leftmark}}}
\fancyhead[OR]{\textsl{\textsf{\rightmark}}}
\fancyfoot[OR]{\thepage}
\renewcommand{\sectionmark}[1]{ \markboth{\thesection{} #1}{\thesection{} #1} } 
\renewcommand{\subsectionmark}[1]{ \markright{\thesubsection{} #1} }
\renewcommand{\headrulewidth}{0.5pt} 
\renewcommand{\footrulewidth}{0pt} 
\fancypagestyle{plain}{%
\fancyhead{} 
\renewcommand{\headrulewidth}{0pt} 
\fancyfoot[R]{\thepage} } 

\setcounter{page}{1} 

\begin{spacing}{1.1}
\tableofcontents
\end{spacing}

\clearpage
\thispagestyle{empty}
\section*{Abbreviations}

\begin{tabular}{l@{\hspace{2cm}}l}
  QM       & Quantum Mechanics \\
  MD       & Molecular Dynamics \\
  MM       & Molecular Mechanics \\
  AA       & All-Atom \\
  CG       & Coarse-Graining \\

  GAP      & Gaussian Approximation Potential \\[0.75ex]

  CV       & Collective Variable \\
  PMF      & Potential of Mean Force \\
  MF       & Mean Force \\
  ICF      & Instantaneous Collective Force \\[0.75ex]

  DBI      & Direct Boltzmann Inversion \\
  IBI      & Iterative Boltzmann Inversion \\
  IMC      & Inverse Monte Carlo \\
  MSCG/FM  & Multiscale Coarse-Graining/Force Matching \\[0.75ex]

  COM      & Centre Of Mass \\
  RDF      & Radial Distribution Function \\
  ADF      & Angular Distribution Function \\
  RMSE     & Root Mean Squared Error
\end{tabular}
\clearpage
\thispagestyle{plain}

\clearpage

\chapter{Introduction}

\section{Molecular Simulations}

Understanding processes on an atomic or molecular scale is important in a wide range of fields from materials science to molecular biology. This includes diverse topics, from studying the formation and propagation of cracks or elucidating the properties of novel ``nano'' materials to investigating the behaviour of proteins, lipids, and nucleic acids or studying the interactions between small molecules and binding sites to aid in the development of new drugs. %
Ultimately, only experiments can give definite, authoritative answers to questions about these processes. %

However, there are many questions that cannot directly be answered by experiments alone. This can be for a number of reasons: We do not always have the tools available to access all the properties in which we are interested. Experiments might only provide limited resolution without being able to access microscopic details of structure or dynamics. The experimental probe might require a modification of, or otherwise interfere with, the system we want to study, possibly changing its behaviour.
Moreover, experiments tend to be very expensive and require a large amount of direct involvement and human time. Consequently, the number of experiments performed is often small, resulting in comparatively large errors and uncertainty, especially in biological studies.

Computer simulations of molecular systems can help us fill these gaps. They allow us to access \emph{all} properties of the system within the simulation, and let us see the effect of manipulating any part of the system. This extends to manipulations that would be difficult or impossible in the real world (for example, we can simply replace one type of molecule with another). In approaches that simulate the dynamics of a system, we can also study its \emph{dynamic} behaviour and its response to virtual probes (such as stretching a molecule to determine its force--extension curve).

It is important to note that simulations do not make experiments redundant. A simulation is just a model, and its accuracy and its predictions, ultimately, need to be validated by experiments\cite{generalMDreference1}.
In fact, computer simulations can often be complementary to experiments, providing a bridge between real world and theory. Simulations can assist us in devising and refining experiments, and can help us understand experimental observations\cite{kermode2008crackpropagation}.

To be useful, a model should accurately represent those features of the real system in which we are interested. Moreover, it should make quantitative predictions that can be compared, where possible, with experimental results. Only then is a simulation a valuable tool.%

\subsection*{All-atom simulations}

The most accurate theory for describing processes on an atomistic scale is quantum mechanics (QM).
By solving the time-dependent Schr\"odinger equation for a molecular system, we can obtain the time evolution of the probability distributions for the positions of electrons and nuclei.
However, this is rarely feasible except for the smallest systems.
Instead, we usually consider approximations to the full quantum-mechanical description of a system.
The nuclei are much heavier than the electrons; thus, in many cases we can consider their motion separately (Born--Oppenheimer approximation\cite{BornOppenheimerApproximation}) and often the motion of the nuclei is described well by classical mechanics (Newton's laws of motion).
This is the basis of Molecular Dynamics (MD), one of the main approaches to molecular simulations, which calculates the dynamic trajectories of discrete particles based on the forces acting on them\cite{UnderstandingMolSim}.
We will explain this approach in more detail in \cref{ch:md}.

For an MD simulation of atoms as classical particles at the positions of the nuclei, we require the forces acting on them, as a function of their configuration. Using the Hellman--Feynman theorem\cite{Feynman}, we can calculate the forces from a quantum-mechanical description of the electrons. Common computational approaches for this are Quantum Monte Carlo\cite{QuantumMonteCarlo} or Density Functional Theory (DFT)\cite{parr1989density}. Quantum Monte Carlo provides high accuracy, but also has a high computational cost that limits it to comparatively small systems containing on the order of a thousand electrons. DFT considers only the electron \emph{density}, and can be considered the state-of-the-art approach for routine QM calculations.
While DFT codes can now compute forces in systems containing tens of thousands of atoms\cite{OnetepForces}, quantum-mechanical calculations are still too expensive for simulations of the dynamics of systems on biologically relevant scales.
\begin{figure}[tbp]
  \centering
  \includegraphics[width=\textwidth]{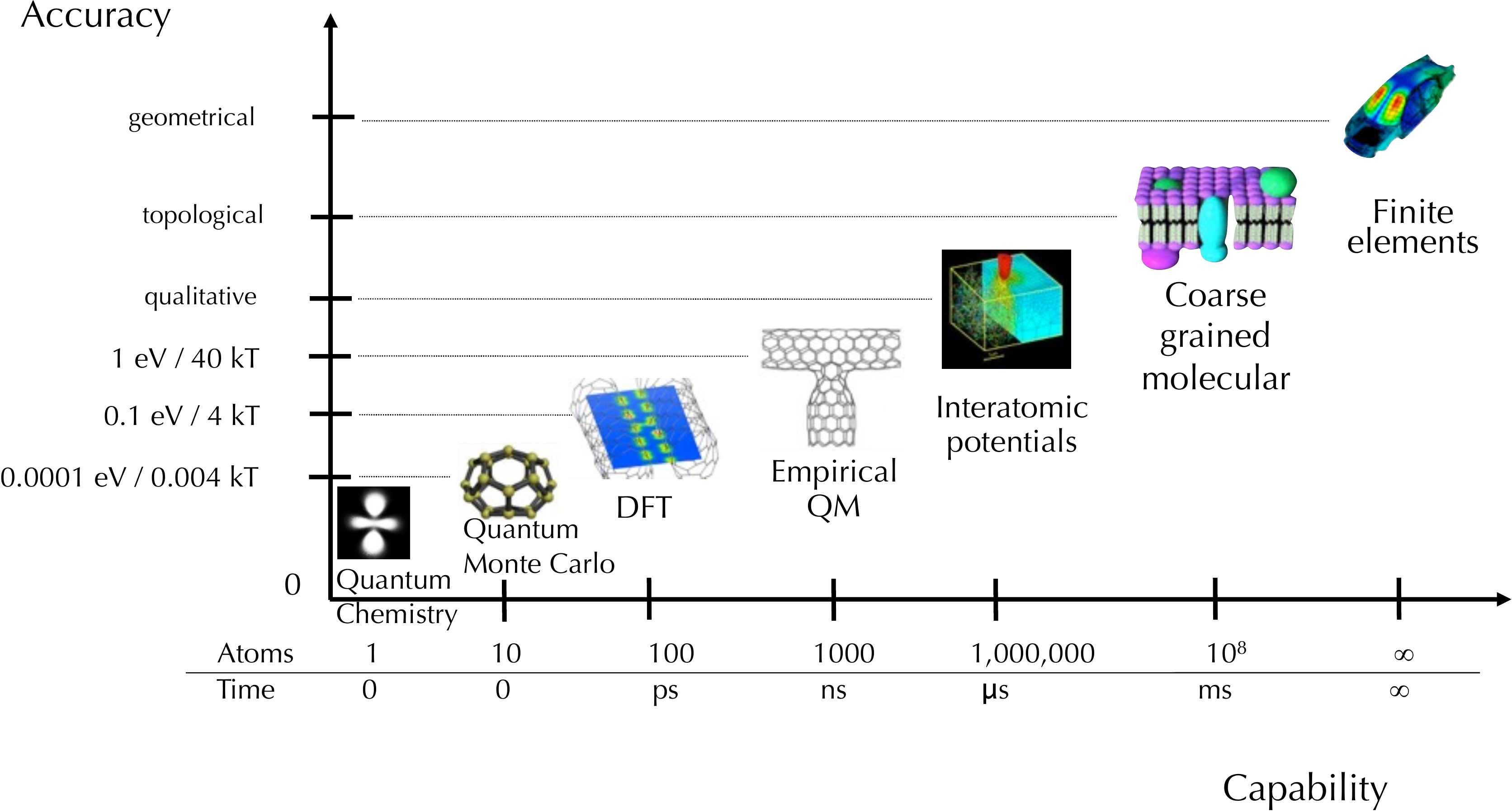}
  \vspace{1ex}
  \caption{Scales of atomic and molecular simulations. \textit{Figure courtesy of G\'abor Cs\'anyi.}}
  \label{fig:multiscale}
\end{figure}

In practice, molecular simulations are usually based on empirical force fields that approximate the true quantum-mechanical forces by a sum of independent interaction terms. Each term only depends on the positions of a few atoms within a cutoff and generally has a fixed functional form (see \cref{sec:md:forcefield}). This approach is referred to as ``molecular mechanics'' (MM). (We note that it is also possible to perform combined QM/MM simulations that treat certain parts of a model quantum-mechanically and use MM approximations for the remainder of the system.) The scales and accuracies of the different approaches are illustrated in \cref{fig:multiscale}. %

Molecular mechanics is routinely used to simulate molecular systems. On commodity machines with typically \num{16} cores, it is possible to simulate systems containing several thousand atoms, reaching ratios of model time to computing time that are on the order of $\SI{10}{\ns\per\day}$.

Using highly specialised computers\cite{ANTON}, simulations based on molecular mechanics have been extended up to the millisecond time scale for small systems such as ubiquitin with approximately ten thousand atoms in the solvated model\cite{bigMD:DEShaw:longubiquitin}; on \num{512} nodes this achieved a performance of \SI{15}{\micro\second\per\day}.
On general-purpose high-performance computing clusters, one of the largest all-atom MD simulations to date was of the HIV-1 capsid\cite{bigMD:largeAllAtom}, with 64 million atoms in the solvated system. Obtaining dynamic trajectories of \SI{0.1}{\us} required two weeks of computing time on \num{128000} cores, corresponding to an average of \SI{7}{\ns\per\day}. A recent large-scale simulation of the influenza A virion with 5 million particles achieved a \SI{5}{\us} trajectory\cite{bigMD:largeCG} (no performance details published).
These systems span length scales of up to a few hundred nanometres. We note that such simulations are very expensive and cannot be repeated routinely.

The human drive to study larger and larger systems far outpaces the development of high-performance computing. Systems of interest in soft matter typically span micrometres, contain hundreds of millions of atoms and show behaviour on much longer time scales than is accessible in routine simulations\cite{LargeScaleMD}.
Even with modern supercomputers, we are far away from being able to simulate systems such as an entire living cell.%

\subsection*{Limitations of all-atom simulations}

Accessing large systems remains a computational challenge. The number of particles in the simulation grows with the volume and hence cubically with the length scale.
Moreover, with an increase in the length scale the \emph{time} scale of the relevant motions also increases: On larger scales, it takes longer for the effect of a perturbation to ripple across the system. Furthermore, entropic effects become more important; these are generally softer and hence result in slower motions (consider, for example, the large-scale fluctuations of a lipid membrane \vs the small-scale fluctuations of individual lipid molecules).
This means that not only do we have to simulate more particles, but we also have to carry out the simulations for longer.
The reach of all-atom simulations is limited by the available computational resources.
The differential equations governing the motions of the atoms must be discretised, and the integration time step needs to be small enough to represent the fastest motion in the system (usually vibrations of bonds involving hydrogen atoms, if present in the system, due to their small mass; this is generally the case in biomolecular systems).
For each time step, we need to calculate the forces on all the atoms based on their current positions. This is the bottleneck in molecular simulations.

\section{Coarse-Graining}

\subsection*{Accessing larger scales}

In many cases we do not care about the fluctuations on the scale of individual atoms, but are interested in the behaviour of a system on a larger scale. In these cases we can extend the reach of our simulations to larger systems and longer time scales by employing lower-resolution models.
This can be achieved by grouping together several atoms and representing them as single particles. This process is referred to as ``coarse-graining'' (CG), and the particles are usually referred to as ``CG sites''\cite{voth:cgbook}.

We note that speeding up simulations is not the only application of coarse-graining. Another motivation is being able to only keep those degrees of freedom which (we think) are relevant for certain processes or behaviours, and to tweak the interactions at that scale, thereby allowing us to study the effects of eliminating degrees of freedom or changing the interactions at different scales.

Coarse-graining can take place on a wide range of scales. On the smallest scale, there are ``united-atom'' models in which heavy atoms are grouped with their bonded hydrogen atoms and described by new effective atom types, while all other atoms keep their identity. On the largest scale, a single CG site can represent several residues or molecules at once\cite{DNAsuperlowres1,MARTINICG:4water}.

\subsection*{Why coarse-graining speeds up simulations}%

Coarse-graining of a system leads to faster simulations for several reasons\cite{CGpower}:

First, coarse-graining reduces the number of particles that have to be simulated. Hence, the number of interactions that have to be computed can be much smaller, decreasing the computational effort required for each time step.
This is an especially prominent advantage if we are only interested in the behaviour of a solute, such as a protein solvated by water. For this we can consider a ``solvent-free'' CG model in which we do not include the solvent molecules: their influence is only captured implicitly through the effective coarse-grained interactions between solutes, vastly reducing the number of interactions that need to be calculated in each time step.
Moreover, longer-range interactions such as electrostatics are usually screened by the solvent, and hence the effective interactions between CG sites can have a much shorter range, further reducing the number of interactions that need to be calculated.

Second, the small-scale fluctuations on the atomic scale are no longer represented in the coarse-grained model, and high-frequency motions, such as covalent bond vibrations, are filtered out. Hence, CG simulations can employ a larger time step.
Third, fluctuations on the atomistic scale give rise to friction; for example, the atoms of two molecules may ``interlock'' and be unable to slide past each other immediately. Hence the CG model, without these fluctuations, evolves faster in time than the all-atom model. However, this actually changes and distorts the dynamic behaviour of the system. In general, one would like to retain this friction and its effect on the dynamics. (We note that in some cases, such as searching for stable states, the distortion of the dynamics may be irrelevant.) How to obtain accurate dynamics in a CG model is an area of ongoing research; this is the challenge of ``real dynamics''.

\myvspace
Coarse-graining is especially useful if models at different scales can be integrated\cite{ScaleBridging}. An example of this would be a simulation of the active binding site of a protein with atomistic resolution, while the rest of the system is modelled at a much coarser level.%
Like any simulation, coarse-graining has to be \emph{accurate} to be useful. The CG simulation needs to reproduce the relevant properties of the system. This is a relative measure, and depends on the properties in which one is interested.

\subsection*{Limitations of coarse-graining}

Coarse-graining faces three major challenges:

\paragraph{Transferability.}
Coarse-graining can be considered an average over fluctuations on the atomic scale, arriving at an effective CG interaction (this will be discussed in more detail in \cref{ch:cg}). This makes the CG interactions dependent on the state point (temperature, density, pressure). CG interactions parametrised for one state point cannot be assumed to reproduce the properties of the system at another, different state point. Hence, a given CG model is specific to the state point for which its interactions have been parametrised, and it cannot be assumed to be applicable in other contexts. The transferability of a CG model across state points needs to be confirmed empirically\cite{CGtransferability} or modelled explicitly in the interactions between CG sites\cite{MSCG:transferringTemperature,MSCG:densityDependent,CGtransferabilityExtendedEnsemble}.

\paragraph{Real dynamics.} As noted previously, fluctuations on the atomic scale create friction between atoms and molecules. By coarse-graining, we average over these fluctuations. This means that the CG particles experience less friction, and motions in the CG model occur in less simulation time than in the all-atom model (corresponding to a scaling of the time axis).
This could be seen as an advantage, as it provides even larger speedups.
However, the faster time evolution is an artefact of artificial dynamics. Even if the CG model accurately reproduces structural distributions, the relation between ``CG time'' and real time is generally unknown, and cannot be assumed to be linear.
To get an accurate coarse-grained simulation of the \emph{dynamics} of a system, we have to account for the missing friction and for the impact of small-scale fluctuations by explicitly including drag and stochastic forces in the CG model\cite{voth:RealDynamics}.

\paragraph{Representability.}
To date, coarse-grained models generally assume that the CG interactions can be separated into a sum of simple terms, similar to those of the molecular mechanics force fields used in all-atom simulations. Specifically, interactions between non-bonded CG sites are commonly assumed to be pairwise additive and radial (isotropic) ``pair potentials''.
Based on this assumption, there are many different approaches that aim to derive effective pair potentials for a CG model, optimising for different properties of the system, such as structures, forces, or thermodynamics.

However, the interactions in the CG model are \emph{effective} interactions resulting from an average over many atom--atom interactions. This leads to correlations between CG sites that, in general, cannot be decomposed into pairwise interactions.\footnote{In fact, even interactions between atoms can never be exactly represented by pair potentials\cite{AAL:representability1}.}
In \cref{ch:cg} we will discuss the connection between all-atom and coarse-grained models within a statistical mechanics framework. This leads to a definition of ``consistency'' for CG models based on matching probability distributions of the atomistic ensemble. A consistent CG model will reproduce all ensemble averages of an all-atom model that can be described in terms of the CG coordinates.
To obtain a consistent CG model for $N$ CG sites we need, in principle, an $N$-body potential to describe the CG interactions.
It has been shown that it is impossible to reproduce both structural and thermodynamic properties of a system using only pair potentials\cite{AAL:representability2}.

\myvspace
Hence, CG models that rely on pair potentials necessarily trade off structural against thermodynamic accuracy. This is known as the representability problem: pair potentials cannot represent the ``true'' CG interactions. This issue will be investigated in this thesis, with the aim of overcoming this challenge. %

\section{Many-Body Potentials}

A full $N$-body potential is generally infeasible because its computational cost scales exponentially with the number of sites.
To obtain interactions that can be used in practice (and to realise CG simulations that are faster than the all-atom model), we need to approximate the $N$-body potential by separating it into a sum of simpler interaction terms.
This leads to a trade-off between simplicity of the terms and accuracy of the resulting overall potential.

Whereas a coarse-grained model based on pair potentials suffers from the representability problem and is limited in the accuracy it can provide, this could be alleviated by considering more complex interactions that go beyond pair potentials.
Both all-atom and CG models typically include angle and dihedral terms, which are three- and four-body interactions, respectively. However, these terms are usually limited to bonded short-range interactions with simple functional forms. They do not constitute a general many-body potential. In this thesis we focus on the non-bonded interactions that, so far, are predominantly described by pair potentials alone.

In the context of coarse-graining, many-body terms for non-bonded interactions have generally been neglected. They are often considered to be too computationally expensive and, as the primary aim of coarse-graining is a speedup of the simulation, only the simplest and cheapest interactions tend to be considered.
Thus there has been only limited previous research into non-bonded many-body interactions in coarse-graining. There have been a few studies using three-body terms \cite{voth:threebody,MSCG:3body}; however, these studies only considered one-site models of homogeneous liquids. 
Moreover, interaction terms are generally assumed to be a function of independent sites: when a molecule is described by multiple sites, this neglects the correlation between the sites describing a single molecule.

\myvspace
In this thesis we consider coarse-grained models for molecules that retain some internal structure and are described by more than one CG site, and we base the coarse-grained interactions on \emph{molecular} terms that take into account all CG sites belonging to a molecule. This leads to more complex interactions that cannot be described by simple functional forms. To capture the interactions, we transfer the Gaussian Approximation Potential (GAP) approach to coarse-graining. Originally developed for atomistic modelling, GAP\cite{GAP:thesis,GAP:PRL,GAP:tutorial} is a very general approach that is based on Gaussian process regression\cite{MacKay:itprnn,GaussianProcessesMachineLearning} to fit a flexible interaction potential to observations of energies and forces, without imposing a specific functional form on the potential.

We will show that we can successfully employ GAP to generate CG interactions based on forces sampled from the atomistic model, whilst still reproducing its full structure (as measured by a variety of structural distributions and correlations). This demonstrates that we can indeed create \emph{consistent} CG models.
Although GAP models are computationally expensive, for solvent-free CG models, which are relevant in many biomolecular simulations, these models can still provide a significant speedup compared to all-atom simulations, while being much more accurate than current approaches could hope to be.

\clearpage
\section{Outline}

In the first part of this thesis, we give an overview of classical molecular dynamics with a focus on all-atom models (\cref{ch:md}) and review current approaches to coarse-graining (\cref{ch:cg}).
In \cref{ch:gap} we introduce Gaussian process regression (\cref{sec:gp:regression}) and discuss the locality approximation and the descriptors that take into account our knowledge about the properties of physical systems and enable the Gaussian Approximation Potential (GAP) to provide effective and flexible many-body interactions (\cref{sec:gap}). In \cref{sec:gap:cgspecific} we show how GAP can be used as a new, many-body approach to coarse-graining (GAP-CG). Inferring interactions only from force observations is a novel application of GAP, and we develop solutions to the challenges this brings up.
In \cref{ch:gapcg} we present the application of GAP-CG to methanol clusters and bulk systems of methanol and benzene. We show that the flexible molecular terms that are available in the GAP approach can capture atomistically consistent CG interactions, allowing a representation of both complex structural distributions \emph{and} forces, and hence thermodynamic properties. Furthermore, we study how the accuracy of force predictions depends on the hyperparameters.
In \cref{ch:waterbenzene} we discuss solvent-free CG models and demonstrate for the example of water-solvated benzene that GAP-CG can have clear computational benefits while being significantly more accurate than current approaches to coarse-graining.
In \cref{ch:conclusion} we discuss our results, put them into the context of performance and scaling of CG models, and give an outlook to future work.

\chapter{Classical Molecular Dynamics}
\label{ch:md}

The atomistic and coarse-grained simulations we consider in this thesis are based on Molecular Dynamics (MD). %
In MD, we model our system in terms of discrete (typically structureless, point-like\footnote{In principle, simulations can also be carried out with anisotropic particles or rigid bodies, but this leads to more complicated equations, and point-like particles are most commonly used.}) particles.

Each particle, indexed by the subscript $i$, has a mass\footnote{In atomistic simulations, this is usually the average mass across the isotopes of the atom.} $m_i$ and is described by its position $\vec{r}_i$ and velocity $\vec{v}_i = \inlinederiv{\vec{r}_i}{t}$.
The time evolution of positions and velocities is given by the equations of motion, which are integrated numerically; this will be discussed in \cref{sec:md:equationsofmotion}. We thereby obtain a dynamic trajectory of the system, allowing calculations of both structural and dynamic properties.

The behaviour of the system depends on the interactions between the particles, defined in terms of the forces acting on them. The collection of the interaction terms is called the ``force field'' and will be described in \cref{sec:md:forcefield}.

In practice, many systems of interest are not isolated, but are in thermal equilibrium with the environment. In statistical mechanics, this is described by the canonical ensemble. To sample the canonical ensemble, the simulation needs to include a thermostat; this will be discussed in \cref{sec:md:thermostat}.

In thermal equilibrium, the system exchanges energy with is environment, and hence the internal energy (kinetic + potential energy) is no longer a constant of motion. Instead, the \emph{free} energy becomes important to characterise the state of the system. Generally, we are interested in the (conditional) free energy as a function of one or more ``collective variables'' that characterise the configuration of the system on a larger scale; this is also referred to as the Potential of Mean Force (PMF). The negative gradients of the conditional free energy with respect to collective variables are \emph{mean} forces. Free energy/PMF, collective variables and mean forces will be discussed in \cref{sec:md:freeenergy}.

\section{Equations of Motion}
\label{sec:md:equationsofmotion}

In atomistic simulations, each atom is modelled by a point-like particle at the position of the nucleus. The electrons are only modelled implicitly through the interactions they mediate between nuclei. Due to the large masses of the nuclei, quantum effects can generally be neglected, and their motion can be described by classical mechanics. This also holds for coarse-grained simulations, where the particle masses are even larger.
The dynamics are then governed by Newton's laws of motion. The acceleration $\vec{a}_i$ of particle $i$ %
is given by
\begin{equation}
  m_i \vec{a}_i = m_i \deriv{\vec{v}_i}{t} = m_i \deriv[2]{\vec{r}_i}{t} = \vec{f}_i \big( \{\vecr_j\} \big),
  \label{eq:md:newton}
\end{equation}
where $\vec{f}_i$ is the force acting on the particle. In general, the force can be a function of the positions of all particles. When the forces are known, this differential equation can be integrated to obtain the positions of all the atoms as a function of time.

In general, the integration needs to be done numerically, by discretising \cref{eq:md:newton}. The most commonly used symplectic (phase-space conserving) integrator is the Velocity Verlet algorithm\cite{velocityVerlet}. The new position and velocity after a time step $\Delta t$ are calculated as
\begin{equation}
  \begin{split}
    \vec{r}(t + \Delta t) &= \vec{r}(t) + \vec{v}(t)\, \Delta t + \frac{\vec{a}(t)}{2} \, (\Delta t)^2 , \\
    \vec{v}(t + \Delta t) &= \vec{v}(t) + \frac{\vec{a}(t) + \vec{a}(t + \Delta t)}{2} \, \Delta t ,
  \end{split}
  \label{eq:md:verlet}
\end{equation}
where for clarity we suppressed the index $i$.%

The time step needs to be small enough to be able to describe the fastest motions in the system. Otherwise, discretisation errors would cause instabilities.
In atomistic models, the fastest motion is generally the vibration of bonds with hydrogen atoms, if present in the system, due to their small mass. This requires a time step $\Delta t \lessapprox \SI{1}{\fs}$\cite{AllenTildesley}.

\Cref{eq:md:verlet} describes Hamiltonian dynamics. When the forces are conservative, this corresponds to a constant-energy simulation of the microcanonical ensemble (also called the $NVE$ ensemble, as the number of particles $N$, the volume $V$, and the energy $E$ are fixed). To simulate systems at a constant temperature $T$, a thermostat algorithm (see \cref{sec:md:thermostat}) can be used to amend these equations to sample the canonical (or $NVT$) ensemble, though their general structure remains the same.

\section{Force Fields}
\label{sec:md:forcefield}

The equations of motion~\eqref{eq:md:verlet} depend on the forces acting on all the particles. These are determined by the force field, which gives the forces as a function of the configuration of the system. We generally assume conservative forces, which can be written as the negative gradient of the potential energy $U$,
\[
  \vec{f}_i = - \nabla_i U\big(\{\vecr_j\}\big) ,
\]
where $\nabla_i$ is the gradient operator with respect to the position $\vecr_i$ of particle $i$. %
This applies to both all-atom and coarse-grained models.
We now focus on all-atom simulations; the interactions in coarse-grained models will be discussed in \cref{ch:cg}.

In principle, the potential $U$ can be a general function of the coordinates of all the atoms in the system.
For example, this could be the quantum-mechanical (QM) total energy, obtained from (approximately) solving the QM Hamiltonian (for instance, using Density Functional Theory, Quantum Monte Carlo, or quantum chemistry methods such as Coupled Cluster Theory).

For simulations of larger systems such as biomolecules or complex materials, this would not be computationally feasible. Hence, it is often assumed that the potential energy of the system can be separated into a sum of individual terms that only depend on the positions of small subsets of the atoms. Moreover, these terms commonly have fixed functional forms describing the various interactions. %
This simplification is generally called ``molecular mechanics'' (MM).\footnote{It is also possible to treat a selected small region quantum-mechanically, and simulate a combined QM/MM model.}

In typical molecular mechanics force fields\cite{AMBERff,ff99SB,forcefieldOPLS,forcefieldCHARMM,forcefieldGROMOS,forcefieldGROMOS2011}, the potential energy is made up of bonded (covalent) and non-bonded terms,
$
  U = U_\text{bonded} + U_\text{non-bonded} .
$

The bonded part usually consists of terms involving the bond lengths $b_i$ between two atoms, the bond angles $\theta_j$ between three atoms, and the dihedral angles $\chi_k$ between four atoms. A representative functional form is
\begin{equation*}
  U_\text{bonded} = \! \sum_i^\text{bonds} \! \half K^b_i (b_i - b_{0,i})^2
      + \! \sum_j^\text{angles} \! \half K^\theta_j (\theta_j - \theta_{0,j})^2
      + \! \sum_k^\text{dihedrals} \! K^\chi_k \big[1 + \cos(n_k\chi_k - \chi_{0,k})\big] ,
\end{equation*}
where $K^b,K^\theta,K^\chi$ and $b_0,\theta_0,\chi_0,n$ are the parameters defining the interactions.

The bond and angle terms are commonly described by harmonic ``springs''. Terms involving dihedral angles can describe both actual dihedral angles formed by three consecutive bonds as well as so-called ``improper'' dihedral angles that do not correspond to consecutive bonds but are included to fix the structure, for example, to ensure an aromatic ring remains planar.

The non-bonded part usually consists of electrostatic (Coulomb) and van-der-Waals (dispersion) interactions between pairs of atoms:
\begin{equation*}
  U_\text{non-bonded} = \sum^\text{atoms}_{\substack{i,j \\ i < j}} \frac{q_i q_j}{4\pi\epsilon r_{ij}}
  + \sum^\text{atoms}_{\substack{i,j \\ i < j}} \left(\!\frac{A_{ij}}{r_{ij}^{12}} - \frac{B_{ij}}{r_{ij}^6}\right) \! ,
\end{equation*}
where $r_{ij}$ is the distance between two atoms, $q_{i}$ are partial (point) charges, $\epsilon$ is the permittivity, and $A_{ij}$ and $B_{ij}$ are parameters of the Lennard--Jones potential that describes the van-der-Waals interactions.
(In recent years there have been extensions to polarisable force fields\cite{PolarizableFF1,PolarizableFF2,PolarizableFF} and reactive force fields\cite{ReactiveFF:EVB1,ReactiveFF:ReaxFF,ReactiveFF:RWFF}, but these are beyond the scope of this thesis.)

The parameters such as spring force constants, Lennard--Jones coefficients, and so forth are determined by a combination of fitting to results from quantum-mechanical calculations, experimental data (\eg structural data derived from X-ray diffraction), and further \textit{ad-hoc} adjustments to reproduce specific experimental properties\cite{generalMDreference1}.

Even though the molecular mechanics potential is a poor approximation of the true quantum-mechanical energy,
this approach has been used successfully in vast numbers of simulations of materials and biomolecular systems from small molecules up to proteins, nucleic acids, and lipid membranes\cite{MMexampleProtein1,MMexampleLipidPeptide,MMexampleLipids}.%

\myvspace
Typically, we will be interested in simulations of bulk systems.
To get rid of unwanted surface effects, periodic boundary conditions are imposed on the simulation box. The system size needs to be sufficiently large so that the interactions of the periodic mirror images of a particle with themselves do not play a significant role.

The bonded interactions, by their nature, generally involve only a small number of groups of atoms, and their contribution to the total computational cost of force evaluations is typically negligible. In contrast, the evaluation of the non-bonded interactions takes most of the computing time. Whereas the Lennard--Jones potential is usually cut off at around \SI{10}{\angstrom} without sacrificing accuracy, the Coulomb interaction only decays with the inverse of the distance and cannot be cut off at short range. Algorithms such as the Particle Mesh Ewald method\cite{ParticleMeshEwald} make it feasible to calculate electrostatic interactions in periodic simulation boxes. However, %
the evaluation of the non-bonded forces is still the bottleneck for many all-atom simulations.

Often, we are interested in the behaviour of a small number of solute molecules within a large volume of solvent. For example, to simulate biomolecules in their natural environment, we need to include a description of the water surrounding them. This can be achieved most accurately by including the water molecules and dissolved ions explicitly.
In such models, the solvent molecules typically contribute about \SI{90}{\percent} of all atoms, and most of the computing resources are spent on calculating the solvent--solvent non-bonded interactions.

However, a dipolar solvent such as water and the dissolved ions will strongly screen even the electrostatic interaction between charged solutes, and the \emph{effective} interactions in bulk simulations generally have a finite, comparatively short range. In physiological conditions, this effective range is around one nanometre, based on the Debye--H\"uckel screening length for typical ion concentrations\cite{israelachvili2011intermolecular}.
This motivates the use of implicit solvent models in which the solvent contribution to solute--solute interactions is captured in effective interactions. In typical implicit solvent approaches for all-atom models, the charge-screening effects of water are modelled by an infinite continuum medium\cite{ImplicitSolventOverview}. However, this does not capture hydrophobic effects properly\cite{chandler2005hydrophobicity,ImplicitSolventFails}. %

\section{Constant-Temperature Simulations and Thermostats}
\label{sec:md:thermostat}

As discussed previously, evolving a system in time using \cref{eq:md:verlet} with a conservative force field yields a constant-energy simulation that samples the microcanonical ensemble.
This would be appropriate for a system that is completely isolated.

In practice, it would be very difficult to completely isolate a system from its environment, and systems of interest typically exchange heat with their surroundings.
We can model this by introducing a heat bath connected to the system. The system exchanges energy with the heat bath, and thereby maintains a constant \emph{temperature} $T$. This is described by the canonical ($NVT$) ensemble. To describe a system at constant temperature, the simulation needs to include a thermostat.

A thorough overview and comparison of thermostat algorithms is given by H\"unenberger\cite{Thermostat-review}.
The two most common approaches are the Nos\'e--Hoover thermostat and Langevin dynamics. %

The Nos\'e--Hoover thermostat is a fully deterministic approach. It introduces an additional, artificial degree of freedom to the system which represents the heat bath.
However, this thermostat has issues related to ergodicity\cite{Legoll2007,Leimkuhler2009},
and it can lead to distorted dynamics, especially in small systems.%

Langevin dynamics is fully ergodic and samples the canonical ensemble properly, and hence this is the approach we adopt in the present work. It is a stochastic approach that modifies the equations of motion~\eqref{eq:md:newton} by adding a friction (damping) term, proportional to the velocities of the atoms, as well as a stochastic noise term (based on the fluctuation--dissipation theorem\cite{FluctuationDissipationTheorem}):
\[
  m \vec{a} = \vec{f} - \gamma \vec{v} + \sqrt{2 \gamma \kT} \, \vec{\eta}(t) ,
\]
where $\gamma$ is the friction coefficient (typically scaled with the particle mass), $\kB$ is the Boltzmann constant, and $\vec{\eta}(t)$ describes white noise [with zero mean, $\average{\vec{\eta}(t)} = 0$, and correlation $\average{\eta_\mu(t) \eta_\nu(t')} = \delta_{\mu\nu} \delta(t-t')$, where $\delta(t-t')$ is the Dirac delta].
In the limit of $\gamma\to 0$, we recover Newtonian MD. The value of $\gamma$ needs to be chosen large enough for the thermostat to work effectively and small enough not to overdamp low-frequency modes; a reasonable choice is $\gamma = m \times \SI{5}{\ps^{-1}}$, where the friction coefficient is proportional to the mass $m$ of each particle\cite{LangevinGamma}.

A similar approach which also considers the history of the system and correlations in the noise can be used to obtain ``real dynamics'' in CG models, where parameters like $\gamma$ and the noise correlations are derived from underlying atomistic simulations\cite{voth:RealDynamics}.

\section{Free Energy}
\label{sec:md:freeenergy}

When a system is kept at a constant temperature $T$ by letting it exchange energy with a heat bath, its total internal energy $E$ (kinetic + potential energy) is no longer a conserved quantity. Instead, a crucial role is played by the \emph{free} energy $A$, related to the total energy through $ A = E - T S $, where $S$ is the entropy. The dynamics of the system act to minimise the free energy such that, in thermodynamic equilibrium, the free energy takes its minimum value.

Statistical mechanics defines the free energy at a microscopic level as
\[
  A = -\kT \log z ,
\]
where $z$ is the canonical partition function. For a system of $N$ classical point particles with positions $\vec{r}^N$ and momenta $\vec{p}^N$, the partition function is given by
\[
  z = \frac{1}{N!h^{3N}} \intdd{^{N}\vec{p}} \dr[^{N}\vec{r}] \expp{-\beta H(\vec{r}^N,\,\vec{p}^N)},
\]
where $H$ is the Hamiltonian and $\beta = 1/\kT$.
With the Hamiltonian separating into a sum of kinetic and potential energy, the integral over the momenta can be carried out explicitly. For identical particles, this results in
\[
  z = \frac{z_N}{N! \Lambda^{3N}} \, ,
\]
where $z_N$ is the classical configuration integral,
\begin{equation}
  z_N = \intdd{^N\vecr} \expp{-\beta U(\vecr^N)} ,
  \label{eq:md:configurationintegral}
\end{equation}
and $\Lambda = h / \sqrt{2\pi m \kT}$ is the thermal de Broglie wave length.
For a given temperature, the contribution of the momentum integral to the free energy is constant. Therefore, we generally focus on the configurational part, and in the following we will suppress the subscript $N$ and refer to configurational integral and partition function interchangeably. %

\subsection{Collective Variables and the Potential of Mean Force}
\label{sec:md:collectivevariables}
\label{sec:md:pmf}

The integral~\eqref{eq:md:configurationintegral} cannot be evaluated analytically except for the simplest systems. All practical approaches rely on numerical methods.
While there have been attempts to calculate the \emph{absolute} value of the free energy\cite{AbsoluteEntropy},
in practice this is rarely feasible, and usually we only care about the free energy differences between states. This can be due to different types of atoms, corresponding to a ``transmutation'' or alchemical transformation of one molecule into another\cite{AlchemicalFreeEnergy}, or due to a transition along a reaction coordinate\cite{FreeEnergyCalculationsBook}. For this reason we are often interested in the free energy as a function of parameters that can characterise different states of interest on a larger scale than that of individual atoms.

These parameters are often referred to as Collective Variables (CVs); they are a function of the atomic configuration, and hence denoted as $\xi(\vecr^N)$. A simple example would be the distance between a selected pair of atoms, $\xi(\vecr^N) = \norm{\vecr_j-\vecr_i}$. However, a collective variable does not have to be a geometric parameter; it could be any function of the atomic coordinates, such as a coordination number or the potential energy\cite{PMFlib}.

For a given collective variable $\xi$, we can define a ``conditional'' free energy,
\begin{equation}
  A(\xi) = -\kT \log z(\xi) ,
  \label{eq:md:pmf}
\end{equation}
where
\[
  z(\xi') = \intdd{^N\vecr} \expp{-\beta U(\vecr^N)} \delta\big( \xi' - \xi(\vecr^N) \big)
\]
is the conditional partition function, corresponding to the configurational integral over all degrees of freedom orthogonal to the collective variable $\xi$.
The probability for the system to be in a state with a certain value of the collective variable $\xi$ is given by
\[
  p(\xi) = \frac{z(\xi)}{z} = \frac{1}{z} \exp\!\big[\! - \beta A(\xi) \big],
\]
in which the total partition function $z = \intdd{\xi} z(\xi)$ acts as a normalisation factor.
These definitions straightforwardly generalise to a multidimensional set of CVs, $\vec{\xi}(\vec{r}^N) = \big\{ \xi_i (\vec{r}^N) \big\}$.

The conditional free energy is commonly referred to as the Potential of Mean Force (PMF): its negative gradient, $-\nabla_\xi A(\xi)$, describes the mean force (as an average over atomic forces) acting on a collective variable. %
The PMF describes the free energy surface as a function of the CVs, which can allow us to gain a better understanding of the stable states of a system and the transition pathways between them. A common use-case for this is a Ramachandran plot\cite{Ramachandran} of the dihedral backbone angles in peptides which indicates equilibrium configurations and transitions from one to another. For this reason, di- and trialanine are often used in benchmarking of free energy methods\cite{stecher2014freeenergygp,LetifsPaper}.

\myvspace
The free energy surface or PMF of a system can be computed by a variety of methods. The most popular methods are the gradient-based Umbrella Integration\cite{UmbrellaSamplingReview,UmbrellaIntegration}, the Blue Moon method\cite{BlueMoon,BlueMoon2}, the Adaptive Biasing Force (ABF) method\cite{ABF,ABF2}, and the histogram-based Metadynamics\cite{MetadynamicsFirst,MetadynamicsPRL,MetadynamicsReview}.
Gradient-based methods typically estimate the mean forces on a grid and then integrate these results to obtain the PMF.  This means that their computational cost scales exponentially with the number of variables; hence, they are usually only applied to up to a few CVs at once.
Histogram-based methods directly estimate the probability density function of the CVs (or its logarithm) by accumulating the visited CV values. In Metadynamics, this is based on a history-dependent biasing potential built up of Gaussian peaks that reflects the free energy surface; it can access somewhat larger sets of CVs, though this is typically still limited to no more than six CVs. In a recent study, related to the work in this thesis, ABF and Metadynamics were combined with Gaussian process regression to carry out grid-free multidimensional integration of derivative observations of the free energy\cite{LetifsPaper}. We will review Gaussian process regression in \cref{ch:gap}.
In the following section we describe how we can obtain mean forces.

\subsection{Mean Forces}

The mean forces (MF) for a given configuration of collective variables can be obtained directly in restrained or constrained MD simulations.

In a restrained MD simulation, a restraint potential $U_\text{rst}$ for the collective variable is added to the potential energy of the system. This restraint potential is typically harmonic:
\[
  U_\text{rst}(\vecr^N)  = \half k_\xi \big(\xi(\vecr^N) - \xi_0\big)^2,
\]
where $\xi_0$ is the centre of the restraint and $k_\xi$ is the restraint force constant. This keeps the value of the CV confined around the position of the restraint. (Confusingly, this was historically referred to as a ``harmonic constraint''.) We then perform a MD simulation to obtain the thermal average of the CV, $\average{\xi}$, in the presence of the restraint force. This is related to the MF acting on the collective variable $\xi$ by
\[
  - \pderiv{A}{\xi} \big(\average{\xi}\big) = k \big( \average{\xi} - \xi_0 \big) .
\]
Note that this calculates the MF at the \emph{average} (biased) value of the CV, and not at the minimum $\xi_0$ of the restraint potential.%
\footnote{It is also possible to calculate the mean force at $\xi_0$\cite{UmbrellaIntegration}, but the ``unbiasing'' from $\average{\xi}$ to $\xi_0$ leads to a much larger error in the value of the mean force\cite{UmbrellaIntegrationError}, and in preliminary investigations we found this unsuitable for our purposes.}
As the average value of the CV is shifted proportionally to the magnitude of the mean force, it is difficult to obtain accurate mean forces using restrained MD for unfavourable configurations of CVs. We could reduce $\average{\xi}-\xi_0$ by increasing the strength of the restraint potential, but this would require smaller time steps and increase the error in the measurement.

In a constrained MD simulation, the values of the collective variables are held exactly constant. (This formally corresponds to an ``infinite'' restraint force.)  A constraint solver calculates the Lagrange multipliers that ensure that in each integration step the values of the CVs remain unchanged, resulting in instantaneous ``constraint forces''.
To derive the mean forces from the constraint forces requires, in general, a Jacobian correction\cite{ConstraintForce}. For Cartesian CVs, the correction vanishes and the mean forces can be calculated as the negative of the time average of the constraint forces in a constrained MD simulation\cite{UnderstandingMolSim}. This is the method on which we rely in this thesis.

Alternatively, it is possible to obtain noisy samples of the free energy gradient by calculating the so-called Instantaneous Collective Forces (ICF) acting on the CVs. We will discuss this in more detail in \cref{sec:cg:forcebased}.

\section{Summary}

In this chapter we discussed the principles underlying classical molecular dynamics in the context of all-atom simulations. We introduced the free energy that characterises the state of a system at constant temperature, relevant to typical experimental conditions.
Often, we are interested in the properties of a system not as a function of the atomic positions, but as a function of ``global'' parameters, depending on the atomic positions, that may be more appropriate for describing the configuration on a larger scale. This motivated the introduction of the conditional free energy/potential of mean force as a function of collective variables. %
In the next chapter, we make the connection between coarse-graining and the free energy by considering the coordinates of the CG model as collective variables of the atomistic model. %

\chapter{Coarse-Graining Approaches}
\label{ch:cg}
As discussed in the introduction, coarse-graining (CG) aims to increase the length and time scales that can be reached in molecular simulations by describing the system at a lower resolution than individual atoms. This reduces the number of interactions that need to be calculated and allows for a larger integration time step, and can thereby speed up the simulation. %

The coarse-graining process is necessarily an approximation, and as such it will always require a compromise between various concerns such as speed and use of computational resources, accuracy with respect to different properties (depending on what the modeller is interested in), and transferability between different systems and state points.

The construction of a CG model consists of two tasks. First, we need to define the CG sites and their topology\footnote{The topology describes the structure of the molecules; it defines the bonded interactions between sites. Two CG sites are generally considered ``bonded'' when there are covalent bonds connecting atoms from both sites.}.
Second, we need to specify the interactions between CG sites.
The definition of CG sites depends closely on the system of interest and the questions that the model should answer. There is a wide range of options for how to define the sites;
a CG site can represent a single atom, a ``united atom'' that describes a heavy atom together with all the hydrogen atoms bound to it, a small group of atoms, a single side chain, a residue or whole molecule, or even several molecules at once.
The CG mapping influences the properties that can be reproduced by the CG simulation, and hence this forms an important part of the modelling process. So far, there has been only limited research into a systematic determination of CG sites\cite{cgmapping:optimalnumber,cgmapping:minimizingmemory,NoidShell:CGinformation}, and CG sites are predominantly chosen by physical/chemical intuition.
The choice of CG sites depends on the level of description we seek, and should consider several factors\cite{noid:cgbio:review,vanGunsterenReview}:
First, the phenomena in which we are interested need to be representable by the coarse-grained description, and the CG sites should be able to capture the slow, large-amplitude motions in the system.
Second, to speed up the simulations, the choice of CG sites should eliminate unnecessary detail and filter out the fast, high-frequency but low-amplitude motions.
Moreover, the choice of CG sites impacts how well the effective coarse-grained interactions are able to reproduce properties of the atomistic system. There should be a one-to-one mapping between distinct molecular states in the all-atom model and in the CG description\cite{NoidRudzinskiMappings}, and coarse-graining will produce more accurate results when the CG degrees of freedom are less correlated\cite{NoidShell:CGinformation}. (Hence, increasing the resolution of the CG mapping may actually lead to reduced accuracy\cite{MSCG:X,CGtransferabilityExtendedEnsemble}.)

In this thesis, we will assume the definition of CG sites as given. We focus on the second task in creating a CG model, that is, specifying the interactions between CG sites. Whereas in all-atom modelling, empirical force fields are now being routinely applied in a production setting, for each new CG model the CG interactions need to be parametrised again (as they depend on the choice of sites and on the state point). How to generate CG interactions is still a topic of active research.

There are many different approaches to coarse-graining. They can generally be classified into more phenomenological ``top-down'' models and ``bottom-up'' models that are based on an underlying all-atom model. This distinction will be discussed in \cref{sec:cg:topdownvsbottomup}.
Bottom-up models rely on a mapping from the all-atom to the coarse-grained description; we will introduce this mapping in \cref{sec:cg:mapping}.
The mapping between all-atom and CG model allows us to consider the \emph{consistency} of the CG model in terms of probability distributions.
In \cref{sec:cg:consistency} we will discuss this and see how the Potential of Mean Force (PMF) for the coarse-grained coordinates (\cf \cref{sec:md:pmf}) provides a CG potential that exactly reproduces all structural and thermodynamic properties of the underlying all-atom model\cite{RudzinskiNoidEntropyForcesStructure,NoidShell:CGinformation}.
However, using the many-body PMF as the interaction potential for the CG model is generally not feasible.

In practice, coarse-graining approaches rely on simpler potentials that describe effective interactions, optimised to reproduce a subset of the properties of the underlying system.
In the second half of this chapter we review and compare the current approaches to determining effective coarse-grained interactions that can be found in the literature.
We discuss structure-based approaches in \cref{sec:cg:structurebased} and force-based approaches in \cref{sec:cg:forcebased}. We conclude with a summary in \cref{sec:cg:summary}.

\section{Top-Down \vs Bottom-Up Models}
\label{sec:cg:topdownvsbottomup}

CG models are commonly classified into whether they follow a ``top-down'' or a ``bottom-up'' approach.
Here we outline and compare these two coarse-graining philosophies.
A more extensive discussion (including other, more specialised approaches such as ``knowledge-based'' network models) can be found, for example, in \textcite{noid:cgbio:review}.

\subsection*{Top-down models}

In the top-down approach, interaction terms between CG sites are proposed based on physical principles (with no connection to an underlying atomistic model). Typically, the interactions are given by fixed, analytical functional forms. Their parameters can be optimised to reproduce quantities that are measured in experiments, such as thermodynamic properties, or are simply specified based on physical intuition.
Top-down models can be considered ``fine-grained'' models, parametrised on (experimental) data from larger scales. %

Top-down models generally focus on reproducing thermodynamic and other large-scale properties.
They are often seen as a ``minimal model'' to elucidate what types of interactions need to be considered to reproduce specific properties and emergent behaviours in a system.
(However, different underlying atomistic interactions can lead to the same patterns on a larger scale, so top-down models cannot explain how or which atomistic properties lead to these patterns.) %

Top-down models have been successfully used in practice for a wide range of systems\cite{CGpower}. They are well suited for studies of large changes, such as DNA duplex formation\cite{Ouldridge:thesis,Ouldridge2011}. Moreover, they tend to be comparatively transferable between state points\cite{CGpower}.
However, top-down models are often not quantitatively accurate. Generally, they cannot be improved systematically, and their relation to more detailed models and properties at the atomistic scale is undetermined.

\subsection*{Bottom-up models}

Alternatively, we can base a CG model on an underlying, more detailed ``fine-grained'' model, with the aim of inferring the CG interactions from simulation results of the more detailed model; this is called the bottom-up approach.
This takes advantage of all the effort that went into building and validating the fine-grained model and removes many of the arbitrary choices that are made in the development of top-down models.
It is common, but not necessary, that the underlying fine-grained model is an all-atom model.
In principle, this allows a hierarchical approach to modelling, starting on the atomistic scale and going up to a very large-scale CG model.
In the following, we assume that the fine-grained model is an all-atom model.

The defining feature of bottom-up coarse-graining is that there is a \emph{mapping} function from the all-atom model to the CG model. This is a many-to-one mapping: multiple all-atom configurations correspond to the same coarse-grained configuration. We consider this mapping function in more detail in Section~\ref{sec:cg:mapping}.
The explicit mapping makes it possible to systematically connect models at different scales (``multiscale coarse-graining'').

A disadvantage of the bottom-up approach is that the CG model can only contain information that has been sampled in the atomistic model: If the all-atom trajectory that is used to construct the CG model does not explore the full phase space but is caught in a free-energy basin, the CG model cannot represent the behaviour of the atomistic model outside this basin (irrespective of the approach used to determine the CG interactions).
Generally, this means the CG model will only reproduce the behaviour inside this basin, even if, on \emph{long} time scales, there would be fluctuations that take the all-atom system out of this basin. %
Hence, bottom-up models are often biased towards the structures for which they were parametrised. In principle, they are primarily suited to describe fluctuations around the equilibrium structure.
(We note that there are free energy methods such as Metadynamics\cite{MetadynamicsReview} and ABF\cite{ABFEverything} with which  we can run \emph{biased} simulations that enable the model to escape local free energy basins and explore wider parts of the free energy surface, though this is beyond the scope of this thesis.)

\subsection*{Comparison}

Noid\cite{noid:cgbio:review} characterises the distinction between these types of approaches as follows: top-down models are \emph{underconstrained} by experimental data --- there is a large degree of flexibility in terms of what functional forms and parameters to choose, and there are not enough different experimental observables to uniquely determine them ---, whereas bottom-up models are \emph{overconstrained} by all the data available from simulations of the underlying all-atom model.
Bottom-up models tend to focus on structural properties, whereas top-down models tend to focus on large-scale emergent phenomena.

Top-down modelling has generally been used for low-resolution models (with each CG site describing a large part of a residue or even several residues at once), whereas bottom-up models have historically been used only for ``soft'', limited coarse-graining (with a few atoms per CG site). However, recently there has been more interest in bottom-up approaches to drastic coarse-graining of molecules like lipids where each CG site encompasses thirty to fifty atoms\cite{voth:lipidhcg}. This is where we expect the GAP-CG approach that we introduce in this thesis to be most useful.

\myvspace
We are interested in a consistent bridge between scales, following on from previous research to connect atomistic molecular dynamics with quantum-mechanical potentials\cite{GAP:PRL}. A link between CG simulations and the underlying atomistic structure is crucial, for example, for developing new materials with specific properties\cite{MaterialsGenome}. For this reason, we will focus on bottom-up approaches to coarse-graining.

In the following, we first discuss the mapping between the underlying all-atom model and the coarse-grained description (\cref{sec:cg:mapping}). We then consider a definition of \emph{consistency} between all-atom and CG model (\cref{sec:cg:consistency}). In the remainder of this chapter, we discuss different approaches to determining CG interactions from simulations of the underlying all-atom model, based on its structural distributions (\cref{sec:cg:structurebased}) or on the atomistic forces (\cref{sec:cg:forcebased}).

\section{Mapping from All-Atom to Coarse-Grained Model}
\label{sec:cg:mapping}

The defining feature of bottom-up CG models is their explicit connection to an underlying (usually atomistic) model, given by a mapping from the all-atom model to the CG model. Here we show how this connection is formally made.

In the remainder of this chapter, variables denoting the atomistic model are designated by lower-case characters, and variables denoting the CG model are designated by upper-case characters.

The state of the atomistic system is specified by the set of positions and momenta of all $n$ atoms.
We collectively describe this state by the $3n$-dimensional vectors for position, $\vecr^n$, and momenta, $\vec{p}^n$:
\begin{equation*}
  \vec{r}^n=\{\vec{r}_1,\dots,\vec{r}_n\} , \qquad
  \vec{p}^n=\{\vec{p}_1,\dots,\vec{p}_n\} .
\end{equation*}
Correspondingly, the state of the CG model is defined by the positions and momenta of the $N$ CG sites, where we assume that the CG sites are defined as structureless, point-like particles:
\begin{equation*}
  \vec{R}^N=\{\vec{R}_1,\dots,\vec{R}_N\} , \qquad
  \vec{P}^N=\{\vec{P}_1,\dots,\vec{P}_N\} .
\end{equation*}
We now introduce the mapping function from the atomistic configuration $\vec{r}^n$ to the CG configuration $\vec{R}^N$:
\begin{equation}
  \vec{R}^N = \vec{M}^N_{\vec{R}} (\vec{r}^n) ,
  \label{eq:cg:mapping:coordinates}
\end{equation}
where $\vec{M}^N_{\vec{R}} = \{ \vec{M}_{\vec{R}1}, \dots, \vec{M}_{\vec{R}N} \}$.
Note that this is a many-to-one mapping: multiple atomistic configurations correspond to the same CG configuration.

In principle, \cref{eq:cg:mapping:coordinates} could describe a non-linear mapping, but in the following we only consider linear mappings, as is prevalent in the literature. This simplifies many equations and makes it easier to reason about and compare the different approaches to coarse-graining. A linear mapping is given by
\begin{equation}
  \vec{M}_{\vec{R}I} (\vec{r}^n) = \sum_{i=1}^{n} c_{Ii} \vec{r}_i
  \qquad\text{for } I = 1, \dots, N .
  \label{eq:cg:mapping:coordinates:linear}
\end{equation}
We require translational invariance: if we shift all atomic coordinates by a certain vector, the CG coordinates should shift by the same vector. This imposes the constraints $\sum_{i=1}^n c_{Ii} = 1$ for all $I$.
\newcommand*{\involvedset}{\ensuremath{\mathcal{I}_I}\xspace}
\newcommand*{\specificset}{\ensuremath{\mathcal{S}_I}\xspace}
It will be useful to introduce the set \involvedset of all atoms that are \emph{involved} in the definition of a CG site $I$, that is, all atoms $i$ for which $c_{Ii} \neq 0$. Similarly, the set \specificset contains all atoms \emph{specific} to a site $I$, that is, all atoms which are not involved in any other CG site.
Common choices for the definition of CG sites are the centre of mass ($c_{Ii} = m_i / \sum_{j \in \involvedset} m_j$) or the centre of geometry ($c_{Ii} = 1 / \sum_{j\in\involvedset} 1$) of a group of atoms.

Corresponding to the mapping function~\eqref{eq:cg:mapping:coordinates} for the coordinates, there is a similar mapping from the momenta of the atoms to the momenta of the CG sites,
\begin{equation*}
  \vec{P}^N = \vec{M}_\vec{P}^N (\vec{p}^n) ,
\end{equation*}
where $\vec{M}_\vec{P}^N = \{ \vec{M}_{\vec{P}1}, \dots, \vec{M}_{\vec{P}{N}} \}$. Coarse-grained position and momentum are connected by $\vec{P}_I = M_I \inlinederiv{\vec{R}_I}{t}$, where $M_I$ is the mass of CG site $I$.
Hence, the linear mapping~\eqref{eq:cg:mapping:coordinates:linear} induces the following relation for the CG momenta:
\begin{equation*}
  \vec{M}_{\vec{P}I} (\vec{p}^n) = M_I \sum_{i=1}^{n} c_{Ii} \vec{p}_i / m_i
  \qquad\text{for } I = 1, \dots, N .
\end{equation*}
For a typical CG model, $N \ll n$, and thus it has far fewer degrees of freedom than the corresponding all-atom model.
We call the $3(N-n)$ degrees of freedom that are lost in the coarse-graining the ``internal'' degrees of freedom.
The reduction in the number of degrees of freedom brings with it an entropy difference between CG and all-atom model\cite{RudzinskiNoidEntropyForcesStructure}.
In the following section, we consider consistency between all-atom and coarse-grained description, and the conditions this imposes on the CG model.

\section{Consistency and the Many-Body Potential of Mean Force}
\label{sec:cg:consistency}
\label{sec:cg:pmf}

Having defined a mapping function, we can now analyse the relation between the atomistic and the coarse-grained scales within a statistical mechanics framework. Here we consider the phase-space probability distribution of both models, following Noid \etal\cite{MSCG:I}.
A more detailed discussion can be found in \textcite{Noid2013}.

\newcommand*{\CGexpur}{\exp\!\big[\!-u(\vec{r}^n)/\kT\big]}
\newcommand*{\CGexpUR}{\exp\!\big[\!-U(\vec{R}^N)/\kT\big]}

The phase-space probability distribution of the atomistic system is determined by its Hamiltonian $H_\text{AA}$; we assume that it can be written as the sum of momentum-dependent kinetic energy and position-dependent potential energy:
\[
  H_\text{AA} = \sum_{i=1}^n \frac{\vec{p}_i^2}{2 m_i} + u(\vec{r}^n) .
\]
Then the phase-space probability distribution factorises into independent contributions from coordinates and momenta,
\begin{equation*}
  p_{rp}(\vec{r}^n,\vec{p}^n) = p_r(\vec{r}^n) p_p(\vec{p}^n) ,
\end{equation*}
where
\begin{subequations}
\begin{align}
  \label{eq:cg:prob:coord:atomistic}
  p_r(\vec{r}^n) &\propto \CGexpur \qquad\text{and} \\
  \label{eq:cg:prob:momentum:atomistic}
  p_p(\vec{p}^n) &\propto \exp\!\big(\!-\!\sum_{i=1}^{n} \vec{p}_i^2/(2m_i\kT)\big)
\end{align}
\end{subequations}
are the probability distributions of coordinates and momenta, respectively (defined up to a normalisation factor).

Analogously, the CG model is described by the Hamiltonian
\[
  H_\text{CG} = \sum_{I=1}^N \frac{\vec{P}_I^2}{2 M_I} + U(\vec{R}^N) ,
\]
and the probability distribution of the CG model factorises as
\begin{equation*}
  P_{RP}(\vec{R}^N,\vec{P}^N) = P_R(\vec{R}^N) P_P(\vec{P}^N) ,
\end{equation*}
where
\begin{subequations}
\begin{align}
  \label{eq:cg:prob:coord:cg}
  P_R(\vec{R}^N) &\propto \CGexpUR \qquad\text{and} \\
  \label{eq:cg:prob:momentum:cg}
  P_P(\vec{P}^N) &\propto \exp\!\big(\!-\!\sum_{I=1}^{N} \vec{P}_I^2/(2M_I\kT)\big) .
\end{align}
\end{subequations}
\newcommand*{\CGMR}{\vec{M}_{\vec{R}}^N (\vec{r}^n)}
\newcommand*{\CGdeltaMR}{\delta\big( \CGMR - \vec{R}^N \big)}
\newcommand*{\CGdeltaMI}[1][I]{\delta\big( \vec{M}_{\vec{R} #1} (\vec{r}^n) - \vec{R}_{#1} \big)}
\newcommand*{\CGdeltalinear}{\delta\big( \! \sum_{i\in\involvedset} c_{Ii} \vec{r}_i - \vec{R}_I \big)}
We now make use of the mapping function and consider the probability of finding a given CG configuration within the atomistic ensemble, $p_R(\vec{R}^N)$. This is obtained by integrating out the internal degrees of freedom, averaging over their fluctuations:
\begin{equation}
  \label{eq:cg:prob:coord:atcg}
  p_R(\vec{R}^N) = \intdd{\vec{r}^n} p_r(\vec{r}^n) \, \CGdeltaMR ,
\end{equation}
where
\begin{equation*}
  \CGdeltaMR \equiv \prod_{I=1}^N \CGdeltaMI
\end{equation*}
is the $3N$-dimensional Dirac delta function, ensuring that we only include those atomistic configurations that map to the given CG configuration.

Similarly, for the CG momenta we obtain
\begin{align}
  \notag
  p_P(\vec{P}^N) &= \intdd{\vec{p}^n} p_p(\vec{p}^n)
      \,\delta\big( \vec{M}_{\vec{P}}^N (\vec{p}^n) - \vec{P}^N \big) \\
  &\propto \intdd{\vec{p}^n} \Big(\prod_i \exp\!\big(\!- \vec{p}_i^2 / (2 m_i \kT) \big) \! \Big) \, \prod_I \delta\big( M_I \sum_{j} c_{Ij} \vec{p}_j / m_j - \vec{P}_I \big) .
  \label{eq:cg:prob:momentum:atcg}
\end{align}
The configuration and momentum terms are independent of each other and can be treated separately. We discuss the momentum term in \cref{sec:cg:momentum} and the configuration term in \cref{sec:cg:configuration}.

\subsection{Momentum Space}
\label{sec:cg:momentum}

We consider a CG model to be consistent with an underlying atomistic model if the probability distribution in the CG model matches the probability distribution of the CG sites in the atomistic model, as obtained through the corresponding mapping function.
That is, for consistency in momentum space we require \eqref{eq:cg:prob:momentum:atcg} and \eqref{eq:cg:prob:momentum:cg} to be equal:%
\begin{equation}
  P_P(\vec{P}^N) = p_P(\vec{P}^N) .
  \label{eq:cg:consistency:momentum}
\end{equation}
The left-hand side of this equation can be written as a product of independent zero-mean Gaussians for each $I$,
\begin{equation*}
  P_P(\vec{P}^N) \propto \prod_I \exp\!\big(\!- \vec{P}_I^2 / (2 M_I \kT) \big) .
\end{equation*}
Requiring this to be equivalent to the probability of the CG momenta within the atomistic model imposes two conditions on the CG model:
First, for $p_P(\vec{P}^N)$ to be a product of independent zero-mean Gaussians as well, no atom can be involved in the definition of more than one CG site. %
Second, requiring the second moments of the Gaussians in both probability distributions to be equal places the following condition on the CG masses $M_I$~\cite{MSCG:I}:
\begin{equation*}
  \frac{1}{M_I} = \sum_i \frac{c_{Ii}^2}{m_i} \qquad\text{for all } I .
\end{equation*}
Of all possible definitions of CG sites for a given group of atoms, identifying the site with the centre of mass of the atoms results in the largest mass for the CG site. As this corresponds to slower motions this will often be the preferred choice.

Note that consistency in momentum space only implies that the momentum distribution in the CG model matches the one in the atomistic model. It does not imply that the CG model will produce the correct dynamics (for example, time correlation functions). Obtaining ``real dynamics''\cite{voth:RealDynamics} is beyond the scope of this work, and in the remainder of this thesis we focus on consistency in configuration space.

\subsection{Configuration Space}
\label{sec:cg:configuration}

Analogously to Eq.~\eqref{eq:cg:consistency:momentum}, we consider the CG model to be consistent with the underlying atomistic model in configuration space if the probability distributions of the CG coordinates in the CG model~\eqref{eq:cg:prob:coord:cg} and in the atomistic ensemble~\eqref{eq:cg:prob:coord:atcg} are equal. That is, for consistency we require
\begin{equation}
  P_R(\vec{R}^N) = p_R(\vec{R}^N).
  \label{eq:cg:consistency:coord}
\end{equation}
This is equivalent to
\newcommand*{\CGz}{\intdd{\vec{r}^n} \CGexpur \, \CGdeltaMR}
\begin{equation*}
  \CGexpUR \propto \CGz ,
\end{equation*}
thereby defining the ``optimal'' CG potential $U(\vec{R}^N)$. Note that this is not a potential energy, but also contains an entropic contribution due to integrating over fluctuations in the internal degrees of freedom. Hence, this CG potential actually describes a free energy surface, and it explicitly depends on the temperature $T$.
We can explicitly write $U$ as a conditional free energy:
\begin{equation}
  U(\vec{R}^N) = - \kT \log {z(\vec{R}^N)} + \text{(const.)} ,
  \label{eq:cg:CGpotential}
\end{equation}
where
\begin{equation}
  {z(\vec{R}^N)} \equiv \CGz
  \label{eq:cg:partitionfunction}
\end{equation}
is the constrained partition function (as a function of the CG coordinates).
Hence, \cref{eq:cg:CGpotential} is equivalent to the Potential of Mean Force (PMF) $A(\xi)$ discussed in the previous chapter [Eq.~\eqref{eq:md:pmf} on \mypageref{eq:md:pmf}].
Whereas the PMF is usually considered for a small number of collective variables $\xi$, such as order parameters, reaction coordinates or pair distances, the CG potential corresponds to the \emph{many-body} PMF $A(\vec{R}^N)$ for all $3N$ CG coordinates.

When the CG interactions are given by the PMF for the CG coordinates, both structural and thermodynamic properties of the atomistic model are exactly preserved. However, explicitly evaluating the integral in \cref{eq:cg:partitionfunction} would not be feasible in practice. Moreover, $A(\vec{R}^N)$ describes many-body interactions that, in principle, involve all CG sites simultaneously.
In the following, we review current approaches for obtaining CG potentials that approximate \cref{eq:cg:CGpotential} or otherwise generate effective interactions that aim to reproduce probability distributions of the atomistic model.

\section{Structure-based Approaches}
\label{sec:cg:structurebased}

Structure-based approaches to coarse-graining derive the CG interactions from structural distributions of the underlying system.
We give an overview of the four main approaches that can be found in the literature: Direct and Iterative Boltzmann Inversion (DBI and IBI; Sections~\ref{sec:cg:dbi} and~\ref{sec:cg:ibi}, respectively), Inverse Monte Carlo (IMC; Section~\ref{sec:cg:imc}), and Relative Entropy (Section~\ref{sec:cg:relativeentropy}).

Direct and Iterative Boltzmann Inversion are based solely on one-dimensional distributions of individual collective variables (CVs). Inverse Monte Carlo also takes into account cross-correlations between distributions of different CVs. The Relative Entropy approach considers, in principle, the joint probability distribution of all CG coordinates. %
In any case, typically only pair potentials are considered as the basis for the (non-bonded) CG interactions. If there are only pairwise interactions, the Henderson theorem\cite{HendersonTheorem} shows that the pair potential that reproduces the radial distribution function (RDF) is unique (up to a constant). %
However, it should be noted that in practice a series of completely different potentials may result in coinciding RDFs\cite{BriniReview,VOTCA}.

\subsection{Direct Boltzmann Inversion}
\label{sec:cg:dbi}

When a collective variable $\xi$ is completely uncorrelated with all other degrees of freedom of the system, it appears as an independent factor in the probability distribution of the CG model [\cref{eq:cg:prob:coord:cg}]. In this case, the probability distribution for this CV can be written as a Boltzmann weight,
\begin{equation}
  p(\xi) \propto \exp\!\big(\! -\beta A_\xi(\xi) \big) ,
  \label{eq:cg:dbi:prob}
\end{equation}
where $A_\xi(\xi) = -\kT \log p(\xi)+\text{(const.)}$ is the conditional free energy for $\xi$, and $\beta = 1/\kT$.
Based on this, Tsch\"op \etal\cite{DirectBoltzmannInversion} proposed a CG potential for the variable $\xi$ obtained by ``Boltzmann inversion'':
\begin{equation}
  U_\xi(\xi) = -\kT \log \frac{p(\xi)}{J(\xi)} .
  \label{eq:cg:dbi:potential}
\end{equation}
Note the Jacobian factor $J(\xi)$ that, for the general case when $\xi$ is not a Cartesian coordinate, accounts for the change of volume elements in the integration over orthogonal degrees of freedom.
Commonly used collective variables are the distance $r$ between a pair of (bonded or non-bonded) CG sites, the angle $\theta$ between three CG sites, and the dihedral angle $\chi$ defined by four CG sites. The corresponding Jacobian factors are $J(r) = r^2$, $J(\theta) = \sin(\theta)$ and $J(\chi) = 1$, respectively.
In practice, this approach is used even when a CV does have some correlation with other degrees of freedom. However, when multiple $\xi_i$ are correlated, we cannot just add up the corresponding DBI potentials without introducing a bias.

We illustrate this for the example of the pairwise distance $r = \norm{\vec{R}_J - \vec{R}_I}$ in a one-site model. In this case, the scaled distribution $p(r) / J(r) = p(r)/r^2$ is proportional to the two-body pair correlation function $g^{(2)}(\vec{R}_I,\vec{R}_J) = g(r)$, more commonly known as the Radial Distribution Function or RDF. It is related to the \emph{two-body} potential of mean force $w^{(2)}(r)$, sometimes referred to as the pair potential of mean force (ppmf)\cite{ChandlerStatMech}:
\[
  g(r) = \exp\!\big(\!-\beta w^{(2)}(r)\big) .
\]
Direct Boltzmann Inversion results in the CG pair potential 
\begin{equation}
  U_\text{CG}(r) = -\kT \log g_\text{AA}(r) = w^{(2)}_\text{AA}(r) ,
  \label{eq:cg:dbi:pairpot}
\end{equation}
where for clarity we added the subscript ``AA'' to refer to the all-atom model.
In the limit of infinite dilution (when there are no correlations between different pairs of particles), a CG simulation using the potential~\eqref{eq:cg:dbi:pairpot} will exactly reproduce the all-atom RDF, as in this case $w^{(2)}_\text{CG} = U_\text{CG} = w^{(2)}_\text{AA}$.
However, in general, the ppmf is a sum of the direct interaction potential and environment-mediated interactions. The two-body pair correlation function contains effective contributions due to many-body effects, such as the packing of non-overlapping particles in the liquid phase.

This can easily be understood by considering a CG model that is based on a one-to-one mapping with a single-site hard-sphere liquid. In the fine-grained model, the interactions are given by a pair potential that is zero beyond the hard sphere radius. Because of the one-to-one mapping, the CG potential should be identical to the atomistic potential. However, the close packing of the spheres leads to oscillations in the RDF, and hence the CG potential obtained by DBI will also show an oscillatory behaviour. This is entirely due to the environmental effect of correlations with other sites. (In contrast, for an ideal gas, the RDF is exactly one beyond the hard sphere radius, correctly resulting in the DBI potential being zero in that region.)

When CVs are significantly correlated, the potentials obtained by DBI will not reproduce the distributions in the atomistic model, and hence cannot be used directly.
The assumption of independent degrees of freedom is a good approximation for non-bonded pair potentials when CG sites are \emph{dilute} and for bond potentials when the CG bonds are \emph{stiff}~\cite{noid:cgbio:review}.
In practice, DBI is commonly used for bonded interactions (bonds, angles and dihedral angles),
 and the remaining non-bonded interactions can be obtained by other, more refined methods. %
Non-bonded interactions can only be successfully described by DBI when the CG sites are very dilute. However, the DBI pair potential can provide a useful starting point for other approaches, such as IBI and IMC, that iteratively refine CG pair potentials to improve the reproduction of the atomistic pair correlation function. %

\subsection{Iterative Boltzmann Inversion}
\label{sec:cg:ibi}

As discussed in the previous section, Direct Boltzmann Inversion (DBI) neglects environmental effects. Hence, in general, it does not reproduce the atomistic distributions in the CG model. Reith \etal\cite{IBI} proposed \emph{Iterative} Boltzmann Inversion (IBI) as an extension of DBI to derive \emph{effective} pair potentials that do reproduce atomistic distributions. While this approach also applies to other degrees of freedom, we focus on non-bonded pair potentials and the RDF. Starting from an initial guess $U_{(0)}(r)$ for the CG pair potential (for example, based on DBI of the RDF), in each iteration step a correction is applied to the CG potential:
\[
  U_{(i+1)} (r) = U_{(i)} (r) + \alpha \Delta U_{(i)} (r) ,
\]
where 
$\alpha$ is an optional scaling factor ($0 < \alpha < 1$) to ensure the stability of the iterative procedure.
The correction is given by the difference between effective CG ppmf and all-atom ppmf,
\[
  \Delta U_{(i)} (r) = w^{(2)}_{\text{CG},(i)} (r) - w^{(2)}_\text{AA}(r) = \kT \log{ \frac{g_{\text{CG},(i)}(r)}{g_{\text{AA}}(r)} },
\]
where $g_\text{AA}$ is the reference RDF of the atomistic model. 
This process is repeated until the RDF in the CG model is sufficiently close to the all-atom RDF.

This scheme is motivated by the following observation: when $g_{\text{CG},(i)} (r) > g_\text{AA}(r)$ at a given pair distance $r$, this implies that the ppmf of the CG model at that distance is too attractive, and that it should be more repulsive. As the ppmf is given by the sum of direct pair interaction $U_{(i)}(r)$ and environmental contributions, the simplest way to increase the ppmf is to increase the value of the pair potential at that distance.
Clearly, $g^\text{CG}_*(r) = g_\text{AA}(r)$ is a fixed point of this iteration. Hence, assuming convergence, the IBI procedure results in a pair potential that reproduces the RDF of the all-atom system.

If more than one type of interactions should be derived in this way, it is advantageous to update one potential at a time while keeping the others fixed, as this leads to improved convergence.
This procedure works best by starting out with the stiffest, least correlated degrees of freedom; thus, the typical order would be bond, angle, dihedral, and finally non-bonded interactions\cite{IBI}.
When there are significant cross-correlations between different pair correlation functions, as is the case for the non-bonded degrees of freedom in a multicomponent systems, the IBI procedure may have convergence problems\cite{VOTCA}.
IBI is a local algorithm in the sense that the update to the pair potential at a distance $r$ only depends on the values of CG and atomistic RDF at that distance, and the potential updates for different interactions are independent of each other. This implies that uncertainty in the RDF at a certain distance only has a constant effect on the potential update, independent of the distance.
However, in dilute systems it can be difficult to sample the RDF between sites to sufficient accuracy\cite{PeterCG}. %

\subsection{Inverse Monte Carlo}
\label{sec:cg:imc}
To explicitly account for the correlations due to different types of interactions in multicomponent systems, Lyubartsev and Laaksonen proposed the Inverse Monte Carlo (IMC) approach, also referred to as Reverse Monte Carlo (RMC)\cite{IMC1,IMC2,IMC3}. Like IBI, it is an iterative approach. IMC is based on linear response theory, estimating how a change in the CG pair potential $U_{\xi'}(r')$ for one type of interaction $\xi'$ at distance $r'$ impacts \emph{all} pairwise distance distributions $P_\xi(r)$. This is given by the susceptibility matrix,
\[
  K_{\xi \xi'}(r,r' | U) = \fderiv{P_\xi(r|U)}{U_{\xi'}(r')} ,
\]
defined as the functional derivative of the probability distribution $P_\xi(r)$ with respect to the pair potential $U_{\xi'}(r')$.
The susceptibility matrix can be related to the covariance between different pair distances,
\[
  K_{\xi \xi'}(r,r') = \beta \big( P_\xi(r) P_{\xi'}(r') - P_{\xi\xi'}(r,r') \big) ,
\]
where $P_{\xi\xi'} (r,r')$ is the \emph{joint} probability distribution (cross-correlation) of finding a pair of type $\xi$ at distance $r$ and finding a pair of type $\xi'$ at distance $r'$.
Corrections to the pair potentials induce changes in the pair distance distributions. This is given by
\begin{equation}
  \delta P_\xi(r) = \sum_{\xi'} \intdd{r'} K_{\xi \xi'}(r,r') \delta U_{\xi'}(r') ,
  \label{eq:cg:imc}
\end{equation}
where $\delta P_\xi(r) = p_\xi(r) - P_\xi(r)$ is the difference between the atomistic and CG distribution.
In practice, the pair potentials are discretised, and \cref{eq:cg:imc} leads to a system of coupled linear equations, whose solution provides the corrections to the pair potentials. %
For this we need to invert the susceptibility matrix $K$, which can only be determined approximately. Inverting it could amplify the errors, leading to convergence issues. To alleviate such issues, we can scale the potential correction by a factor $0 < \alpha < 1$ as in IBI, though this may increase the number of iterations required for convergence.
This process is repeated until the atomistic distributions are reproduced sufficiently accurately.
As a covariance matrix, $K$ is positive definite. Hence, in principle, the IMC algorithm will converge to a global optimum.

Inverse Monte Carlo has a single, global update step for all interactions in the system. As this takes into account cross-correlations between different interactions, this is a particular advantage in multi-component systems, where IMC provides more stable updates than IBI. Furthermore, IMC converges faster than IBI with the number of iterations\cite{VOTCA}.

However, whereas IBI only depends on averages of pair distance distributions (the radial distribution functions), IMC also requires us to accurately calculate the cross-correlations between all combinations of pair distances; this necessitates much longer molecular dynamics runs in each iteration step.
Moreover, R\"uhle \etal\cite{VOTCA} also demonstrated that IMC suffers from significant finite-size effects. In contrast to IBI, the influence of errors in the distributions on the IMC updates scales quadratically with the distance; for this reason, the long-range tails of the RDFs must be converged to a very high accuracy with respect to both system size and simulation length. Hence, IMC may not be appropriate for small systems.

In principle, IMC could be applied to three-body potentials, but this comes with a prohibitive computational cost\cite{IMC1,voth:threebody}.

\subsection{Relative Entropy}
\label{sec:cg:relativeentropy}

\newcommand*{\Srel}{\ensuremath{S_\text{rel}}}
\newcommand*{\Smap}{\ensuremath{S_\text{map}}}
Another, recent approach to structure-based coarse-graining was proposed by Shell, based on the relative entropy \Srel{} of a CG model\cite{RelativeEntropy}. The relative entropy measures how well we can estimate the underlying atomistic distribution when sampling from a given CG distribution. It is defined as
\begin{equation}
  \Srel = \intdd{\vecr^n} p_{r}(\vecr^n) \log\! \left( \frac{p_{r}(\vecr^n)}{P_{R}\big( \CGMR \big)} \right) + \average{ \Smap }_{r} ,
\end{equation}
where
\[
  \Smap(\vec{R}^N) = \log \intdd{\vecr^n} \CGdeltaMR
\]
is called the mapping entropy. \Smap{} describes the volume of atomistic configuration space corresponding to a single CG configuration; it depends only on the CG mapping $\CGMR$, not on the CG force field.

The relative entropy is a measure of the overlap between the atomistic and the CG ensembles, related to the Kullback--Leibler divergence between $p_r$ and $P_R$; as such, it implicitly depends on the CG potential $U_\text{CG}$. It can be considered to describe the amount of information that is lost when moving from an all-atom description to the CG model.

According to the Gibbs inequality,
\[
  \Srel[U_\text{CG}] \ge 0 ,
\]
and the relative entropy only vanishes when the CG and all-atom probability distributions match exactly.

This motivates an optimisation of force field parameters based on minimising the relative entropy. Substituting $p_r$ and $P_R$ [using \cref{eq:cg:prob:coord:atomistic,eq:cg:prob:coord:cg}], we obtain
\begin{equation}
  \Srel = \beta \average[\big]{U_\text{CG} - U_\text{AA}}_\text{\!AA} - \beta (A_\text{CG} - A_\text{AA}) + \average[\big]{\Smap}_\text{\!AA} ,
  \label{eq:cg:relativeentropy:avg}
\end{equation}
where the subscripts ``AA'' and ``CG'' refer to the all-atom and coarse-grained models, respectively; $U$ is the potential energy and $A$ is the free energy.
The angular brackets denote the ensemble average.

The relative entropy can be considered to be a function of all free parameters $\lambda$ of the CG potential; we can minimise it numerically without requiring the absolute value, guided by the first derivatives\cite{VOTCA:RE}:
\[
  \pderiv{\Srel}{\lambda} = \beta \Average{\pderiv{U_\text{CG}}{\lambda}}_\text{\!\!AA} - \beta \Average{\pderiv{U_\text{CG}}{\lambda}}_\text{\!\!CG} .
\]
For parameters that enter the CG potential linearly, the curvature implied by the second derivatives of the relative entropy is always positive, and hence there will only be a single minimum. However, the relative entropy can also be minimised for analytic functional forms and parameters that enter non-linearly.

When obtaining CG interactions by minimising the relative entropy, the resulting interactions can reproduce the expectation values of all terms that are included in the potential.
For instance, this means that finely tabulated pair potentials will reproduce the RDF, and in this case relative entropy optimisation is equivalent to IBI or IMC. Moreover, for a CG potential that can accurately represent an $N$-body potential, minimising the relative entropy will reproduce the PMF\cite{RelativeEntropy:CoarsegrainingErrorsNumericalOptimization}.

\section{Force-based Approaches}
\label{sec:cg:forcebased}
\label{sec:cg:mscgfm}

The approaches reviewed in the previous section focus on the reproduction of the \emph{structure} of the atomistic model to determine effective CG interactions.
Alternatively, we can aim to reproduce the mean \emph{forces} acting on the CG sites. This approach is based on the force-matching scheme of Ercolessi and Adams\cite{ercolessiadams} and has been extensively developed by Voth \etal\cite{voth:mscg1,voth:mscg2,MSCG:I,MSCG:II}.
Reproducing the mean forces forms the basis for our own approach to coarse-graining (which will be introduced in \cref{ch:gap}), hence we will discuss the derivation given in \textcite{MSCG:I} in more detail.
\subsection{Instantaneous Collective Forces}
\label{sec:cg:icf}

We start from the ``ideal'' CG potential based on the many-body PMF as defined by \cref{eq:cg:CGpotential}.
The coarse-grained force on site $I$ is given by the negative gradient of the PMF,
\begin{equation*}
  \vec{F}_I(\vec{R}^N) = - \pderiv{U(\vec{R}^N)}{\vec{R}_I} .
\end{equation*}
Carrying out the derivative, we obtain
\begin{equation*}
  \vec{F}_I(\vec{R}^N) = \frac{\kT}{z(\vec{R}^N)} \intdd{\vec{r}^n} \CGexpur \prod_{J(\neq I)}\! \CGdeltaMI[J] \, \pderiv{}{\vec{R}_I} \CGdeltalinear .
\end{equation*}
We integrate this by parts, making use of
\begin{equation*}
  \pderiv{}{\vec{R}_I} \CGdeltalinear = - \frac{1}{c_{Ik}} \pderiv{}{\vec{r}_k} \CGdeltalinear ,
\end{equation*}
where we chose an arbitrary atom $k$ involved in the definition of CG site $I$. If this atom is also involved in the definition of other CG sites, we cannot simply move the partial derivative to $u(\vec{r}^n)$. In this case, we use a linear combination of all specific atoms $j$:
\begin{equation*}
\pderiv{}{\vec{R}_I} \CGdeltalinear = - \sum_{j\in\specificset} \frac{d_{Ij}}{c_{Ij}} \pderiv{}{\vec{r}_j} \CGdeltalinear ,
\end{equation*}
where we require $\sum_{j\in\specificset} d_{Ij} = 1$.
Thus we need at least one specific atom per CG site.
Integration is then straightforward and yields
\begin{equation}
  \vec{F}_I(\vec{R}^N) = \bigg\langle  \sum_{j\in\specificset} \frac{d_{Ij}}{c_{Ij}} \left(- \pderiv{u(\vec{r}^n)}{\vec{r}_j}\right)  \!\!\bigg\rangle_{\!\!\!\vec{R}^N} \;,
  \label{eq:cg:forcefield}
\end{equation}
where
\begin{equation*}
  \average[\big]{ { {\color{blue!60}\bullet} (\vec{r}^n)} }_{\vec{R}^N} \equiv
  \frac{
    \intdd{\vec{r}^n} \CGexpur \, \CGdeltaMR \, { {\color{blue!60}\bullet} (\vec{r}^n) }
  }{
    \intdd{\vec{r}^n} \CGexpur \, \CGdeltaMR
  }
\end{equation*}
is the conditional expectation value (or equilibrium average) in the ensemble of the atomistic system; this expectation value is a function of the CG coordinates $\vec{R}^N$ (which are kept constant while averaging). In practice, this ensemble average will be replaced by a time average over atomistic trajectories.

\Cref{eq:cg:forcefield} can be broken down into simpler parts. The force on an atom $j$ in the atomistic system is given by
\begin{equation*}
  \vec{f}_j(\vec{r}^n) = - \pderiv{u(\vec{r}^n)}{\vec{r}_j} .
\end{equation*}
The linear combination
\begin{equation}
  \vec{\mathcal{F}}_I(\vec{r}^n) = \sum_{j\in\specificset} \frac{d_{Ij}}{c_{Ij}} \vec{f}_j(\vec{r}^n)
  \label{eq:cg:cgforce:inat}
\end{equation}
describes the ``atomistic forces acting on CG site $I$ only''; it is the total (collective) force on a CG site $I$ within the atomistic model. (Note that only atoms \emph{specific} to this site contribute.)
When CG sites are defined as the centres of mass of a group of atoms, we have $c_{Ij} = d_{Ij} = m_j / M_I$, where $M_I = \sum_{i\in\involvedset} m_i$, and \cref{eq:cg:cgforce:inat} simplifies to
\[
  \vec{\mathcal{F}}_I(\vec{r}^n) = \sum_{j\in\specificset} \vec{f}_j(\vec{r}^n) .
\]
The forces $\vec{\mathcal{F}}_I$ defined by \cref{eq:cg:cgforce:inat} are equivalent to the instantaneous collective forces (ICF) that are used, for example, in the Adaptive Biasing Force method.
Here we only consider linear combinations of Cartesian coordinates; however, it is possible to derive ICF for arbitrary collective variables, though this leads to significantly more complex equations\cite{lelievre2010free}.

The ICF can be considered noisy samples of the mean forces; the optimal CG forces that correspond to a consistent CG model as defined by \cref{eq:cg:CGpotential} are derived from the atomistic forces through an equilibrium average over the internal degrees of freedom, with the CG coordinates kept constant:
\begin{equation}
  \vec{F}_I(\vec{R}^N) = \average[\big]{ \vec{\mathcal{F}}_I(\vec{r}^n) }_{\vec{R}^N} .
  \label{eq:cg:consistency:force}
\end{equation}
This describes the mean forces on the CG sites.
If the CG force field has the flexibility to reproduce \cref{eq:cg:consistency:force}, which will be the case for an $N$-body potential, this will reproduce the potential of mean force, and the CG model will be consistent with the atomistic model.

\subsection{Practical Approach: MSCG/FM}

\Cref{eq:cg:consistency:force} defines the atomistically consistent CG forces. In practice, they need to be approximated by the CG force field $\vec{G} = \{\vec{G}_I(\vec{R}^N)\}$, where $\vec{G}_I(\vec{R}^N)$ calculates the force on CG site $I$ as a function of the CG configuration $\vec{R}^N$. Voth \etal{} introduced a practical approach to the optimisation of $\vec{G}$, termed the ``Multiscale Coarse-Graining/Force Matching'' (MSCG/FM) method\cite{MSCG:I,MSCG:II}. This is based on minimising the residual functional,
\begin{equation}
  \chi^2[\vec{G}] = \frac{1}{3N} \Average{ \sum_{I=1}^N \norm{ \vec{G}_I \big(\vec{M}_{\vec{R}}^N(\vecr^n) \big) - \vec{\mathcal{F}}_I(\vecr^n) }^2 } \! ,
  \label{eq:cg:mscgfm:residual}
\end{equation}
where the average is over the equilibrium canonical ensemble for the atomistic model. Given a sample of $K$ configurations $\Gamma_k = [\vec{r}^n]_{k}$ from an all-atom simulation, we can numerically evaluate the residual as
\begin{equation}
  \chi^2(\lambda) = \frac{1}{3NK} \sum_{k=1}^K \sum_{I=1}^N \norm{ \vec{G}_I(\Gamma_k; \lambda) - \vec{\mathcal{F}}_I(\Gamma_k) }^2 ,
  \label{eq:cg:mscgfunctional}
\end{equation}
where $\lambda$ is a set of parameters specifying the CG force field.
The MSCG/FM approach assumes that the CG force field is linear in these parameters. (This includes, for example, harmonic approximations to bonded interactions, as well as more flexible potentials such as step functions on a grid or spline interactions, and can even be extended to three-body interactions\cite{voth:threebody,MSCG:3body}.)
\Cref{eq:cg:mscgfunctional} can then be written as a matrix equation, which can be solved in a least-squares sense using QR decomposition.
To obtain a smooth potential (splines with continuous derivatives at the grid points), this would require a constrained solver.
In practice, the set of all-atom configurations is partitioned into blocks, and an unconstrained solver is used for each block. The results are averaged over all blocks, which leads to an approximately smooth potential.

\subsection{Connection to Relative Entropy}

We can measure the information content in a CG configuration for discriminating between the all-atom model and a CG model with potential $U$ as
\begin{equation*}
  \Phi(\vec{R}^N | U) = \log \! \left[ \frac{p_R(\vec{R}^N)}{P_R(\vec{R}^N|U)} \right] \! .
  \label{eq:cg:cginformation}
\end{equation*}
This can also be considered a configuration-dependent entropy (if both models sample the canonical ensemble):
\[
  \Phi(\vec{R}^N | U)  = - \beta \big(A(\vec{R}^N) - U(\vec{R}^N)\big) + \text{(const.)}[U] ,
\]
where the constant may depend on the CG potential.
The relative entropy can then be written as
\begin{equation*}
  S_\text{rel}[U] = \kB \!\intdd{^{N}\vec{R}} p_R(\vec{R}^N) \Phi(\vec{R}^N | U) .
  \label{eq:cg:relativeentropy}
\end{equation*}
Hence, minimising the relative entropy corresponds to minimising the expectation value of $\Phi$.
Minimising the residual functional~\eqref{eq:cg:mscgfm:residual} of the MSCG/FM approach can be shown to correspond to minimising the expectation value of $\|\nabla \Phi\|^2$\cite{RudzinskiNoidEntropyForcesStructure}.
For a force field that can represent an $N$-body potential, both approaches will arrive at the potential of mean force. However, when using simpler force fields, for example, using molecular-mechanics-style terms such as pair potentials, the two approaches will lead to different effective potentials.

\section{Summary}
\label{sec:cg:summary}

In this chapter we compared different approaches to coarse-graining. For ``bottom-up'' models, we can define a mapping between atomic and CG coordinates. This allows us to define the ``consistency'' of a CG model in terms of matching the probability distributions of the underlying atomistic model. In momentum space, imposing this consistency condition defines the masses of the CG sites. In configuration space, we find that the consistent CG interaction potential should be equal to the $N$-body Potential of Mean Force (PMF) for the CG coordinates in the atomistic system.

In practice, current CG models are usually based on pair potentials, and rely on different approaches to find the ``best'' \emph{effective} pair potential. These approaches can be classified according to whether they are based on matching \emph{structures} or \emph{forces}. Structure-based approaches that aim to reproduce RDFs will all arrive at equivalent pair potentials. In contrast, force-based approaches that aim to approximate the PMF do not necessarily reproduce structural distribution functions.
An advantage of force-based approaches is that they rely on \emph{gradients} of the PMF which are defined locally, and hence they can be combined with biasing algorithms such as ABF\cite{ABFEverything,LetifsPaper}.
When a CG approach that aims to match the mean forces in the all-atom model \emph{does} recover structural distributions of the atomistic model, this is strong evidence for a consistent CG model that captures the PMF and will thus be able to reproduce both structural \emph{and} thermodynamic properties, overcoming the representability challenge of coarse-graining.
In the next chapter we will introduce a new force-based approach to many-body CG interactions with which we hope to achieve this aim. %

\chapter{Gaussian Approximation Potential}
\label{ch:gap}

All the different approaches discussed in the previous chapter are significantly limited by their common assumption that the CG interactions can be separated into terms similar to the molecular mechanics terms introduced in \cref{sec:md:forcefield}, such as individual bond and angle terms and non-bonded terms that are described by pairwise additive isotropic site--site interactions (``pair potentials''). %
This is already an approximation in all-atom simulations\cite{AAL:representability1}, where ongoing research is studying extensions such as polarisable force fields~\cite{Popelier1,Popelier2} and more general many-body approaches\cite{GAP:tutorial,BehlerNNreview1}.\footnote{Many-body potentials such as non-bonded three-body potentials\cite{materialsManyBody:StillingerWeber}, embedded atom models\cite{materialsManyBody:EAMreview}, and bond-order potentials\cite{materialsManyBody:FinnisSinclair} have been used extensively for metallic systems for decades. However, in biomolecular modelling, many-body approaches to non-bonded atomic interactions are only now starting to appear.}
In CG models, only considering molecular-mechanics-style terms is a severe approximation to the true interactions defined by the PMF, 
as each CG interaction is the average of the various interactions in the underlying all-atom model. This is likely to introduce significant correlations between different coarse-grained degrees of freedom, resulting in many-body effects that cannot be captured by simple terms.%
\footnote{With an increasing number of atoms per CG site, the correlations between CG sites are likely to increase as well, and hence the approximation of separating the true interactions into a sum of simple terms, as is typical in coarse-graining, tends to become worse. This is why, historically, (bottom-up) CG models generally grouped just a few atoms within each site.}

Simple site--site potentials may be able to reproduce radial distribution functions, but this does not imply that they can reproduce higher-order correlations such as angular or orientational distribution functions as well, and such more complex correlations have usually been neglected in previous research.

To obtain a more accurate CG model and, thus, a good description of correlations between different CG sites, we need to include many-body interactions.
While there has been some progress in anisotropic potentials\cite{AnisotropicCG}, and a few forays into three-site interactions\cite{voth:threebody,MSCG:3body}, demonstrating that three-body interactions are necessary to reproduce the angular distribution function in a one-site water model, previous research on many-body potentials in CG models has been limited.
IBI is based only on distance distributions and cannot include any form of more complex correlations. IMC includes cross-correlations between different distance distributions; in principle it could take into account three-body interactions, but this would be infeasible in practice\cite{IMC1}. The MSCG/FM approach is based on least-squares fitting and requires interactions to be linear in the parameters; it can be used for general three-body interactions between CG sites, but this has only been applied to one-site descriptions of homogeneous liquids\cite{MSCG:3body}, and interactions involving more than three CG sites would likely be infeasible within this approach. The relative entropy approach can in principle handle any functional form for the CG interactions\cite{RelativeEntropy:CoarsegrainingErrorsNumericalOptimization}, including three-body or more complex interactions, but this has not been applied in practice.

Instead of assuming, as previous approaches, that the PMF can be separated into terms similar to those used in molecular mechanics, the present work is based on a more general many-body approach, the Gaussian Approximation Potential (GAP).
GAP is a machine-learning potential based on Gaussian process regression, a type of nonparametric regression with the flexibility to represent, in principle, \emph{any} function. A Gaussian process ``learns'' a function based on observations of its values and derivatives, allowing us to predict function values at new points even in high-dimensional spaces.
Gaussian process regression also comes with the advantage of allowing us to evaluate the \emph{uncertainty} in a prediction.
In \cref{sec:gp:regression} we will review Gaussian process regression as a general approach to regression.

GAP was originally developed by Bart\'ok \etal\cite{GAP:PRL,GAP:thesis} in the context of atomistic simulations to infer the potential energy surface from energies and forces obtained from quantum-mechanical calculations such as Density Functional Theory. They demonstrated that GAP can provide nearly the same accuracy as quantum-mechanical methods, in a fraction of the time.

In this thesis we transfer GAP to coarse-grained molecular simulations,
with the aim of recovering the many-body PMF $A(\vec{R}^N)$ for the coordinates of the CG model [\cref{eq:cg:CGpotential}].
While possible in principle, a direct interpolation of the PMF in $3N$-dimensional configuration space would be impractical.
Like the potential energy surface in the atomistic system, the PMF is not just an arbitrary function but has specific properties due to its physical nature.
When applying Gaussian process regression to atomic and molecular interactions, it is prudent to take into account our knowledge about the properties of interaction potentials; we will discuss this in \cref{sec:gap}.
In the context of coarse-graining, the GAP approach faces specific challenges that are less relevant when interpolating a potential energy surface. In \cref{sec:gap:cgspecific} we discuss and develop solutions for the following challenges: noise within force-based coarse-graining, the length scale inhomogeneity of interactions, and sampling issues.

\section{Gaussian Process Regression}
\label{sec:gp:regression}

\newcommand*{\GPdata}{\mathcal{D}}
\newcommand*{\GPparam}{\vec{w}}

\newcommand*{\sigmanoise}{\ensuremath{\sigma_\text{noise}}}
Regression analysis is the study of the relationship $y=f(\vec{x})+\varepsilon$ between an input vector $\vec{x}$ and the observed, potentially noisy output $y$. Here we make the common assumption that the noise, described by the stochastic variable $\varepsilon$, is independent of the underlying process in the absence of noise, described by the function $f(\vec{x})$.

We want to infer the unknown function $f$, given a set of $N$ observations $\{y_n\}$ at positions $\{\vec{x}_n\}$ (the ``training points'' or ``data''):
\[
  \GPdata\equiv\{\mat{X},\vec{y}\}=\{(\vec{x}_n, y_n) \,|\, n=1,\dots,N\} .
\]
Our aim is to predict the value of the function at a new input position (the ``test point'') $\vec{x}^*$.
Ideally, we also want to determine the ``trustworthiness'' of the prediction; we want a measure of the \emph{error} of our predictions.
To this end we use Bayesian modelling, a powerful, probabilistic approach to regression analysis.

\subsection{Bayesian Modelling}

In Bayesian modelling, all variables are described by probability distributions rather than a single value. This allows us to explicitly take into account the uncertainty that may generally be present due to approximations, limited sampling, or for any other reasons.

We first consider a parametric model with a fixed functional form $f(\vec{x}|\GPparam)$ that depends on a set of parameters $\GPparam$.
For example, linear regression would be described by the model $f(\vec{x}|\GPparam)=\sum_h w_h x_h$.

We explicitly state our prior assumptions about the functional relationship $f(\vec{x}|\GPparam)$, before making any observations, in the distribution $P(\GPparam)$ over the possible values of the parameters. This is the \emph{prior} distribution.
Note that we always have some prior assumptions; making them explicit in the form of probability distributions allows us to reason about them and make more powerful predictions.

The next step is to calculate the \emph{likelihood} $P(\GPdata|\GPparam)$ of our observations under the model, as a function of the values of the parameters $\GPparam$.
Finally, we use Bayes's rule to determine the \emph{posterior} probability $P(\GPparam|\GPdata)$ of the parameters given our observations:
\begin{align*}
  P(\GPparam|\GPdata)
    &= \frac{P(\GPparam) \, P(\GPdata|\GPparam)}
           {P(\GPdata)} .
\intertext{The marginal likelihood $P(\GPdata)$ is a constant normalisation factor and does not have to be evaluated directly. Hence, we can write}
  P(\GPparam|\GPdata) &\propto P(\GPparam) \, P(\GPdata|\GPparam),
\shortintertext{or schematically:}
  \text{Posterior} &\propto \text{Prior} \times \text{Likelihood} .
\end{align*}
We can predict new function values $f^* = f(\vec{x}^*)$ by integrating over the parameters: %
\[
  P(f^* | \vec{x}^*, \GPdata) = \ints P(f^* | \vec{x}^*, \GPparam) P(\GPparam|\GPdata) \,\dd\! \GPparam.
\]
This is a probability distribution for the unknown function value at the test point $\vec{x}^*$; we could consider this distribution to be our prediction of the function.
In practice, we usually compute the expectation value $\expectval [f^*]$ to obtain a single value for the prediction. We can compute the variance $\variance [f^*]$ as a measure of the ``uncertainty'' of the predicted value of the inferred function at $\vec{x}^*$.

The quality of the prediction depends on the choice of model%
\footnote{We could also include the \emph{model} (that is, the specific functional form) in the ``parameters''. We would then have to specify a prior distribution over models, calculate the posterior, and integrate over all models.}
and the prior distribution of the parameters. 
Even a probabilistic parametric regression remains constrained by the fixed functional form of the chosen model.

\subsection{Distributions over Functions}

Instead of imposing a functional form and specifying a prior distribution over its parameters, we can directly consider distributions over \emph{functions}. This results in a very flexible, nonparametric model. We can write the posterior probability as
\[
  P(f(x)|\vec{y},\mat{X}) = \frac{P\big(\vec{y}|f(x),\mat{X}\big) P\big(f(x)\big)}{P(\vec{y}|\mat{X})}
  \propto P\big(\vec{y}|f(x),\mat{X}\big) P\big(f(x)\big) .
\]
We can picture a distribution over functions as an extension to multivariate distributions, in the limit of infinitely many coordinates:
consider sampling a function $f(\vec{x})$ at positions $\vec{x}_n$, where $n=1,\dots,N$. We denote the function values $\{f_n\} = \{f(\vec{x}_n)\}$ by the vector $\vec{f}$. Taking $N\to\infty$, we can \emph{define} the function by all the values it takes, $f(\vec{x})\equiv\vec{f}$.
Even though functions correspond to vectors of infinite length, in practice there is only ever a finite number of observations and predictions of the function values.

A suitable prior distribution over functions is given by the Gaussian process, a generalisation of the multivariate Gaussian distribution to the infinitely-dimensional vector space of functions.
Of course there are many other distributions over functions, but a Gaussian process can be treated analytically rather than having to approximate it numerically, which results in much higher computational efficiency.

Whereas a multivariate Gaussian distribution is defined by its mean vector and covariance matrix, the Gaussian process is defined by a mean function $m(\vec{x})$ and a covariance function $k(\vec{x},\vec{x}')$ (also referred to as the kernel).
The defining characteristic of a Gaussian process is that, for any finite subset of input vectors $\{\vec{x}_n\}$, the function values $f_n$ follow a multivariate Gaussian distribution, $f_n \sim \mathcal{N}(\vec{m},\matSigma)$, where the mean vector is given by $m_n=m(\vec{x}_n)$, and the covariance matrix is given by $\Sigma_{n n'} = k(\vec{x}_n,\vec{x}_{n'})$.
This implies
\begin{align*}
  m(x) &= \expectval \big[f(x)\big] \qquad\text{and} \\
  k(x,x') &= \variance \big[f(x)\big] = \expectval\big[\big(f(x)-m(x)\big)\big(f(x')-m(x')\big)\big]
\end{align*}
for any $x,x'$.

We can derive the Gaussian process as the extension to infinitely many basis functions of a parametric model that is linear in the parameters. This will be summarised in the following. %
To keep the notation simple, we show how to model a function of a single scalar argument. However, the derivation straightforwardly generalises to functions over $\mathbb{R}^N$.

We can expand the unknown function $f$ in an arbitrary basis set of $H$ basis functions $\phi_h(x)$:
\begin{equation}
  f(x) = \sum_{h=1}^H w_h \phi_h(x) .
  \label{eq:gp:basisexpansion:function}
\end{equation}
A specific choice of weights $\vec{w} = \{w_h\}$ then describes a single function.
The set of basis functions can be chosen such that we can in principle describe any function.

If we consider the weights $w_h$ to be probabilistic variables, \cref{eq:gp:basisexpansion:function} describes a prior distribution over functions, determined by our choice of the distribution of $w_h$.
For a set of input points at positions $\{x_n\}$,
the corresponding function values $\vec{f} = \{f_n\}$ are given by
\begin{equation}
  f_n = f(x_n) = \sum_h w_h \phi_h(x_n) = \sum_h R_{nh} w_h ,
  \label{eq:gp:basisexpansion:vector}
\end{equation}
where $ R_{nh} = \phi_h(x_n) $ is the $N \times H$ matrix of basis functions evaluated at the input points.

We now make the assumption that the weights $w_h$ are independently and identically distributed according to a multivariate Gaussian distribution with zero mean and constant variance $\sigma_w^2$,
\[
  \vec{w} \sim \normaldist{0,\, \sigma_w^2 \identity} .
\]
Then the function values $f_n$, as linear combinations of Gaussian-distributed random variables, also follow a multivariate Gaussian distribution, again with zero mean. Their covariance matrix $Q$ is given by
\[
  Q = \average{\vec{f} \transposed{\vec{f}}}
  = \average{R \vec{w} \transposed{\vec{w}} \transposed{R}}
  = R \average{\vec{w} \transposed{\vec{w}}} \transposed{R}
  = \sigma_w^2 R \transposed{R} ,
\]
and hence the function values are distributed according to
\[
  P(\vec{f}) = \normaldist{0, Q} = \normaldist{0, \sigma_w^2 R \transposed{R}} .
\]
Let us now consider a set of radial basis functions of width $\ell$, centred at positions $h$:
\[
  \phi_h(x) = \exp\!\left( - \frac{(x-h)^2}{2 \ell^2} \right)\! .
\]
The matrix multiplication $R \transposed{R}$ corresponds to a sum over all basis functions:
\begin{equation}
  Q_{n n'} = \sigma_w^2 \sum_h \phi_h(x_n) \phi_h(x_{n'}) .
  \label{eq:gp:covariancebasissum}
\end{equation}
In the limit of infinitely many basis functions, $H \to \infty$, the sum over $h$ becomes an integral. Extending the integration limits to $\pm\infty$ and scaling $\sigma_w^2$ appropriately, we can evaluate Eq.~\eqref{eq:gp:covariancebasissum} analytically. This results in
\begin{align*}
  Q_{n n'} \propto &
  	\int_{-\infty}^{\infty}
  		\exp\!\left( - \frac{(x_n - h)^2}{2 \ell^2} \right)
		\exp\!\left( - \frac{(x_{n'} - h)^2}{2 \ell^2} \right)
	\dd\! h \\
	& = \sqrt{\pi \ell^2} \exp\!\left( - \frac{(x_n - x_{n'})^2}{4 \ell^2} \right)\! .
\end{align*}
Hence we can write the elements of the covariance matrix as
\[
  Q_{n n'} = \delta^2 \exp\!\left( - \frac{(x_n - x_{n'})^2}{2\theta^2} \right)\! ,
\]
where we introduced the new length scale $\theta = \sqrt{2} \ell$ and $\delta^2$ contains the prefactor describing the magnitude scale of the covariances between function values.

This means that the covariance matrix is completely described by a covariance \emph{function} between input points,
\begin{equation}
  C(x_n, x_{n'}) = \delta^2 \exp\!\left( - \frac{(x_n - x_{n'})^2}{2\theta^2} \right)\! .
  \label{eq:gp:sqexpcovariance}
\end{equation}
The covariance function is also called the ``kernel''. Note that there is no longer any dependence on explicit basis functions; the prior is fully specified by the kernel.
The covariance function defined by Eq.~\eqref{eq:gp:sqexpcovariance} is commonly referred to as the squared exponential kernel.
This is by no means the only choice, and we discuss the effects of this choice in the next section.

\subsection{Covariance Function and Hyperparameters}
\label{sec:gp:kernel}

In the previous section we saw that the Gaussian process prior is defined by the covariance function. This relates the covariance between two function values (how much a change in one affects the other) to the ``distance'' between the corresponding input points: $k(\vec{x},\vec{x}')$ provides a measure of similarity between two input feature vectors $\vec{x}$ and $\vec{x}'$. In fact, a distance metric can be defined based on the kernel\cite{KernelDistanceIntro}.
This motivates more complex approaches that directly calculate a similarity measure between atomic environments\cite{SOAP}.
In principle, any function that results in a positive-definite covariance matrix can be used as a covariance function.

The choice of covariance function determines the ``typical behaviour'' of functions drawn from the Gaussian process. In this way we can impose properties such as continuity, smoothness/differentiability, %
periodicity, and so on. Samples drawn from Gaussian processes with different kernels are shown in \cref{fig:gp:kernel-priors} on the following page.
\begin{figure}[htbp]
  \centering
  \input{fig_gp_priors.pgf}
  \mycaption{Samples from Gaussian processes with different covariance functions.}{\\ This shows three randomly drawn samples, illustrating different types of priors that can be used in Gaussian process regression. From left to right: squared exponential kernel~\eqref{eq:gp:sqexpcovariance} with $\delta=1$ and $\theta=1$; periodic kernel with period $4$: $k(x,x')= \exp[-\half \sin(\pi(x-x')/4)^2 ]$; Mat\'ern kernel with $\nu=1/2$: $k(x,x') = \exp(- \abs{x-x'})$.}
  \label{fig:gp:kernel-priors}
\end{figure}

We now focus on the squared exponential kernel introduced in Eq.~\eqref{eq:gp:sqexpcovariance}. Generalised to $d$-dimensional input vectors, the squared exponential kernel is given by
\begin{equation}
  k(\vec{x},\vec{x}') = \delta^2 \exp\!\Bigg( \!- \half \sum_{i=1}^d \bigg( \frac{x_i - x'_i}{\theta_i} \bigg)^2 \Bigg) ,
  \label{eq:gp:sqexpcovariance:ard}
\end{equation}
where we also introduced the possibility of different length scales $\theta_i$ for each input dimension.
\Cref{eq:gp:sqexpcovariance:ard} is infinitely differentiable, which means that it corresponds to a prior distribution of continuous, smooth functions. This is well suited to potential energy surfaces as they are relatively smooth functions of the atomic coordinates; coarse-graining further smoothens the interaction potential surface. Hence, this is the kernel we use in the remainder of this thesis.

\subsubsection{Hyperparameters}
\label{sec:gp:hyperparameters}

Kernels typically depend on \emph{hyperparameters}. This nomenclature distinguishes them from the parameters of a fixed functional form used in a parametric model; the hyperparameters are parameters of the \emph{prior}.
The hyperparameters influence the distribution over functions described by the kernel.
Our choice of kernel function and hyperparameter values contains our beliefs about the behaviour of the function $f$; they depend on the system we want to model, but not on the training data.

The squared exponential kernel~\eqref{eq:gp:sqexpcovariance:ard} has the hyperparameters $\delta$ and $\{\theta_i\}$.
$\delta^2$ describes the signal variance; hence, $\delta$ corresponds to the magnitude of typical fluctuations of the function.
$\theta$ is the length scale over which the correlation between values of the function at different points decays or, equivalently, the length scale over which the function changes appreciably. This is illustrated in Fig.~\ref{subfig:gp:prior} on \mypageref{fig:gp:picture}.
There can be different values of $\theta$ for the different dimensions of the input space.

Ideally, we would marginalise over the space of hyperparameters, for example using Monte Carlo. However, for practical applications this is generally infeasible, and a single value is used for each hyperparameter.
In principle it is possible to infer the optimal values of the hyperparameters from the data\cite{GaussianProcessesMachineLearning}. However, in the applications we consider, this would be too computationally expensive.

Fortunately, for the molecular systems in which we are interested we can estimate the relevant scales from physical/chemical intuition and other considerations, and hence can determine reasonable values for all the hyperparameters without resorting to automatic determination. This is a great advantage of Gaussian process regression.

Note that the influence of the prior on predictions reduces as the volume of training data increases. With a sufficient volume of training data, the Gaussian process regression is stable with respect to changes in the hyperparameters.

\subsection{Predictions}
\label{sec:gp:prediction}

As with any distribution, we can draw samples from a Gaussian process; in this case, each sample is a \emph{function}. In practice, we only sample the set of function values on a finite grid.
Pictorially, we could consider taking a number of samples from the Gaussian process prior. The mean of these sampled functions approximates the prior mean function and their covariance approximates the covariance function.

\begin{figure}[htbp]
  \centering
  \begin{subfigure}{0.48\textwidth}
    \includegraphics[width=\textwidth]{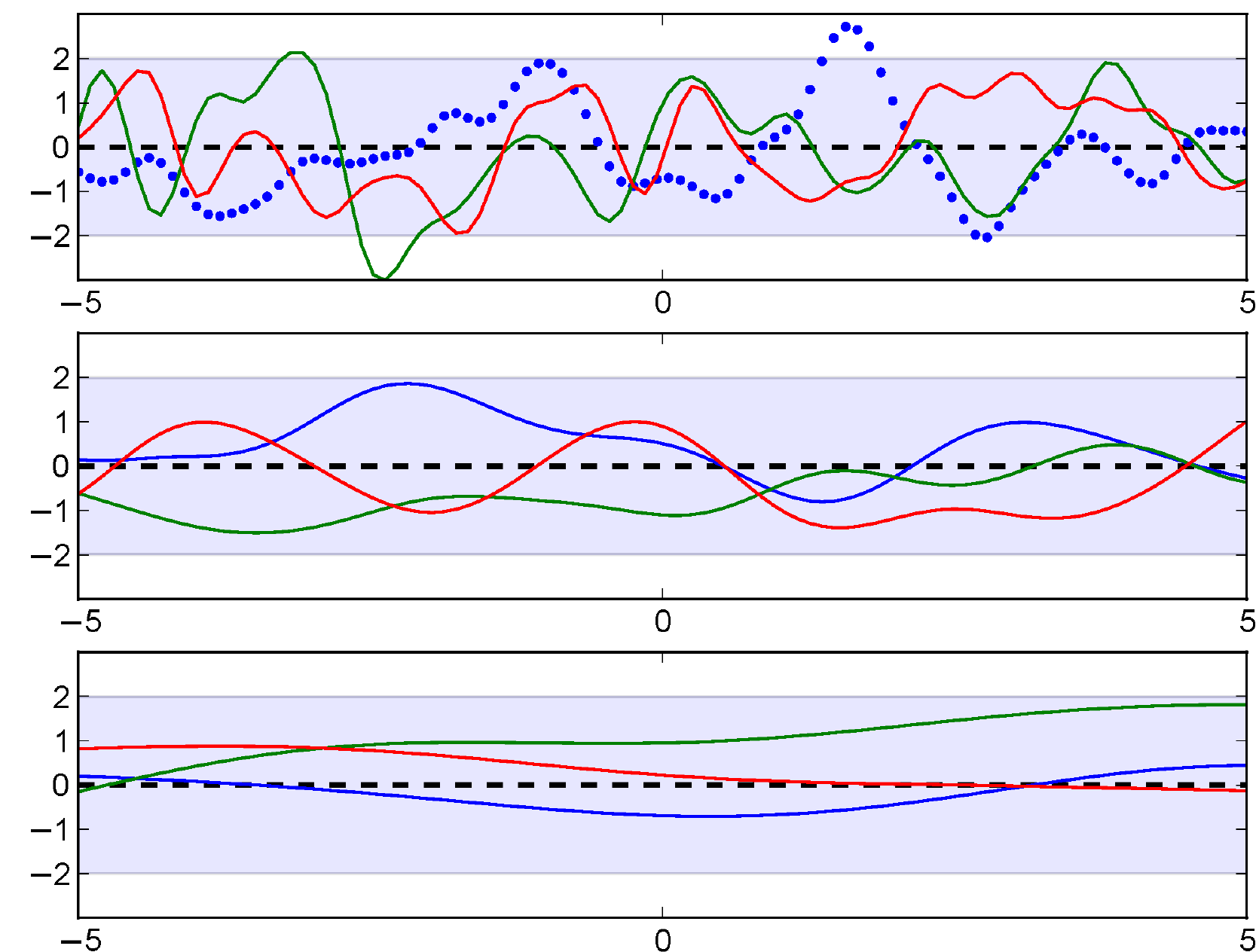} %
    \caption{Prior}
    \label{subfig:gp:prior}
  \end{subfigure}
  \hfill
  \begin{subfigure}{0.48\textwidth}
    \includegraphics[width=\textwidth]{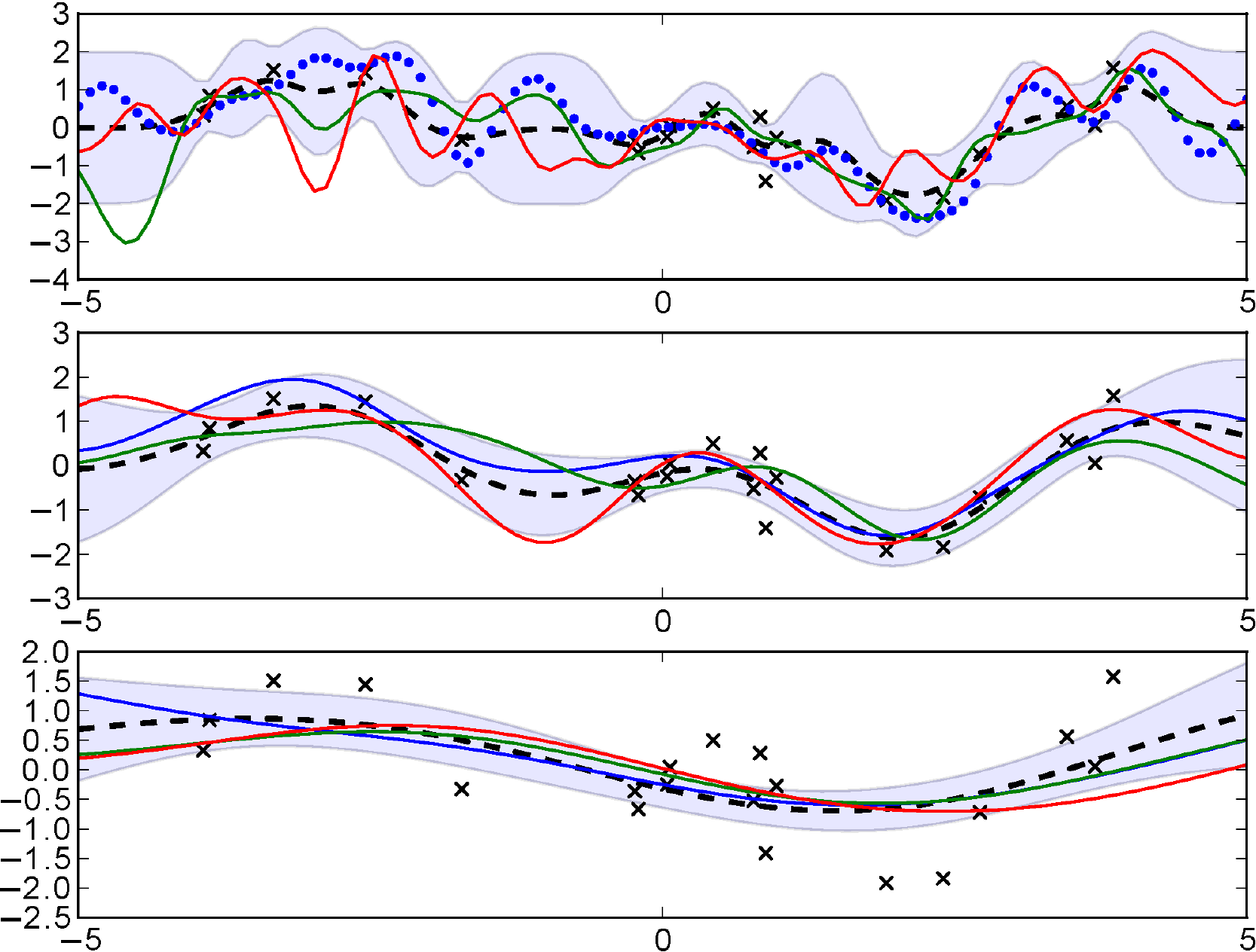} %
    \caption{Posterior}
    \label{subfig:gp:posterior}
  \end{subfigure}
  \mycaption{Illustration of Gaussian process regression in one dimension.}{\\
  \subref{subfig:gp:prior}~Prior distribution using the squared exponential kernel~\eqref{eq:gp:sqexpcovariance:ard} [\cf Section~\ref{sec:gp:kernel} on \mypageref{eq:gp:sqexpcovariance:ard}]. The effect of different values for the length-scale hyperparameter is shown, from top to bottom: $\theta=0.3,\,1,\,3$. The signal variance is $\delta^2=1$.
  The dashed line denotes the mean function, and the shaded area corresponds to the \SI{95}{\percent} confidence interval based on the predicted variance at each point.
  In each case three samples were drawn at random from the Gaussian process: The dots explicitly show the discrete sampling of a function; the other functions were approximated by joining the sampled points.\\
  \subref{subfig:gp:posterior}~Posterior distribution of the Gaussian process after training on the (noisy) data (represented by crosses).
  The training data was generated by drawing a function from a Gaussian process as in~\subref{subfig:gp:prior} for $\theta=1$, with additive Gaussian noise of variance $\sigmanoise^2=0.25$.
  Note that outside the range covered by the data, the variance of the posterior distribution returns towards that of the prior.}
  \label{fig:gp:picture}
\end{figure}

We then take into account the observed data: for each sample we evaluate whether it agrees with the data; this corresponds to calculating the likelihood.
If a sampled function does not describe the data, we discard it.
The set of remaining samples describes the posterior distribution. We can determine the expected function value and its variance at any point by calculating the mean and variance of our samples of the posterior distribution.
This is illustrated in Fig.~\ref{fig:gp:picture}. %

\myvspace
In practice, calculating an accurate mean in this way would require far too many samples to be computationally feasible. Fortunately, for a Gaussian process prior we can determine the mean and variance of the posterior distribution analytically.
The posterior distribution will again be a Gaussian process. In the following we show how to predict the function value at a single test point $\vec{x}^*$.

We consider a Gaussian process prior with covariance function $k(x,x')$ and zero mean function, $m(x)=0$.
This is without loss of generality:
we can introduce a fixed mean function $m(x)$ by considering the Gaussian process regression on the modified relation $\tilde{y}(x) = \big(y(x) - m(x)\big)$, which then has zero mean.
We can obtain predictions of the original function using $y(x^*) = \tilde{y}(x^*) + m(x^*)$.
It is also possible to consider a prior distribution over mean functions and include this in the regression; we refer the interested reader to the literature~\cite{GaussianProcessesMachineLearning}.

The likelihood for the observations $\vec{y}$ at the $N$ training points $\{\vec{x}^n\} \equiv \mat{X}$ is given by the multivariate Gaussian distribution
\[
  \vec{y} | \mat{X} \sim \normaldist{0, \mat{C}_{N}} ,
\]
where $\mat{C}_{N} = \mat{K}(\mat{X},\mat{X})$ is the $N\times N$ matrix of covariances between the training points, with $[\mat{K}(\mat{X},\mat{X})]_{mn} = k(\vec{x}^m,\vec{x}^n)$. %

We now consider the joint distribution under the Gaussian process prior of observing both the training data $(\mat{X},\vec{y})$ and the function value $f^*$ at a test point $\vec{x}^*$. This is again a multivariate Gaussian distribution:
\newcommand*{\yfvector}{\begin{bmatrix}
    \vec{y} \\
    f^*
  \end{bmatrix}}
\begin{equation}
  \left.\yfvector\right| \mat{X}, \vec{x}^*
  \sim \mathcal{N} \!\left( 0,
  \begin{bmatrix}
    \mat{K}(\mat{X},\mat{X}) & \vec{k}(\mat{X},\vec{x}^*) \\
    \transposed{[\vec{k}(\mat{X},\vec{x}^*)]} & k(\vec{x}^*,\vec{x}^*)
  \end{bmatrix}
  \right)\! ,
  \label{eq:gp:covariance}
\end{equation}
where $\vec{k}(\mat{X},\vec{x}^*)$ is the vector of the covariances between the test point $\vec{x}^*$ and each of the training points.
The joint probability for this multivariate Gaussian distribution is given by
\begin{equation}
  P(f^*, \vec{y}) \propto \exp\!\left( -\half \begin{bmatrix}\transposed{\vec{y}} & f^*\end{bmatrix} C^{-1}_{N+1} \yfvector \right)\! ,
  \label{eq:gp:jointprobability}
\end{equation}
where $C_{N+1}$ is the covariance matrix of the joint probability,
\[
  C_{N+1} =
  \begin{bmatrix}
    [ C_N ]			&	[ \vec{k} ] \\
    [ \transposed{\vec{k}} ]	&	[ \kappa ]
  \end{bmatrix}\! ,
\]
with $\kappa = k(\vec{x}^*, \vec{x}^*)$.
We could calculate $C^{-1}_{N+1}$ by brute-force matrix inversion, but in fact the inverse can be calculated more elegantly\cite{barnett1979matrix}.
Writing the inverse as
\[
  C^{-1}_{N+1} =
  \begin{bmatrix}
    [ M ]			&	[ \vec{m} ] \\
    [ \transposed{\vec{m}} ]	&	[ m ]
  \end{bmatrix}
\]
and setting $C_{N+1} C^{-1}_{N+1} = \identity$, we obtain
\begin{align*}
  m &= (\kappa - \transposed{\vec{k}} C^{-1}_N \vec{k})^{-1} , \\
  \vec{m} &= - m C^{-1}_N \vec{k} , \\
  M &= C^{-1}_N + \frac{1}{m} \vec{m} \transposed{\vec{m}} .
\end{align*}
We now want to determine the \emph{conditional} distribution for $f^*$, given $\vec{x}^*$ and the training data $(\mat{X},\vec{y})$. This amounts to marginalising over all other rows in Eq.~\eqref{eq:gp:jointprobability}. %

For multivariate Gaussian distributions this can be calculated exactly~\cite{GaussianProcessesMachineLearning,MacKay:itprnn}. The result is our posterior distribution:
\[
  f^* | \vec{y}, \mat{X}, \vec{x}^* \sim \normaldist{\hat{f}_*, \hat{\sigma}_*^2} ,
\]
corresponding to
\[
  P(f^* | \vec{y}) \propto \exp\!\left( - \frac{(f^* - \hat{f}_*)^2}{2 \hat{\sigma}_*^2} \right)\! ,
\]
where the expectation value or mean of the prediction is
\begin{align}
  \hat{f}_* &= \transposed{\vec{k}} C_N^{-1} \vec{y}
  \label{eq:gp:meanprediction}
\intertext{and the variance, describing the uncertainty in the mean, is}
  \hat{\sigma}_*^2 &= \kappa - \transposed{\vec{k}} C_N^{-1} \vec{k} .
  \label{eq:gp:variance}
\end{align}
$\hat{f}_*$ as a function of $\vec{x}^*$ corresponds to the mean function of the posterior Gaussian process. While the mean function of the prior was assumed to be zero, the mean function of the \emph{posterior} distribution is not zero.
From now on we will identify the result of our prediction with the expectation value, $f^* \equiv \hat{f}_*$.

An important aspect of Gaussian process regression is that we only need to calculate $C_N^{-1}$, and not the inverse of $C_{N+1}$. This allows us to separate the calculation into a ``training'' stage involving the matrix inversion (scaling cubically with the number of training points $N$), and a prediction that scales linearly in $N$.
We can write the expectation value~\eqref{eq:gp:meanprediction} of the predicted posterior distribution as
\begin{equation}
  f^* = f(\vec{x}^*) = \sum_{n=1}^N \alpha_n \, k(\vec{x}^n, \vec{x}^*) ,
  \label{eq:gp:meanlinear}
\end{equation}
where the sum is over all training points.
The coefficients $\alpha_n$ are given by
\begin{equation*}
  \{\alpha_n\} \equiv \vec{\alpha} = C^{-1}_N \vec{y} .
\end{equation*}
This corresponds to a linear combination of the kernel function centred on the training data points, similar to kernel ridge regression\cite{GaussianProcessesMachineLearning}.
Hence, in contrast to parametric regression, the functional form can be considered to change with every new observation.
This is what allows the nonparametric regression using Gaussian processes to, in principle, interpolate \emph{any} function.

\subsubsection*{Noisy observations}
\label{sec:gp:noise}

So far we assumed exact, noise-free observations. However, in practice, observations are generally afflicted with some error.
As mentioned at the beginning of Section~\ref{sec:gp:regression}, we typically consider the noise to be independent and additive to the ``true'' functional relationship $f(\vec{x})$. We further assume that the error in observations follows a Gaussian distribution with variance $\sigmanoise^2$. This can be described by a conditional probability of observations $\vec{y}$ on the true function values $\vec{f}$:
\[
  P(\vec{y} | \vec{f}) \sim \normaldist{\vec{f}, \sigmanoise^2 \identity} ,
\]
where the mean is given by the exact function values, and the standard deviation $\sigmanoise$ describes the noise level.

We can account for this in the Gaussian process regression by including a noise term in the covariance between observations at training points.
Instead of being given directly by the kernel function, the covariance between two observations in the training set is now given by
\[
  [C_N]_{mn} = \covariance(y^m, y^n) = k(\vec{x}^m, \vec{x}^n) + \sigmanoise^2 \delta_{mn} .
\]
Note that the kernel function only depends on the \emph{positions} of the input points, whereas the noise term only depends on the \emph{indices}. This captures the assumption that two independent observations at the same position will still differ due to noise.
As we do not want to add noise to the predictions, the covariance between training points and test point remains unchanged.
Equivalently, we can add the noise covariance to the noise-free kernel matrix:
\[
  C_N = K + \sigmanoise^2 \identity .
\]
The prediction of the Gaussian process regression for the function value at a test point is then given by
\begin{equation}
  \hat{f}_* = \transposed{\vec{k}} [\mat{K} + \sigmanoise^2 \identity]^{-1} \vec{y} .
  \label{eq:gp:meanprediction:noise}
\end{equation}
Here we assumed the same noise covariance for all training points. However, we can also consider a diagonal noise covariance matrix with different noise levels for each training point. Each observation can have its own noise level. %

This way of accounting for noisy observations is the Bayesian approach to regularisation. In contrast to other approaches such as the MSCG/FM method (even when it is based on Bayesian statistics\cite{BayesianMSCG}), Gaussian process regression provides a \emph{physical} noise parameter that has the units of the physical observables. This means that we can directly specify the uncertainty of the training data in terms of physical quantities; we can set the actual noise level which we may be able to determine from the training data.

\subsubsection*{Notes on variance prediction}

It is a key feature of Bayesian modelling that we consider \emph{distributions} over variables, and that we do not obtain just a single prediction value but can also determine, for example, the variance of the predicted distribution, indicating how strongly we believe our prediction.

In Gaussian process regression, the predicted variance is given by Eq.~\eqref{eq:gp:variance}.
Note that this variance only depends on the \emph{positions} of training and test points, not on the values of the predicted function.
In practice, the predicted variance at a given test point is just a measure of the ``distance'' from the training points.
Thus, the value of the predicted variance may need to be treated with caution; for example, if the noise parameter is set too small, the Gaussian process regression will predict a very small variance near the training points, even though this is just a result of overfitting.
Hence it is important to remember that the predicted variance only gives an indication of whether the test point is in a region that is sufficiently well-covered by the training data.

Nevertheless, the predicted variance can be very useful for diagnosing failures of Gaussian process regression and to find out when our model has encountered a part of the input space that was not covered by the training set.

In practice, the calculation of the predicted variance involves a small regularisation constant in the sparsification procedure outlined in \cref{sec:gp:sparsification}. Because of this, in the limit of high data coverage, the predicted variance tends towards this regularisation constant. This will become apparent when considering the GAP-predicted variance in \cref{ch:gapcg}/\cref{fig:methanol:monomer}. In the limit of zero data coverage, the predicted variance tends towards the variance of the prior, $\delta^2$.

\subsection{Sums and Derivatives}

In our application of Gaussian process regression to coarse-graining, we need to be able to train from and predict derivatives and sums of function values.

\subsubsection{Derivative predictions}
\label{sec:gp:derivative:predict}

We can straightforwardly predict derivatives of the inferred function value by taking the derivative of Eq.~\eqref{eq:gp:meanprediction} with respect to coordinates $x^*_i$ of the test point:
\[
  \pderiv{\hat{f}^*}{{x}^*_i} = \pderiv{\transposed{\vec{k}}}{{x}^*_i} C^{-1}_N \vec{y} ,
\]
where the partial derivative on the right-hand side is the derivative of the covariance function.
For the squared exponential kernel~\eqref{eq:gp:sqexpcovariance:ard}, this is given by
\[
  \pderiv{k^n}{x^*_i} = \frac{x^n_i - x^*_i}{\theta_i^2} k(\vec{x}^n,\vec{x}^*) .
\]

\subsubsection{Derivative observations}
\label{sec:gp:derivative:teach}

We can also train the Gaussian process from derivative observations~\cite{gp-derivative-solak,gp-derivative-rasmussen}. Differentiation is a linear operator and hence the derivative of a Gaussian process is again a Gaussian process.

The covariance between a function value observation and a derivative observation is given by
\[
  \covariance\!\left( y^m, \pderiv{y^n}{x^n_i} \right) = \pderiv{C(\vec{x}^m,\vec{x}^n)}{x^n_i} ,
\]
and the covariance between two derivative observations is given by
\[
  \covariance\!\left( \pderiv{y^m}{x^m_i}, \pderiv{y^n}{x^n_j} \right) = \pcrossderiv{C(\vec{x}^m,\vec{x}^n)}{x^m_i}{x^n_j} .
\]
We can simply extend the covariance matrix $C_N$ accordingly, and include the derivative observations in the vector $\vec{y}$ for Eq.~\eqref{eq:gp:meanprediction}.
Being able to consistently include derivative observations is a significant advantage of Gaussian process regression.

\subsubsection{Sum observations}
\label{sec:gp:sums:teach}

Within the Gaussian process regression we can also learn from observations of sums of function values\cite{GAP:thesis}. We can describe the linear combination of function values as
\[
  F_m = \sum_n L_{mn} f(x_n) .
\]
Returning to \cref{eq:gp:basisexpansion:vector} on \mypageref{eq:gp:basisexpansion:vector}, we now write
\[
  F_m = \sum_{n,h} L_{mn} R_{nh} w_h .
\]
This still follows a Gaussian distribution, but now there is a new covariance matrix given by
\[
  Q' = \average{ \vec{F} \transposed{\vec{F}} }
  = \average{ L R \vec{w} \transposed{\vec{w}} \transposed{R} \transposed{L} }
  = \sigma_w^2 L R \transposed{R} \transposed{L}
  = L Q \transposed{L} ,
\]
where $Q=\average{\vec{f} \transposed{\vec{f}}}$ is the covariance matrix for the ``hidden'' function we want to infer from observations of sums of its values.
A similar derivation can be carried out for observations of sums of derivatives.
Predictions of sums are trivially obtained by adding the predictions of the individual terms.

\subsection{Sparsification}
\label{sec:gp:sparsification}

The linear algebra operations (specifically, the matrix inversion) involved in teaching the Gaussian process and determining the coefficients $\alpha_n$ scale as $N^3$ in computing time and as $N^2$ in memory, where $N$ is the number of training observations. Hence the number of training points that can be included in a regression is limited by the available computational resources.
(The prediction at a test point only scales in time as $N$ for the mean and as $N^2$ for the variance.)

In Section~\ref{sec:gap} we will see that in our application of Gaussian process regression the training data has a large amount of redundancy: there will be many similar input points, but as we only learn from \emph{total} forces on all CG sites, we need to include all training points and cannot simply thin out the training set. This would give rise to values of $N$ beyond computational tractability.

\newcommand*{\sparsex}{\bar{\vec{x}}}

When the input points are highly correlated, we can use \emph{sparsification} to reduce the computational cost without sacrificing much accuracy. Instead of including all closely spaced points, we introduce $S$ latent pseudo-inputs $\{\sparsex_s\}$ ($s=1,\ldots,S$), the ``sparse points'', that support the function we want to learn.

Instead of the function predictions being directly related to the observations and covariances between training points, we have a ``hidden layer'' of sparse points, and we can consider this procedure as two sequential Gaussian process regressions: in the first one we obtain the prediction at the sparse points based on the original training points, and in the second one we obtain the prediction at the test point based on the latent values at the sparse points.

There are various different approaches to the sparsification approximation~\cite{quinonero2005unifying}. %
The implementation in GAP relies on the Deterministic Training Conditional (DTC) approximation\cite{seeger2003fast}.\footnote{The more recently developed Fully Independent Training Conditional (FITC) approximation by Snelson and Ghahramani\cite{snelson2005sparse} is theoretically a better approximation, and was originally used in GAP\cite{GAP:thesis}. However, in practice it was found that FITC is much slower than DTC when including derivative observations in the training set, whereas their predictive performance is comparable.}
We calculate the covariances between sparse points:
\begin{align*}
  [K_{S}]_{s s'} &= k(\sparsex_s, \sparsex_{s'}) ,
\intertext{and between sparse and input points:}
  [K_{NS}]_{n s} &= k(\vec{x}_n, \sparsex_{s}) .
\end{align*}
The latent observations $\bar{\vec{y}}$ at the sparse points are given by
\[
  \bar{\vec{y}} = \transposed{K_{NS}} \Lambda^{-1} \vec{y} ,
\]
where $\Lambda = \sigmanoise^2 \identity$ is the noise variance. %
The sparse covariance matrix between the latent observations at the sparse points is then given by
\[
  Q_S = K_S + \transposed{K_{NS}} \Lambda^{-1} K_{NS} .
\]
To predict the function value at a test point $\vec{x}^*$, we now require the covariance between the test point and the sparse points:
\[
  [\bar{\vec{k}}]_s = k(\sparsex_s, \vec{x}^*) .
\]
Finally, the sparse predictions for function value and predicted variance are given by
\begin{align*}
  \hat{f}_* &= \transposed{\bar{\vec{k}}} Q_S^{-1} \bar{\vec{y}} \qquad\text{and} \\
  \hat{\sigma}_*^2 &= k(\vec{x}^*,\vec{x}^*) - \transposed{\bar{\vec{k}}} (K_S^{-1} - Q_S^{-1}) \bar{\vec{k}} .
\end{align*}
This reduces the required computing time to $N S^2$ for the teaching and to $S$ and $S^2$ for prediction of mean and variance, respectively. As $S \ll N$, this significantly reduces the computational cost.

In our application, the sparse points are chosen as a subset of the full set of training points. We will discuss different approaches to the selection of sparse points in \cref{sec:gap:sparsepoints}.

\clearpage
\section{Interaction Potentials}
\label{sec:gap}

In \cref{ch:cg} we saw that the ``true'' interactions for a consistent CG model are given by the Potential of Mean Force (PMF) $A(\vec{R}^N)$ for the CG coordinates $\vec{R}^N$.
Our aim is to infer the coarse-grained interaction potential from derivatives of the PMF.\footnote{In principle, it is possible to measure the value of the PMF directly, for example, using Metadynamics. However, this would require a long time to convergence, is not as accurate as methods based on integrating derivative observations, and current applications of approaches such as Metadynamics are generally limited to no more than six collective variables.}
This can be based directly on observations of mean forces determined by methods such as Constrained or Restrained MD\cite{stecher2014freeenergygp}. Alternatively, we can infer the PMF by averaging over the instantaneous collective forces\cite{LetifsPaper}.
In principle, we could use Gaussian process regression to infer the PMF directly as a function of the coordinates of all the CG sites, $A(\vec{R}_1, \ldots, \vec{R}_N)$, from the observations of mean or instantaneous collective forces.
However, this would not be a practicable approach for interaction potentials, as in coarse-graining the PMF is very high-dimensional. Moreover, we would be neglecting the knowledge we have about the properties of physical systems, such as symmetries: for instance, the PMF is invariant under translations and rotations of the whole system and under permutations of equivalent sites or molecules.
In this section we discuss how to take into account this knowledge. This distinguishes the Gaussian Approximation Potential (GAP) from a direct application of Gaussian process regression.

\subsection{Locality Approximation}
\newcommand*{\neighbourhood}{\mathcal{N}}

The exact $N$-body PMF depends, in principle, on the positions of all the CG sites in the system.
In practice, molecular systems are \emph{localised}: The mean force on any particular site only depends on the relative positions of nearby sites, and does not vary significantly under perturbations of other sites that are sufficiently far away.

This can be demonstrated explicitly in a ``locality test''. For a given site or molecule, we determine all others within a certain cutoff radius; this constitutes the ``neighbourhood'' of the central site/molecule. We keep the positions of all CG sites within the neighbourhood fixed, and measure the variance of the mean force(s) on the central site/molecule as we vary the positions of the CG sites outside the neighbourhood. As the neighbourhood is made larger, this variance decreases. We shall demonstrate this in \cref{sec:gapcg:bulk:locality}.

The locality of the mean forces motivates a separation of the $N$-body PMF $A(\vec{R}^N)$ into a sum of local contributions,
\begin{equation}
  A = \sum_\xi \sum_{i \in \xi} w^\xi(\neighbourhood_i) ,
  \label{eq:gap:pmfneigh}
\end{equation}
where $\xi$ describes different types of local contributions, $i$ iterates over all instances of this type in a given configuration of the system, and $\neighbourhood_i$ describes the ``neighbourhood'' or local environment on which the $i\nth$ local contribution depends.
Such a separation into local terms is a feature of \emph{all} interatomic potentials.\footnote{There may be additional long-ranged terms, such as electrostatic and/or dispersion forces, that do not fall into this schema and would need to be treated explicitly.}
Instead of inferring the full PMF, Gaussian process regression can then be used to infer the local free energy contributions $w^\xi$.

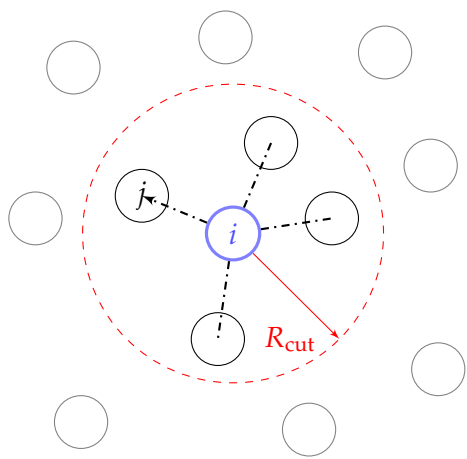
\begin{figure}[hbt]
  \centering
  \begin{tikzpicture}[
  help lines/.style={thin,draw=black!50},
  line/.style={draw,thick},
  myarrow/.style={-latex'},
  site/.style={circle,inner sep=0pt,minimum size=7mm},
  centralsite/.style={site,draw=blue!50,very thick},
  innersite/.style={site,draw=black},
  outersite/.style={site,draw=black!50},
]
  \node (sitei) at (3,3) [centralsite,label={[blue!70]center:$i$}] {};
  \coordinate [label={[label distance=5pt,red]left:$R_\text{cut}$}] (cutoff) at (4.4,1.6);
  \node [draw=red,dashed,circle through=(cutoff)] at (sitei) {};
  \coordinate (sitej) at (1.8,3.5);
  \coordinate (site1) at (3.5,4.2);
  \coordinate (site2) at (4.3,3.2);
  \coordinate (site3) at (2.8,1.6);
  \node at (sitej) [innersite] {$j$};
  \node at (site1) [innersite] {};
  \node at (site2) [innersite] {};
  \node at (site3) [innersite] {};
  \path[line,myarrow,dash dot] (sitei) -- (sitej);
  \path[line,dash dot] (sitei) -- (site1);
  \path[line,dash dot] (sitei) -- (site2);
  \path[line,dash dot] (sitei) -- (site3);
  \path[myarrow,draw=red] (sitei) -- (cutoff);
  \node at (1.0,0.5) [outersite] {};
  \node at (4.0,0.4) [outersite] {};
  \node at (5.7,1.2) [outersite] {};
  \node at (5.6,3.9) [outersite] {};
  \node at (5.0,5.4) [outersite] {};
  \node at (2.9,5.6) [outersite] {};
  \node at (0.9,5.2) [outersite] {};
  \node at (0.4,3.2) [outersite] {};
\end{tikzpicture}
  \caption{The neighbourhood $\neighbourhood_i$ of a site $i$ (blue, centre). Only sites within a cutoff radius $R_\text{cut}$ (dashed red circle) are included. The neighbourhood consists of the set of (vectorial) distances $\vec{R}_{ij} = \vec{R}_j - \vec{R}_i$ to the other sites $j$ (dash-dotted lines) for which $\norm{\vec{R}_{ij}} < R_\text{cut}$.}
  \label{fig:gap:neighbourhood}
\end{figure}

The definition of the neighbourhood $\neighbourhood_i$ depends on the type $\xi$, but it can be considered to contain all or a subset of the positions of the CG sites within a certain cutoff radius; for example, this can be one central site and all other sites within a cutoff distance (illustrated in \cref{fig:gap:neighbourhood}), or the sites describing a pair of molecules whose centres of mass are within a cutoff.

To make use of this approach in Gaussian process regression, we still need to specify how the local free energy contribution $w^\xi$ depends on the neighbourhood.
We note that the molecular-mechanics-style interaction terms commonly used in coarse-graining approaches can all be described within this framework, though this is often not discussed explicitly. There, $\xi$ would specify the different types of bonds, angles, dihedral angles, and pairs of non-bonded sites. However, in those cases the concept of ``neighbourhood'' is very limited.

For example, the neighbourhood for a CG bond between two sites $i$ and $j$ would consist of just these two sites. The CG bond would then be associated with a contribution to the PMF given by $w^\text{bond}(\vec{R}_i, \vec{R}_j)$.
But the physical properties of the bond do not depend on its orientation in space; the dependence of the contribution to the free energy needs to be invariant under the symmetries of the system. We could impose this symmetry on the local free energy contribution by considering $w^\text{bond}\big(\norm{\vec{R}_i-\vec{R}_j}\big)$.
This motivates the definition of a function $d(\vec{R}_i,\vec{R}_j) = \norm{\vec{R}_{i} - \vec{R}_{j}}$ that converts the positions of the sites within a neighbourhood into an argument to a local interaction potential term, while incorporating invariance under translation, rotation, and permutation of sites. We call this a ``descriptor'', in the following denoted $\vec{d}^\xi$.
More generally, this allows us to write \cref{eq:gap:pmfneigh} as
\[
  A = \sum_\xi \sum_{i \in \xi} w^\xi\Big(\vec{d}^\xi\big(\{\vec{R} \in \neighbourhood_i\}\big)\Big) .
\]
All interaction terms can be considered as a combination of a local free energy contribution and a descriptor.
For the molecular-mechanics-style interaction terms, the descriptor value is one-dimensional, and the local free energy term could be fitted just as well by a spline. However, in general the descriptors may return a multidimensional vector.

The choice of descriptors used in modelling a system corresponds to our assumptions about how the PMF can be separated into various local contributions.
As discussed throughout this thesis, only considering a separation into one-dimensional descriptors such as non-bonded site--site interactions can be too restrictive for accurate CG simulations. %

For molecular systems, a much less restrictive approximation is a separation of the PMF into terms corresponding to whole molecules, to pairs of molecules, and to triplets of molecules. This amounts to assigning an internal free energy contribution to each monomer configuration, and free energy contributions to all two- and three-body interactions, where ``body'' refers to the \emph{entirety} of a molecule (not just one of its individual sites).%

Here we have neglected long-range electrostatic and dispersion/van-der-Waals interactions. 
These could be treated separately,
either by including them in the fitting procedure or by setting partial charges based on the underlying atomistic system.
But in biological and other solvated systems, these long-range interactions are often screened by the solvent, and in the following we will not consider explicit long-range interactions. In this work, we will only include their contribution to the effective short-range interactions within the locality cutoff.

The forces (instantaneous collective forces or mean forces) that we can observe are gradients of the \emph{total} potential of mean force $A$. The Gaussian process can be trained on sums of derivatives and thereby infer the ``hidden'' local free energy contributions $w^\xi$ (which do not have a direct physical representation).
Together with the compactness of the neighbourhoods, this is what allows us to apply Gaussian process regression to larger systems, for which learning the PMF $A(\vec{R}^N)$ directly would be infeasible.
As the local free energy contributions only depend on local neighbourhoods, the number of interactions only grows linearly with the number of CG sites in the system. This allows GAP to be a viable approach to coarse-graining.
In the following section we discuss the descriptors that we use in the present work in more detail.

\subsection{Descriptors}
\label{sec:gap:descriptors}

A descriptor is a function that transforms the neighbourhood defined in terms of the Cartesian coordinates of CG sites into a (smaller) set of variables, the ``features'', taking into account relevant symmetries. These features are then used as the input positions in the Gaussian process regression.
In our approach, each descriptor is associated with a local free energy contribution, which is a function of the descriptor value.

Descriptors can be one-dimensional scalars, such as the distance between two sites, or a set of values, such as the three distances defining a three-site monomer, or more generally the set of all distances between CG sites that describe a monomer or a pair (dimer) or triplet (trimer) of molecules. Descriptors can also represent more complex features such as an expansion of the relative positions of all sites within the neighbourhood into spherical harmonics.

A descriptor is the representation of certain degrees of freedom of a system.
Hence, the descriptor values should be invariant under transformations that do not change physical aspects of the system. Specifically, the local interaction terms have to be invariant under rotations and translations of the set of all coordinates that are used in a descriptor.

Descriptors should be ``unique'': similar neighbourhoods should correspond to similar descriptor values, and distinct neighbourhoods should correspond to distinct descriptor values.
Together with the covariance function, descriptors have to provide a similarity measure. Physically equivalent neighbourhoods, which differ at most by a transformation that leaves the local free energy contribution unchanged, should map to the same descriptor values.

\myvspace
We now discuss the descriptors that will be used later in this thesis. This includes the site--site distance descriptor that is equivalent to splines or tabulated pair potentials used in many previous approaches to coarse-graining as well as \emph{molecular} descriptors that describe entire molecules (monomers) and their combinations as pairs and triplets.

\subsubsection{Pairwise site--site distance}

\newcommand*{\descriptorname}[1]{\texttt{#1}\xspace}
\newcommand*{\descdist}{pairwise distance\xspace}

GAP based on the \descdist descriptor is equivalent to the commonly used pairwise splines between sites. This descriptor only depends on two sites at a time and simply returns the distance between the two sites. It is useful for describing the repulsive core potential between sites and as an independent verification that GAP can generate potentials that are equivalent to those obtained by the MSCG/FM approach.

\subsubsection{Monomer-based terms}
\label{sec:gap:generalx}

\newcommand*{\descmonomer}{monomer\xspace}
\newcommand*{\descdimer}{dimer\xspace}
\newcommand*{\desctrimer}{trimer\xspace}

In our work, we focus on \emph{molecular} rather than site-based descriptors, with local contributions to the free energy associated with each single molecule (internal or monomer free energy), pairs of molecules (two-body or dimer interaction), and triplets of molecules (three-body or trimer interaction). It is important to distinguish these molecular terms from site-based terms; for example, for a three-site CG description, the molecular dimer term already corresponds to 6-body interactions between CG sites.

The molecular descriptors can be considered as a cluster expansion of the PMF:
\begin{multline*}
  A( \vec{R}_1, \ldots, \vec{R}_N ) = A( \vec{R}_{\{i,\alpha\}} )
  \approx \sum_j w_\text{mono}(\vec{R}_{j,\{\alpha\}})
  + \sum_{j,k} w_\text{dimer}(\vec{R}_{j,\{\alpha\}}, \vec{R}_{k,\{\beta\}}) \\
  + \sum_{j,k,l} w_\text{trimer}(\vec{R}_{j,\{\alpha\}}, \vec{R}_{k,\{\beta\}}, \vec{R}_{l,\{\gamma\}}) ,
\end{multline*}
where Roman letters index molecules and Greek letters index sites within each molecule.
Here we only consider systems with a single type of molecule; however, the descriptors are not limited to such systems, and we could just as easily consider, for example, a \desctrimer descriptor for three different types of molecules.

The monomer descriptor considers all monomers in the system. The dimer and trimer descriptors consider pairs and triplets of monomers that are within a cutoff based on the centre-of-mass distances.
In each case, the descriptor is constructed from the list of all pairwise distances between the sites involved in a monomer, pair, or triplet:
\[
  \{ d_{ij} \;|\; i=1,\ldots,n;\; j=i+1,\ldots,n \}
\]
where $d_{ij} = \|\vec{R}_j - \vec{R}_i\|$.
The descriptor dimension $D$ is given by the number of distances, $D = k (k-1) /2$, where $k$ is the number of sites involved in a descriptor. For a homogeneous system, the descriptor dimension as a function of the number of CG sites per molecule is given in \cref{tab:descriptordimension}.
\begin{table}[tbp]
  \centering
  \begin{tabular}{lS[table-number-alignment=center,table-format=1]S[table-number-alignment=center,table-format=2]S[table-number-alignment=center,table-format=2]}
    \toprule
    \multirow{2}{*}{\specialcell[l]{CG description\\of molecule}} & \multicolumn{3}{c}{descriptor dimension} \\
    \cmidrule(l){2-4}
    & {monomer} & {dimer} & {trimer} \\
    \midrule
    one site    & \centercell{---} &  1 &  3 \\
    two sites   &                1 &  6 & 15 \\
    three sites &                3 & 15 & 36 \\
    \bottomrule
  \end{tabular}
  \caption{Dimensions of molecule-based descriptors for one-, two- and three-site CG descriptions.}
  \label{tab:descriptordimension}
\end{table}

Note that this can be an over-complete description, with the number of distances included in the descriptor being larger than the number of degrees of freedom (which is given by $3 k - 6$, where we have excluded the ``external'' degrees of freedom corresponding to overall translation or rotation of the monomer/dimer/trimer). For example, a two-site trimer is described by 15 distances but only has 12 degrees of freedom.\footnote{It is also possible to consider only a subset of all the distances, though we do not make use of this feature in the present work.} %

It would be appealing to base the molecular descriptors directly on certain degrees of freedom, such as the centre-of-mass (COM) distance and orientation of the monomers with respect to each other, instead of including all intermolecular distances. However, it is impossible to describe orientations in such a way that the descriptor has continuous derivatives everywhere. Because of this, the interaction potential surface in the descriptor space would not be smooth and hence this could not be captured well by the Gaussian process regression.

In a one-site model, there is no monomer term, and the dimer term is equivalent to a pairwise spline. Together with a simple three-site term, this is already described well in the MSCG/FM approach. GAP becomes particularly relevant for descriptions of molecules with several CG sites, where higher-order terms are not well represented by combinations of site-based interactions. However, the dimensionality of the feature space grows very quickly with the number of sites. Hence, GAP will be primarily useful when we want to describe individual residues with a comparatively small number of sites, as in \textcite{voth:lipidhcg}, where one lipid molecule of \num{144} atoms is described by 3 to 5 sites.

A tetramer descriptor would be computationally infeasible for several reasons.
First, the number of descriptor dimensions increases factorially with the number of sites; we would require \num{28} distances for a two-site model or \num{66} for a three-site model. Not only does this increase the time taken for each evaluation of the covariance, but we would also need many more training and sparse points to accurately describe the interaction potential in this high-dimensional descriptor space, making both the GAP training and predictions much more costly. Second, in larger systems there can be many more combinations of four molecules than of pairs or triplets, vastly increasing the number of descriptors that are found in each configuration, and slowing down the force calculation even further.

\subsubsection{Other approaches}

Based on the observation that the kernel in the Gaussian process regression measures the ``similarity'' between two input vectors, there are other approaches such as SOAP\cite{SOAP} that, instead of combining a descriptor with the squared exponential kernel, directly calculate the similarity between two environments based on all positions within a cutoff radius. This has been shown to work very well for recovering local atomic energies.

However, preliminary investigations quickly showed that the use of molecule-based terms made it much easier to reproduce the mean forces. This suggests that it is advantageous to associate local free energy contributions with whole molecules and pairs/triplets of molecules rather than with individual CG sites, whereas SOAP would associate a local free energy with each CG site.
We note that SOAP might be useful in a solvent-free CG description of ions, where hydration effects and the polarisation of the water close to the ions lead to three-body effects\cite{sfcgions}.

\subsubsection{Covariance cutoff}

To ensure that the number of descriptor instances only grows linearly with the number of sites in the model, each descriptor has an explicit cutoff and depends at most on the positions of sites within that cutoff. The cutoff distance may be measured from a central site, or between centres of mass; it is not necessarily a site--site distance.

We explicitly impose locality on the descriptors by multiplying the covariance by a cutoff function. We use a cutoff function based on the centre-of-mass distance between monomers; this cutoff function goes to zero as the distance approaches the cutoff radius.
Hence the covariance for descriptors beyond the cutoff will be exactly zero.
To ensure smooth behaviour of the derivatives (forces), the cutoff function smoothly goes from one to zero over a certain transition width; in practice, a transition width of at least \SI{3}{\angstrom} results in smooth forces.

The cutoff is directly related to the locality of the potential of mean force. In practice, typical values for the descriptor cutoffs are \SIrange{3}{12}{\angstrom}. For monomer-based descriptors, the cutoff can be chosen based on minima of the COM RDF. For dimer interactions, this may extend to the second or third minimum. For trimer interactions, the first minimum is usually sufficient. Increasing the cutoff will generally increase the resulting accuracy, at a higher computational cost. Hence, we can check whether a cutoff is large enough by comparing results with those using an increased cutoff. If there is no significant difference, the smaller cutoff was sufficient.

\subsubsection{Permutational symmetry}

The features given by the descriptors, such as the ordered list of all pairwise distances, already incorporate the invariance under rotation or translation of the group of molecules involved.
However, this may not yet account for the invariance of a local free energy contribution under permutations of sites.
For example, a dimer of two identical monomers with two different types of sites is invariant under exchange of the monomers. If the sites within a monomer are of the same type, then the descriptor is also invariant under exchange of the two sites within one monomer.

For small descriptor dimensions, permutational symmetry can be imposed directly within the descriptor. For larger numbers of descriptor values, it is easier to enforce the invariance under permutations at the level of covariances.
Given the unsymmetrised covariance $k(\vec{d}^m, \vec{d}^n)$,
we can turn this into a symmetric covariance by summing over the set $\mathcal{P}$ of all permutations that leave the system invariant:
\[
  k^p(\vec{d}^m, \vec{d}^n) = \sum_{p \in \mathcal{P}} k\big(\{{d}^m_{1,2,\ldots,n}\}, \{{d}^n_{p_1,p_2,\ldots,p_n}\}\big) .
\]

\subsection{Core Potential}
\label{sec:gap:core}

When building a new interaction potential using GAP, it is possible to use any pre-existing potential as a ``core potential''. This corresponds to giving the Gaussian process prior a fixed mean function. In practice, the forces predicted by the core potential for the training points are subtracted from the observed forces, and a Gaussian process is trained on these differences. In the final prediction the results from core potential and difference-taught GAP are added together again.
There are two distinct applications for core potentials.

First, this allows us to make use of previous results, for example, if we already have a good monomer potential and now want to create a dimer potential. Introducing the monomer potential as a core potential for the training of the dimer descriptor reduces the variance or ``ruggedness'' of the energy landscape that has yet to be described. Hence, the function that needs to be inferred is smoother, which makes it easier to learn and can increase the accuracy of the overall potential.

Second, a core potential can be used to cover regions for which there is otherwise not sufficient training data to accurately represent the free energy surface in that region. This will likely be the case for configurations with high free energy, typically containing CG sites at short range due to the strong short-range repulsion between the atoms involved. The issue of insufficient training data will be discussed further in \cref{sec:gap:insufficientsampling}.

\subsection{Multiple Descriptors}
\label{sec:gap:multipledescriptors}

In practical applications, it may not be possible to provide separate training sets for different descriptors such as monomers and dimers. This may be due to technical reasons, or simply because it may not be meaningful to isolate monomers or dimers. For example, a lipid molecule can be expected to have very different properties and interactions in water and inside a lipid membrane. For this it is important that within the GAP approach we are able to teach multiple descriptors simultaneously from the same training data. %

Different descriptors are independent of each other. Hence, the covariance between local free energy contributions associated with different descriptors is always zero. This means that the full covariance matrix will be block-diagonal, with one block per descriptor type.
The values of the $\delta$ hyperparameters that describe the scale of fluctuations of functions determine how much of the variance of the target function ends up being assigned to the different descriptors.

For example, when describing a system of three molecules, the \desctrimer descriptor has the same flexibility for describing free energy surfaces by itself as when it is combined with monomer and dimer descriptors. Hence, we need to set the ratios of the $\delta$ hyperparameters according to the fraction of the free energy contribution that we expect to be due to monomer, two-body and three-body terms.
The advantage of including monomer and dimer contributions explicitly is that, in our applications, the largest contribution to the forces on CG sites is due to monomer terms; the monomer term alone already predicts most of the force on a site. This means that the left-over contribution to the forces, due solely to higher-body interactions, is smaller: this corresponds to a successive flattening out of the part of the free energy surface that is left to predict, which makes it easier to fit the trimer term, thus requiring less data and fewer sparse points.

\subsection{Selection of Sparse Points}
\label{sec:gap:sparsepoints}

When we discussed sparsification of Gaussian process regression in \cref{sec:gp:sparsification}, we assumed the set of sparse points as given. We now discuss methods for constructing this set.
Given the potentially high dimensionality of the space of descriptor values and the large number of training and sparse points we use in GAP, the selection of sparse points should, ideally, be automatic, without requiring user intervention.

The chosen sparse points need to be able to accurately represent the input data as given by the training points.
Predictions become less accurate the further away a test point is from the nearest sparse points. Hence, the sparse points need to cover that subspace of descriptor values that is \emph{accessible} by the model.
Note that there are large parts of the space of descriptor values that will have no training points at all and will never be relevant to a simulation. This can be for two reasons; firstly, because the descriptor may be overcomplete, hence some values of the descriptor cannot correspond to any configuration (for example, because a set of distances does not fulfil the triangle inequality), and secondly, because descriptor values can correspond to configurations that are completely unphysical, such as two overlapping molecules.
To accurately represent the functional relationship of the local free energy contributions, there should be a higher density of sparse points where the function value changes faster.
We now present a range of different sparsification methods. Their performance in practical applications will be compared at the end of \cref{ch:gapcg}.

\subsubsection{Random selection}
The simplest but least effective sparsification method is a random selection of sparse points from the entire training set.
This selection is uniform over training points, and hence directly equivalent to the distribution of training data. If we simply take unbiased samples from an all-atom simulation (which will be the case for learning from ICF as in the MSCG/FM method), this corresponds to a Boltzmann-weighted distribution. This results in a large number of sparse points in the minima of the PMF where the mean forces (MF) are actually zero; much fewer sparse points will correspond to configurations of molecules at short range for which the MF are large and repulsive. For this reason, random selection may not provide enough sparse points at the ``edges'' of the training set.

\subsubsection{Uniform selection}

Instead of a random selection of sparse points, which is sensitive to the distribution of the training points, we can attempt to create a uniform selection across the whole descriptor space.

For this we construct a $d$-dimensional grid, where $d$ is the number of dimensions of the descriptor. The number of bins per grid dimension is given by $n = \ceil{S^{1/d}}$, where $S$ is the number of sparse points. The bins in each grid dimension are evenly distributed between the minimum and maximum value in that dimension across all training points. The training points are then binned into this grid, and one point selected from each bin that is not empty. If there are less than $S$ non-empty bins, a non-empty bin is selected at random and another point from this bin added to the set of sparse points; this process is repeated until $S$ sparse points have been collected.

This technique is better at making sure there is an even distribution of sparse points; this has the advantage that it avoids selecting points that are close to each other, which would give rise to a high condition number of the covariance matrix. If there can be enough bins per grid dimension, uniform selection works best; this is certainly the case in few dimensions. However, uniform selection may struggle in higher dimensions. For the 15-dimensional 3-site dimer, there are at most two separate bins per dimension for any number of sparse points less than $2^{15} = 32768$. Interestingly, though, uniform selection of sparse points still performed best even in this case.

\subsubsection{CUR decomposition}

CUR decomposition can be used as a low-rank matrix approximation of an $m\times n$ matrix $A$, similar to Singular Value Decomposition (SVD).
SVD decomposes the matrix $A$ into a product of three matrices: the $m\times m$ unitary matrix of left-singular vectors of $A$ (eigenvectors of $A \transposed{A}$), the $m\times n$ diagonal matrix of singular values, and the $n\times n$ unitary matrix of right-singular vectors of $A$ (eigenvectors of $\transposed{A} A$).

The CUR decomposition approximates $A \approx C U R$, where the $m\times c$ matrix $C$ contains $c$ columns of $A$ and the $r\times n$ matrix $R$ contains $r$ rows of $A$. The $c\times r$ matrix $U$ is defined as $U = \transposed{C} A \transposed{R}$.
This is not a unique decomposition; in the present work, we implemented the algorithm by Mahoney and Drineas\cite{CURalgorithm}, which aims to select those columns and rows from the data that result in the best low-rank fit of the data matrix.

CUR decomposition is less accurate than SVD, but easier to understand: 
Whereas the left and right singular vectors represent the data in a rotated space and do not have a straightforward interpretation,
the selected rows in $R$ directly correspond to those data points that lead to the best approximation of the data matrix.
In the sparsification procedure, we take advantage of this by determining which training points are most relevant to approximate the full $N \times D$ matrix of descriptor values for the whole set of training points. These points are then used as the sparse points.
Our current implementation of CUR decomposition does not take into account the permutational symmetry of descriptors; however, this does not seem to significantly impede the performance of this selection method in practice.
CUR decomposition may be more appropriate than uniform selection specifically for higher-dimensional descriptors or when the number of sparse points is comparatively low.

The selected columns in $C$ correspond to those features that best distinguish between data points; while we do not make use of this yet, this could be used to reduce the number of descriptor dimensions.

\subsubsection{Covariance-based selection}

It is also possible to ``greedily'' select sparse points directly based on the covariance function. For this, we consider a ``working set'' of sparse points, initially filled with a single, arbitrarily chosen training point. We then calculate the predicted variance for the remaining training points. (Note that the variance predicted by the Gaussian process regression only depends on the \emph{positions} of observations, not the values.) We add to the working set that training point with the largest variance, that is, the point whose function value would be least explained by the working set of sparse points. This procedure is iterated until we have selected the desired number of sparse points.

While this approach may provide a good selection of sparse points, its disadvantage is that the calculation of the ``intermediate'' variance takes increasingly long as the size of the working set grows. For a large number of sparse points, this selection procedure becomes prohibitively slow.

A related, but significantly more efficient approach to selecting sparse points has recently been published by Schreiter \etal\cite{NovelSparsification}. Implementing and testing this approach in the context of GAP remains the subject of future work.

\clearpage
\section{Application to Coarse-Graining: GAP-CG}
\label{sec:gap:cgspecific}

\begin{figure}[p]
  \centering
  \tikzstyle{line} = [draw, thick]
\tikzstyle{myarrow} = [-latex']
\tikzstyle{block} = [rectangle, draw, fill=blue!10,
    text width=14em, text centered, rounded corners, minimum height=4em]
\tikzstyle{smallblock} = [rectangle, draw, fill=blue!10,
    text width=10em, text centered, rounded corners, minimum height=4em]
\tikzstyle{cloud} = [draw, ellipse,fill=red!20, 
    minimum height=4em]

\begin{tikzpicture}[node distance=2cm and 2cm, auto]
  \node (CVs)    [block]                                 {select CVs (CG coordinates)\\
                                                          $\vec{R}^N = \vec{M}_{\vec{R}}^N (\vec{r}^n)$ };
  \coordinate [left of=CVs, left=5cm] (leftofCVs);
  \node (AAMD)   [block, below of=CVs]                   {all-atom MD\\
							sampling of $\{\vec{r}_i\}$, $\{\vec{f}_i\}$ };
  \node (sample) [smallblock, right=3.5cm, below of=AAMD]            {select CG \\configurations};
  \node (ICF)    [smallblock, below=1cm, left=3.5cm, below of=AAMD]             {collect ICF \\ $\{\vec{\mathcal{F}}_I(\vec{R}^N)\}$};
  \node (MF)     [block, below of=sample]                {calculate MF $\{\vec{F}_I\}$\\
    (\eg Constrained MD)\\ $\vec{F}_I(\vec{R}^N) = \average{\vec{\mathcal{F}}_I}_{\vec{R}^N}$ };
  \coordinate [below of=MF, above=0.5cm, left=3.5cm] (aboveGAP);
  \node (GAP)    [cloud, above=0.5cm, below of=aboveGAP]               {GAP};
  \node (desc)   [smallblock, left=2cm, left of=GAP]             {select descriptors\\ $\vec{d}^\xi(\vec{R}^N)$};
  \node (PMF)    [block, below of=GAP]                   { sum of local \\free energy contributions \\ $\sum_\xi w^\xi\big(\vec{d}^\xi(\vec{R}^N)\big)$ };
  \node (CGpot)  [smallblock, below of=PMF]                   {CG potential\\$U_\text{CG}(\vec{R}^N) \approx A(\vec{R}^N)$};

  \path[line] (CVs) -- (AAMD);
  \path[line,myarrow] (AAMD.west) -| node[left] {Option 1} (ICF);
  \path[line,myarrow] (AAMD.east) -| node[right] {Option 2} (sample);
  \path[line,myarrow] (sample) -- (MF);
  \path[line] (MF.south) |- (aboveGAP.east);
  \path[line] (ICF.south) |- (aboveGAP.west);
  \path[line,myarrow] (aboveGAP.south) -- (GAP);
  \path[line,dotted]  (CVs.west) -- (leftofCVs);
  \path[line,dotted,myarrow]  (leftofCVs) |- (desc.west);
  \path[line,myarrow] (desc) -- (GAP);
  \path[line,myarrow] (GAP) -- (PMF);
  \path[line,myarrow] (PMF) -- (CGpot);
\end{tikzpicture}
  \caption{Workflow for GAP-CG}
  \label{fig:gap:workflow}
\end{figure}
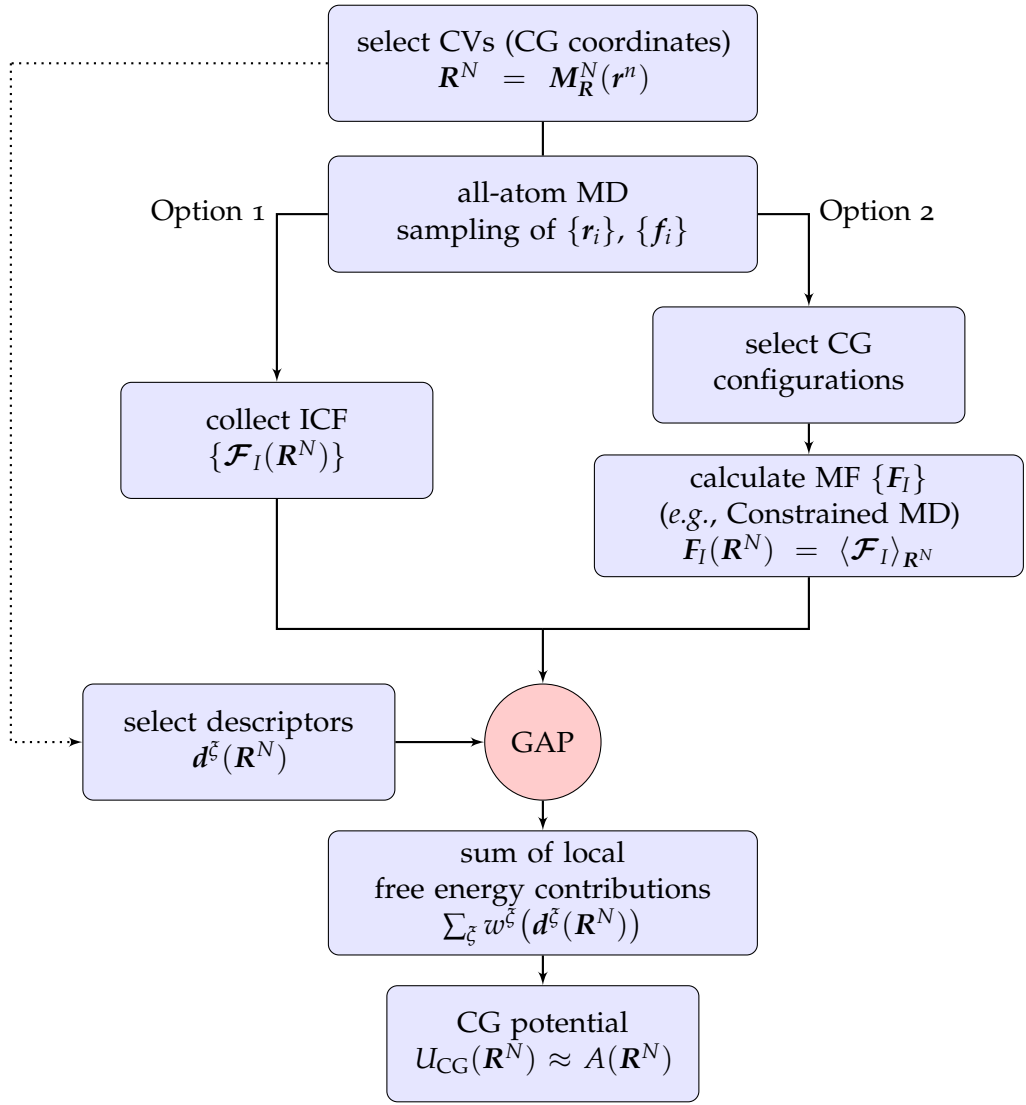

As discussed previously, the Potential of Mean Force (PMF) depends on the state point, and this carries over to the local free energy contributions that are associated with the descriptors. For example, the dimer interaction between two molecules will depend on their environment: we cannot expect a vacuum dimer potential to accurately describe the bulk model. We need to train GAP descriptors in the same environment in which we want to use them.
However, due to the locality of the mean forces, it should be possible to use the local free energy functions trained on data from small all-atom bulk systems to subsequently simulate the CG model in a larger volume of the same density, using the local free energy contributions as ``building blocks'' for the PMF of larger systems.
The typical workflow for using the Gaussian Approximation Potential in Coarse-Graining (GAP-CG) is illustrated in \cref{fig:gap:workflow}.

When using GAP to obtain interactions for coarse-grained models, observations of the function values are generally not available.
GAP can be trained solely on derivatives, and the Gaussian process regression automatically incorporates the integration of the derivative observations to the free energy surface of the CG system. However, this brings up several challenges that are not encountered as prominently when developing interatomic potentials for which we can easily evaluate the total energy.
Here we discuss these challenges, relating to noise and averaging in the GAP approach to coarse-graining (Section~\ref{sec:gap:noisy}), how to deal with inhomogeneous length scales (Section~\ref{sec:gap:transferfunction}), and how to deal with issues resulting from insufficient sampling (Section~\ref{sec:gap:insufficientsampling}).

\FloatBarrier
\subsection{Noise}
\label{sec:gap:noisy}

Gaussian process regression is an inherently regularised approach that can automatically embody Occam's Razor\cite{OccamsRazor}: we should always consider the simplest model that describes our observations. Correspondingly, a smoother function that does not overfit the observations is better at generalising to unknown test points.

Using the noise level hyperparameter $\sigmanoise$, we can tune the Gaussian process regression between closely interpolating the training data and averaging noisy observations. Note that we can directly specify the noise level as a \emph{physical} quantity; this is a significant advantage over other approaches. In many cases, other methods such as least squares minimisation also include some kind of regularisation such as Tikhonov regularisation; however, the available regularisation parameter usually does not have any direct physical interpretation. The Voth group also investigated a Bayesian extension to their MSCG/FM method\cite{BayesianMSCG}; however, their approach relies on a precision hyperparameter without direct physical meaning that needs to be optimised numerically, resulting in a high computational cost.

We can teach a CG GAP from mean forces that have been explicitly obtained by methods such as Constrained or Restrained MD. When possible, this is the favoured approach, as it is always easier to learn from data with smaller error. Such methods typically provide the estimated uncertainty in the resulting mean forces.
When learning from instantaneous collective forces (ICF), the noise level must be set appropriately so that GAP can correctly infer the underlying average. In principle, it is possible to measure the noise level in the training data by binning the ICF and measuring the variance in each bin. Note that this binning refers to the \emph{descriptor space}; this will not be considered in this thesis except for the simplest one-dimensional case, and determining the noise level from ICF for more complex descriptors remains for future work.

To prevent overfitting, it is generally better to use a larger noise level hyperparameter than one that is too small.
When a test set of mean forces is available, the noise level can be gradually reduced from above until the prediction error on the test set no longer decreases. However, this can only be relied on when the test set is completely statistically independent of the training set and at the same time contains a sufficient sample of configurations. Hence, in practice, it is preferable not to ``optimise the noise level'' and to use a somewhat larger noise level hyperparameter instead.

It is important to note that, apart from the stochastic noise in derivative observations, there are also systematic errors involved in using the descriptors to represent the PMF. As discussed in the previous section, we aim to reproduce the PMF using a sum of local free energy contributions that are associated with descriptors. With the exception of small model systems, for which it is possible to use descriptors with the same dimensionality (for example, a trimer cluster described by the trimer descriptor), this is an approximation; note that this applies to \emph{all} coarse-graining approaches that aim to reproduce the PMF. This inherent (and unknown) approximation is given by the discrepancy between the actual shape of the free energy surface in the $3N$-dimensional configuration space and the kinds of hyperdimensional surfaces that can be represented by the descriptors.
The mean forces are the gradients of the \emph{actual} PMF, but it may be impossible to describe the exact PMF as a function of all CG coordinates using the available descriptors. Hence, with respect to the ``approximate representation'' of the PMF using descriptors, even fully-converged mean forces with negligible stochastic error can still contain noise when considered from the point of view of the descriptors. This needs to be taken into account when setting the noise level for the Gaussian process regression.

As descriptors are generally unable to \emph{exactly} describe a system, they only capture an approximation of the full PMF. This means that the choice of training data can bias the resulting interaction potential; this is the case, for example, when the training set contains both trimer and tetramer configurations. Hence, the training set needs to consist of unbiasedly sampled configurations consistent with the environment which the CG potential shall describe.

\subsection{Inhomogeneous Length Scales}
\label{sec:gap:transferfunction}

Commonly, the length scale parameter $\theta$ is assumed to be constant throughout the whole range of a descriptor coordinate.
In the present work, we focus on descriptors that are based on distances between sites.
However, the interactions between atoms --- and, by extension, CG sites --- do not have a single length scale.
Typically, in the long-range regime at large intermolecular separations, the forces due to two- or higher-body interactions are relatively weak and slowly changing, and eventually go to zero. In contrast, at short range the electrostatic repulsion of overlapping electron clouds leads to a strong and quickly increasing repulsion.
This means that the length scale of the interaction changes as a function of the descriptor values.

If we use a constant length scale appropriate to the behaviour at short range, $\theta = \theta_\text{SR}$, then at larger distances we will overfit the noise. Moreover, we will lose predictive power, as this reduces the correlation length and hence more sparse points are needed to sufficiently cover the descriptor space.

If we use a constant length scale appropriate to the behaviour at long range, $\theta = \theta_\text{LR}$, we induce correlations between force observations that are actually very different from each other; this results in the strong short-range repulsion leaking over into longer range and leads to oscillations (``ringing'') in energies and forces. This effect becomes increasingly large as the value of $\theta$ is increased. The ringing can be reduced by increasing the noise level hyperparameter, at the cost of reducing the overall prediction accuracy.

The problem of inhomogeneous length scales in Gaussian processes was discussed by Gibbs\cite{gibbs1997thesis}.
One way to address this issue would be to introduce a non-stationary covariance in which the length scale depends on the values of the descriptors. However, this would require significant changes to and an increase of the complexity of the code base.
A much simpler and just as effective solution to the problem is to introduce a \emph{transfer function} that stretches a coordinate before presenting it to the Gaussian process regression. We can thus ``bend down'' the steep short-range repulsion, thereby effectively increasing the scaled distance over which the function varies at short range, increasing the \emph{effective} length scale at short range and thereby bringing it closer to the length scale at long range. This results in a more uniform effective length scale in the transformed coordinate. The transfer function is much more straightforward to implement than a non-stationary covariance, as the core implementation of Gaussian process regression (such as the covariances) does not need to be changed.

To minimise the number of additional hyperparameters, we make the assumption that the length-scale changes of different distances within the descriptor are independent of each other. This allows us to use a one-dimensional transfer function on each coordinate separately. However, we can still use different transfer functions for the coordinates; for example, we would want to distinguish between bonded and non-bonded distances.

In the following, we consider a single descriptor coordinate $x$.
Introducing the transfer function $T(x)$ leads to a new effective descriptor $\tilde{x}$, given by the transformation
$ x \to \tilde{x} = T(x) $.
Without the transfer function, the aim of the Gaussian process regression is to infer the local free energy term $w(x)$. With the transfer function, the Gaussian process has to learn a different function in the transformed input space, $\tilde{w}(\tilde{x})$. The free energy should not depend on the transfer function and, hence, we require
\[ w(x) = \tilde{w}\big(T(x)\big) . \]
This implies a transformation of the gradients $\inlinederiv{w}{x}$ of the free energy term:
\[
  \deriv{w}{x} = \deriv{\tilde{w}}{\tilde{x}} T'(x),
\]
where $T'(x) = \inlinederiv{T}{x}$ is the derivative of the transfer function and $\inlinederiv{\tilde{w}}{\tilde{x}}$ is the derivative observation used in the Gaussian process regression.
To ensure that the inferred free energy contributions remain smooth, the first derivative of the transfer function must be continuous.

We now consider the functional form of $T(x)$.
Our aim is to have one, longer length scale at long range and another, shorter (effective) length scale at short range, and a smooth transition from one to the other. A multiplicative factor on the distance will stretch the axis as a whole. Hence, we want an ``effective'' multiplicative factor that depends on the distance and smoothly changes from $1$ in the long-range part to the scaling factor $\alpha = \theta_\text{LR} / \theta_\text{SR}$ at short range. This multiplicative factor corresponds to the derivative of the transfer function, and an appropriate effective factor is given by the $\tanh$ function:
\begin{equation}
  \deriv{T}{x} = (1 - \alpha) \half\big[\!\tanh\big((x-r_0)/w\big) + 1\big] + \alpha .
  \label{eq:gap:transferfunctionderivative}
\end{equation}
For $x \ll r_0$, this goes to $\alpha$, and for $x \gg r_0$, this goes to $1$.
We can obtain the actual transfer function by integrating \cref{eq:gap:transferfunctionderivative}. This results in
\begin{equation}
  T(x) = (1 - \alpha) \half\big[w \log\big(2\cosh((x-r_0)/w)\big) + x + r_0\big] + \alpha x .
  \label{eq:gap:transferfunction}
\end{equation}
This transfer function depends on three parameters: $r_0$ is the mid-point for the transition between the length scales, $w$ is the extent of the transition region either side of $r_0$, and $\alpha$ is the stretching factor.

In the present work, we apply the transfer function only to the non-bonded distances in the \descdimer descriptor.

\subsection{Sampling Issues}
\label{sec:gap:insufficientsampling}

All coarse-graining approaches that are based on sampling of an underlying system encounter problems when the sampling is insufficient. This is particularly apparent in methods that rely on learning interactions only from derivative observations.

We can distinguish two types of insufficient sampling that are relevant to coarse-graining. The first one is the failure to sample a completely different state of the system when there are other free energy basins that, due to free energy barriers, were not encountered during the sampling of the underlying all-atom model. Addressing this problem is the focus of various free energy methods such as ABF sampling\cite{ABF2} and Metadynamics\cite{MetadynamicsPRL,MetadynamicsReview}, and will not be considered further in this work. The second type of insufficient sampling, discussed in the following, corresponds to insufficient sampling of configurations with large free energy.
Specifically, this refers to neighbourhoods or local environments associated with descriptors with large local free energy contributions.

The probability of finding the atomistic system in a certain configuration is given by the Boltzmann factor $\exp(-\beta u)$, where $u$ is the potential energy. Thus, configurations with a high potential energy --- often due to the highly repulsive core interaction between two (or more) atoms that are very close to each other --- are exponentially suppressed.
Such atomic configurations map to CG configurations that should similarly have a low probability, corresponding to a large free energy.

The strong core repulsion between atoms (due to the electrostatic repulsion between overlapping electron clouds) means that configurations in which atoms are at close range are highly penalised; the system spends very little time in such configurations.
This carries over to the CG description: for two CG sites to be close to each other, the atoms involved need to be close to each other. Correspondingly, there are few CG configurations that contain descriptors at short range. Hence, there will be few or even zero samples of regions of the descriptor spaces that are associated with large free energy contributions.
In these regions, the CG interaction should likewise be strongly repulsive. However, in an absence of training data, the Gaussian process regresses to the mean of the prior, \ie zero. Because of this regression to zero, as two CG sites are brought closer to each other, the free energy will reach a maximum and then return to zero once the corresponding descriptor ends up outside the sampled region. Hence, instead of providing a strong short-range repulsion, the predicted mean force will decrease in magnitude and eventually turn attractive, trapping the system in an unphysical minimum.
One could employ longer length scale hyperparameters in the hope that this would provide sufficient correlation with strongly repulsive force observations at distances where there is still reasonable sampling. However, this would lead to the problem of artificial ``ringing'' of the force magnitude over the whole range of the descriptor as discussed in the previous section.

When inferring the CG interactions from mean forces, we could improve sampling by choosing the CG configurations from high-temperature simulations, as long as the mean forces are evaluated in simulations at the target temperature. However, this only \emph{reduces} the likelihood that the CG model encounters undersampled short-range configurations, and is not feasible when collecting instantaneous collective forces.

To systematically overcome the issues arising from insufficient sampling in regions of high free energy, we need to provide a core potential that assigns a large free energy to configurations with CG sites at close range, so that the CG system is repelled and cannot fall into an unphysical minimum beyond the initial short-range repulsion.

The easiest and most appropriate core potential is a repulsive site--site pair potential with a short-range cutoff. As this is described by a one-dimensional descriptor, it is easy to provide sufficient training data, there needs to be only a small number of sparse points, and it is easy to visualise the resulting potential to ensure its validity.
As the system only rarely visits configurations with sites at close range, and given the small number of sparse points needed to accurately describe one-dimensional potentials, including such a repulsive short-range core potential adds no significant computational cost.

While we could choose a fixed functional form (such as a Lennard--Jones potential) and appropriately set its parameters, we can also obtain the short-range repulsion from the available training data.
For example, we can determine the repulsive short-range potential by fitting a \descmonomer descriptor together with a \descdist descriptor with a short-range cutoff. Then the \descdist descriptor will pick up only the (repulsive) short-range contribution to the interactions between non-bonded CG sites. These pairwise short-range interactions can be visualised and manually adjusted beyond the coverage by training data to ensure that they provide an appropriate short-range repulsion. %
We then use this repulsive short-range potential as a core potential for teaching the full interaction potential.

\section{Summary}

In this chapter we gave an overview of Gaussian process regression and how it can be combined with local descriptors to derive interaction potentials; this combination constitutes the Gaussian Approximation Potential (GAP). We transferred this approach to CG interactions inferred from derivative observations of the Potential of Mean Force (PMF).  Using an appropriate choice of descriptors, we can infer local contributions to the free energy surface; these are ``building blocks'' of the PMF for molecular CG models.

GAP can infer the PMF directly from observations of either mean or instantaneous collective forces. In this regard, GAP-CG is similar to the MSCG/FM approach. However, the descriptors available in GAP provide for a much more flexible, powerful approach. GAP only depends on very few settings or hyperparameters, most of which have a physical meaning and can be related to our prior knowledge about systems. Moreover, GAP allows us to quantify our confidence in the predictions, immediately highlighting issues related to a lack of coverage by training data.

In the following chapter, we will present the results of applying GAP-CG to molecular systems, demonstrating that we can recover the PMF to a high accuracy. We will also discuss the influence of the hyperparameters on the accuracy of the resulting potentials.

\chapter{GAP-CG: Recovering the Potential of Mean Force}
\label{ch:gapcg}
In Chapter~\ref{ch:cg} we showed that a CG model is consistent with the underlying atomistic potential when the CG potential $U_\text{CG}$ is equal to the many-body PMF $A(\vec{R}^N)$, a function of the coordinates of all $N$ CG sites.
A full $N$-body potential is generally infeasible, so the practical challenge is to find a way to represent the PMF.

In Chapter~\ref{ch:gap} we introduced the Gaussian Approximation Potential (GAP) as a general, flexible many-body approach to interaction potentials. GAP infers the interaction potential from observations of the forces acting on the sites in the model. In the GAP-CG approach we make the assumption that the PMF can be separated into a sum of \emph{molecular} terms, assigning local free energy contributions to internal configurations of molecules (monomer terms) and to dimers and trimers of molecules.

In this chapter we show results of applying GAP-CG to various test systems.
These results support our assumption that the PMF can be separated into a sum of molecular terms and demonstrate that this leads to much more accurate CG models than the more common approximation of separating CG interactions into terms based on pairwise radial interactions between CG sites.
To this end, we compare GAP to approaches relying on pair potentials, both structure-based (\cf Section~\ref{sec:cg:structurebased}) and force-based (\cf Section~\ref{sec:cg:mscgfm})%
.

We note that studies of coarse-graining approaches tend to analyse only radial distribution functions (RDFs) and, in a few cases, angular distribution functions (ADFs).\footnote{\textcite{Voth:dipole} is a rare example of a paper that considers distributions that are more complex than RDF and ADF, in this case the distribution of dipole orientation \vs distance.}
This can provide a reasonable characterisation for CG models in which each molecule is described by a single site.%
\footnote{For one-site CG models, the pairwise splines between sites correspond to molecular two-body interactions, and Larini \etal{} showed that explicitly including three-body interactions in a one-site model of liquid water ``can dramatically improve the reproduction of structural properties of the original atomistic system'' as measured by RDF and ADF\cite{voth:threebody}.}

In the following, we consider the more interesting case of molecules that retain internal structure in their CG description and, hence, have a defined orientation. As an example, we show results for a two-site description of methanol and a three-site description of benzene.
We discuss their chemical composition and the CG mapping in Section~\ref{sec:gapcg:systems}.

Including the orientation within the CG description allows us to consider more complex distributions to characterise the structure of the system than just RDF and ADF.
Structure-based approaches recover the RDF by construction. However, reproducing the RDF does not imply that the model will also reproduce more complex structural properties such as collective variables of the CG system, or correlations between them. In fact, we will see that this is not the case, in general.
Force-based approaches have no guarantee of reproducing the structure. While this can be a severe limitation for pair potentials, we will see that the more complex descriptors available in the GAP approach can reproduce the structure better than even structure-based approaches.

Using the \desctrimer descriptor, GAP can \emph{exactly} represent the full $N$-dimension\-al PMF of systems containing up to three molecules.
In Section~\ref{sec:gapcg:fullpmf} we demonstrate that GAP is able to recover the PMF from force observations for small methanol clusters (monomers, dimers, and trimers).
In contrast, CG models based on pairwise radial interactions fail to describe the overall structure correctly --- even if they reproduce individual distance distributions to a high accuracy.

For larger systems, GAP would still be exact \emph{if} the PMF actually separates into a sum of molecular monomer/dimer/trimer terms. In \cref{sec:methanol:tetramer} we demonstrate that this is a much better approximation for the methanol tetramer than only considering pairwise site-based interactions.

Clusters are just toy systems to demonstrate the GAP approach; there is no practical need for coarse-graining small systems. Coarse-graining only becomes relevant for large systems.
Hence, we also consider the application of GAP-CG to bulk systems.
We show that in bulk the decomposition of the PMF into contributions due to pairs and triplets of molecules provides a very good approximation to the results of all-atom calculations.
In bulk, a large contribution to the effective interactions is simply due to the packing structure. Even a potential that only models the core repulsion can already qualitatively reproduce various structural distributions.
This makes it easier for structure-based approaches such as IBI to reproduce not just the RDF but also other distributions. However, this does not mean that such pair potentials accurately represent the CG interactions.
We show that GAP, based on matching forces, reproduces the RDF as well as the pair potentials obtained by the structure-based IBI approach and even provides a \emph{better} description of the correlation between relative orientation and distance at large molecular separations.
We discuss this in Section~\ref{sec:gapcg:bulk}, where we consider both methanol and, to show that our approach can easily be applied to different systems, benzene.

In \cref{sec:gapcg:factors} we discuss how hyperparameters, sparsification and the noisiness of the training data influence the generation of CG GAP potentials.

\section{Model Systems: All-Atom Description and CG Mapping}
\label{sec:gapcg:systems}

We use methanol and benzene as the test systems for comparing GAP and competitive CG approaches.
In this section we briefly discuss these molecules and their properties.

\subsection{Methanol}
\newcommand*{\siteCHthree}{\chem{CH_3}\xspace}
\newcommand*{\siteOH}{\chem{OH}\xspace}
\newcommand*{\pairCC}{\siteCHthree--\siteCHthree}
\newcommand*{\pairCO}{\siteCHthree--\siteOH}
\newcommand*{\pairOO}{\siteOH--\siteOH}

Methanol is a small molecule with the chemical formula \chem{CH_3OH}. It is the simplest alcohol, consisting of a methyl group (\chem{CH_3}) and a hydroxy group (\chem{OH}). Its structure is shown in \cref{fig:methanolmonomerstructure}.
\begin{figure}[h]
  \centering
  \begin{tikzpicture}[help lines/.style={thin,draw=black!50}]
    \node at (2,2) {
      \chemfig{C(<:[:110]H)(<[:150]H)(-[:230]H)-O(-[:55]H)}
    };
    \draw[blue!80] (1.1,2) ellipse [x radius=0.8,y radius=1.5];
    \draw[rotate around={55:(3.1,2.4)},blue!80] (3.1,2.4) ellipse [x radius=1.1,y radius=0.3];
    \draw[red] plot[mark=x] (1.2,1.9) -- plot[mark=x] (2.95,2.2);
  \end{tikzpicture}
  \mycaption{Structure of the methanol molecule.}{The blue outlines indicate the grouping of atoms to CG sites; the CG coordinates are given by the centres of mass of all the atoms in each group and are indicated by red crosses.}
  \label{fig:methanolmonomerstructure}
\end{figure}

The oxygen in the hydroxy group is considerably electronegative and carries a partial charge of $q_\text{O} = -0.65e$. This is primarily taken from the hydrogen directly bound to it, which ends up with $q_\mathrm{H_O} = +0.42e$. Methanol molecules can form hydrogen bonds between their hydroxy groups, significantly influencing the structures they form.
Each methanol molecule can be an acceptor and a donor of a hydrogen bond, so each molecule tends to form two hydrogen bonds.

\subsubsection{CG mapping}

In previous research on the coarse-graining of liquid methanol, each molecule has usually been represented by a single site\cite{VOTCA,MSCG:II}, neglecting all information about the orientation of the molecules.\footnote{In \textcite{voth:mscg2}, Izvekov and Voth employ a two-site description equivalent to the one used in this work, but they only compare site--site RDFs with the all-atom model.} Here, we will use a two-site description, with one site at the centre of mass of the methyl group and a second site at the centre of mass of the hydroxy group.

Note that this does not allow us to model the highly directional nature of the hydrogen bond explicitly, as this is determined by the position of the hydrogen atom relative to the oxygen--carbon bond.
However, we will see that GAP-CG can recover the hydrogen-bonded structure of the methanol even without an explicit description of the hydrogen bond direction.

\subsection{Benzene}
\label{ref:benzene:structure}

Benzene (\chem{C_6H_6}) forms an aromatic six-membered carbon ring, with one hydrogen attached on the outside of each carbon atom. Due to its delocalised $\pi$ orbitals, this molecule has a very rigid planar structure. The structure of benzene is illustrated in \cref{fig:benzenemonomerstructure}.

\begin{figure}[tbhp]
  \centering
  \begin{tikzpicture}[help lines/.style={thin,draw=black!50}]
    \node at (2,2) {\chemfig{H-C**6(-C(-H)-C(-H)-C(-H)-C(-H)-C(-H)-)}};
    \foreach \phi in {0, 120, 240}
      \draw[rotate around={\phi:(2,2)},blue!80] (2,1) ellipse [x radius=1,y radius=0.25];
    \pgfplothandlermark{\usepgfplotmark{x}}
    \draw[red] plot[mark=x] (2,1) -- plot[mark=x] (2.8660254037844388,2.5) -- plot[mark=x] (1.1339745962155612,2.5) -- (2,1);
  \end{tikzpicture}
  \mycaption{Structure of the benzene molecule.}{The blue outlines indicate the grouping of atoms to CG sites; the CG coordinates are given by the midpoints between pairs of carbon atoms and are indicated by red crosses. The hydrogen atoms are not included in the CG description.}
  \label{fig:benzenemonomerstructure}
\end{figure}

\subsubsection{CG mapping}

To be able to represent the orientation of the planar benzene molecule, we employ a three-site CG description.\footnote{As a further simplification, we could approximate the six-fold rotational symmetry with continuous rotational symmetry and describe a benzene molecule with two sites along the plane normal axis. However, this could not be considered within the linear mapping framework of \cref{eq:cg:mapping:coordinates:linear} and would require more complicated calculations to obtain the forces on the CG sites.}
We place a CG site on the centre of mass of each pair of neighbouring carbon atoms. This choice is consistent with two different CG descriptions for each molecule in an all-atom configuration, though they are physically equivalent.
This means that our CG description only has three-fold rotational symmetry, even though the benzene ring actually has six-fold symmetry. In Section~\ref{sec:benzene:sixfold} we will show that GAP is able to recover the six-fold symmetry of the all-atom interactions in the CG model.
The hydrogen atoms are not part of our CG description. Their influence is only captured through the forces they exert on the carbon atoms.

\subsection{Simulation Protocol}

All simulations were carried out at $T=\SI{300}{\kelvin}$. At this temperature, methanol and benzene are in the liquid phase, which is relevant to a wide range of applications in biology and chemistry.
For the all-atom simulations, we use the \AMBER MD package\cite{AMBER12}, with the ff99SB (Hornak \& Simmerling) force field\cite{ff99SB}.
Letif Mones refined the force field for benzene using \GAUSSIAN\cite{gaussian} for geometry optimisation and calculating the partial charges. To ensure accurate MD simulations, we choose a time step of \SI{0.5}{\femto\second} for all atomistic simulations.

It is important to note that here and in the following the aim is not to reproduce real-world properties of the systems we discuss, but to investigate how well CG models can reproduce the properties of the all-atom model on which they are based.

Unless specified otherwise, we derive the force-based potentials from mean forces, so that we can distinguish issues relating to the averaging procedure from issues arising from the approximation of the PMF as a sum of molecular terms. We will discuss how GAP can include the averaging and be trained on instantaneous collective forces in Section~\ref{sec:gapcg:rawmean}.
As discussed in \cref{sec:gap:insufficientsampling}, we are specifically interested in obtaining mean forces where their magnitude is large, and hence all mean force calculations in this thesis are based on Constrained MD, for which we use the \PMFlib toolkit\cite{PMFlib}. For methanol clusters, the Constrained MD simulation for each CG configuration was run for \SI{10}{\ns}. For bulk systems, Constrained MD for all CG coordinates is much more expensive and mean forces were collected in runs of only \SI{100}{\ps}.

Coarse-grained simulations were run with \QUIP\cite{QUIP} (the in-house code for GAP\cite{GAP:PRL,SOAP}\footnote{The GAP code is available for non-commercial use from \url{www.libatoms.org}.}) and \LAMMPS\cite{LAMMPS} (for pair potentials).

\section{Methanol Clusters}
\label{sec:gapcg:fullpmf}
\label{sec:gapcg:clusters}

When the PMF can be separated into a sum of molecular terms described by the descriptors, GAP can recover the full PMF [as defined by \cref{eq:cg:CGpotential}]. This is proven by the CG MD distributions being indistinguishable from the CG view of the all-atom distributions.
This will certainly be the case when a single descriptor can completely characterise the system configuration. With the \desctrimer descriptor this is the case for up to three molecules. Here we consider clusters of one to four methanol molecules.

By considering in turn monomer, dimer, and trimer, we can use the potential from the smaller system as the core potential for the next one. This means we only need to learn genuine two-body effects in the dimer, and only genuine three-body effects in the trimer. This allows us to optimise the parameters for each descriptor separately, simplifies the modelling process and makes better use of available training data.

For a cluster of four methanol molecules, we are no longer able to reproduce the PMF exactly. However, we will see that GAP based on molecular descriptors is a much better approximation to the all-atom simulation than approaches based on pair potentials.

\subsection{Monomer}
\label{sec:methanol:monomer}

In our two-site CG model of methanol, the monomer has a single (internal) degree of freedom, the CG bond length. This allows us to visualise all properties of the model, which is no longer possible for more complex systems. Moreover, this makes it very easy to capture the CG interaction.

In our CG description, the CG bond between the \siteCHthree and \siteOH sites is very similar to the carbon--oxygen bond in the underlying all-atom simulation. The mean distance is somewhat larger, as the inclusion of the hydrogen atoms shifts the centre of mass away from carbon and oxygen atom positions. The CG force--distance relation is almost linear. Whereas the carbon--oxygen bond in our all-atom model is a perfectly harmonic spring, the CG bond also contains contributions from other interactions. For example, bending of the angles will slightly shift the CG bond length, leading to a (very small) anharmonicity. Nevertheless, the CG bond is described well by a harmonic approximation, and the corresponding force constant is comparable to the C--O all-atom force constant of \SI{23.98}{\eV\per\angstrom^2}, corresponding to \SI{928}{\kT\per\angstrom^2} at $T=\SI{300}{\kelvin}$.

\subsubsection{CG potentials}

We can obtain the CG bond potential in various ways, both structure-based and force-based, which we now describe.

\paragraph{Structure-based (Boltzmann Inversion).}

As there are no other degrees of freedom, \cref{eq:cg:dbi:prob} accurately describes the distribution of the bond length, and Boltzmann Inversion is, in principle, exact. Inaccuracies are due to finite sampling. We derive the Boltzmann-inverted potential from \num{2.6e6} samples of the CG bond length in the all-atom model. The probability distribution is determined by a histogram of the bond lengths with 100 bins between \SIlist{1.36;1.65}{\angstrom}. \Cref{eq:cg:dbi:potential} then gives the CG bond potential. In order to calculate the derivative, we fit a cubic spline to the resulting potential.

The bond force is almost perfectly linear, corresponding to a Gaussian-distributed CG bond length, and the bond potential is well described by the approximation of a harmonic potential. We calculate mean $r_0 = \expectval[p(r)/r^2]$ and variance $\sigma_r^2 = \variance[p(r)/r^2]$ of the Jacobian-corrected distribution of CG bond lengths. In the harmonic approximation, the CG bond potential is given by $U(r) = \half k (r - r_0)^2$, with the force constant $k = \kT / \sigma_r^2$. For our model, this results in $r_0 = \SI{1.5061}{\angstrom}$ and $k = \SI{1007}{\kT\per\angstrom^2}$. %

\paragraph{Force-based (GAP).}

We calculate the mean force (MF) between the two CG sites for a number of equally spaced distances on a grid. We consider two grids, one from \SIrange{1.34}{1.66}{\angstrom} and an extended grid from \SIrange{1.28}{1.85}{\angstrom}, with grid points spaced \SI{0.002}{\angstrom} apart.
For a two-site model, there is no distinction between the \descmonomer term used in GAP and a spline-based interaction between the two sites. Hence, GAP and MSCG/FM give equivalent results. For the monomer, we will only show the GAP results.

Using the grid MF, we teach the \descmonomer descriptor with the following hyperparameters: length scale $\theta = \SI{0.3}{\angstrom}$, energy scale $\delta = \SI{5}{\eV} \approx \SI{200}{\kT}$, 10 sparse points (using uniform selection), and a noise level of $\sigma_f = \SI{1}{\meV\per\angstrom} \approx \SI{0.04}{\kT\per\angstrom}$ for force observations.

For comparison, we also teach the GAP from \num[retain-unity-mantissa = false]{1e5} samples of instantaneous collective forces (ICF).
With the exception of the noise level, all hyperparameters relate to the true PMF and do not need adjustment.
For the methanol monomer we can plot the distribution of CG bond forces as a function of CG bond length, shown in \cref{fig:methanol:monomer:icf} on the following page. This allows us to \emph{measure} the noise in the ICF, given here by the standard deviation of the residuals. Hence, this directs us to use $\sigma_f = \SI{0.25}{\eV\per\angstrom} \approx \SI{10}{\kT\per\angstrom}$.
\begin{figure}[H]
  \centering
  \input{thesis_figures/methanol/monomer/icf_monomer.pgf}
  \mycaption{Methanol monomer training data.}{\textit{Top:} CG bond force as a function of bond length for instantaneous collective forces (ICF, blue) and mean forces (MF, red). \textit{Bottom:} residuals of ICF and MF from a linear fit to the ICF. Note that the ICF have a much higher variance than the MF, but that their average reproduces the MF. The standard deviation of the ICF residuals is \SI{8.7}{\kT\per\angstrom}.} %
  \label{fig:methanol:monomer:icf}
\end{figure}

\begin{figure}[htbp]
  \centering
  \input{thesis_figures/methanol/monomer/monomer.pgf}
  \mycaption{Methanol monomer CG potentials.}{
    From top to bottom (as a function of the CG bond length): CG bond force predictions; prediction error, as measured by the deviation from the reference mean forces (MF); GAP-predicted variance; distribution of the CG bond lengths in the all-atom simulation.

    We compare the following potentials with the reference MF: GAP taught from MF (on grids with two different extents), GAP taught from instantaneous collective forces (ICF), linear fit to grid MF, and Direct Boltzmann Inversion (DBI) of the bond length distribution as well as its harmonic approximation.
    
    The GAP-predicted variance clearly indicates the range covered by the training data. Note that its absolute value is meaningless: it is smaller for the ICF-taught GAP simply because this involved many more training points. Beyond the extent of the training data, it can be seen that the GAP regresses to the prior mean function (that is, zero).}
  \label{fig:methanol:monomer}
\end{figure}
In \cref{fig:methanol:monomer} we show the CG forces resulting from different approaches to obtaining the bond potential, and their deviation from the MF reference values.
These results show clearly that the ICF-trained GAP only provides accurate results where there is sufficient data coverage. There seems to be a slight systematic difference to the MF. %
The results for ICF- and MF-taught GAP demonstrate that the GAP-predicted variance is a useful measure for checking adequate data coverage, but that its absolute value needs to be treated with caution. As the ICF-trained GAP is based on a much larger number of training points, the predicted variance is smaller, even though the force prediction is generally more accurate when training from mean forces. Note that in the limit of high data coverage, the predicted variance goes towards the regularisation constant, in this case \num{e-6}.

\subsubsection{The monomer in other environments}
While the methanol monomer is not particularly interesting in itself, it provides a useful basis for generating GAP potentials for larger systems.

In our work, we found that the monomer (bonded) forces are significantly larger than the (non-bonded) forces due to two- or three-body effects.
The energy scale of the monomer PMF (based on the range of GAP predictions for the monomer term) is of the same order of magnitude as that of the dimer (based on the well depth of the centre-of-mass PMF).
However, the length scale associated with the monomer, corresponding to fluctuations in the CG bond length, is much smaller [on the order of \SI{0.1}{\angstrom}] than the typical length scale for dimer interactions [on the order of \SI{1}{\angstrom}]. Hence the magnitude of the forces produced by the monomer term is much larger, and can potentially mask the signal of dimer and trimer interactions.\footnote{This is true for our models of methanol and benzene. This may not be the case in other systems, for example when considering highly coarse-grained models (such as in \textcite{voth:lipidhcg}), where the scale of distance fluctuations within a monomer is comparable to the intermolecular distances.}
Hence, a good monomer term is essential for learning accurate dimer and trimer potentials.

The monomer bond length distribution is a useful consistency check in CG simulations to help us determine whether something has gone wrong. Obtaining an accurate bond length distribution is a necessary (though not sufficient) condition for an accurate simulation.

The monomer term does not change significantly between systems, whether in vacuum, small clusters or bulk, and so there is little change in the distribution of the CG bond length between different methanol systems.
In all cases the bond length has an average of approximately \SI{1.507}{\angstrom} and a standard deviation of \SI{0.0315}{\angstrom}, with a small trend towards shorter bond lengths in bulk.
The relative change of the average bond length between vacuum and bulk is less than \num{5e-4}.
In all cases, the harmonic approximation to Direct Boltzmann Inversion gives a force constant of approximately $\SI{26.0}{\eV\per\angstrom^2} \approx \SI{1000}{\kT\per\angstrom^2}$, with no significant difference between environments. 
Differences between the monomer in vacuum and the monomer term in dimer, trimer, and larger systems could be captured by the \descdimer or \desctrimer descriptor. However, for systems such as the bulk it might be preferable to directly train the bulk-induced monomer together with dimer and trimer terms.

\subsection{Dimer}
\label{sec:methanol:dimer}

\subsubsection{Atomistic structure}

Two methanol molecules in vacuum form a bound dimer state through the hydrogen bond between the two hydroxy groups.
In the CG model, the hydrogen bond is best described by the vector connecting the two \siteOH sites. In the CG view, this leads to a sharp peak in the \pairOO distance distributions that extends to about \SI{3.4}{\angstrom}, with its maximum at \SI{2.7}{\angstrom}.

Due to random fluctuations, the dimer can dissociate. In a true vacuum simulation, the dissociated monomers would drift apart with vanishing probability of encountering each other again.  The state with two free-floating monomers is of no further interest. We could place the dimer inside a periodic box, so that the two molecules will eventually encounter each other and be able to form a bound dimer again. However, this is a finite-size effect; in an arbitrarily large box, the probability of re-forming the dimer state goes to zero. Instead, to improve the sampling of the bound dimer state, we impose a half-open harmonic restraining potential on the distance $r_\text{COM}$ between the centres of mass of the molecules,
\begin{equation}
  U_\text{rst}(r_\text{COM}) = \begin{cases}
    0  &  r_\text{COM} \le r_\text{rst} , \\
    \half k_\text{rst} (r_\text{COM} - r_\text{rst})^2  &  r_\text{COM} \ge r_\text{rst} .
  \end{cases}
  \label{eq:gapcg:restraintpotential}
\end{equation}
We let the restraint begin at $r_\text{rst} = \SI{10}{\angstrom}$, and we set the force constant of the restraining potential to $k_\text{rst} = \SI{50}{\kcal\per\mole\per\angstrom^2} \approx \SI{84}{\kT\per\angstrom^2}$. The same restraining potential is applied to all-atom and CG simulations.

In the two-site CG model, the methanol dimer has six degrees of freedom. These can be completely described by the two bonded and four non-bonded site--site distances as in the \descdimer descriptor. To compare the different models, we will show distributions of collective variables of the CG model that are easier to understand, such as the angle between a CG bond and the \pairOO separation vector and the dot product between the two normalised bond directions.

\subsubsection{CG potentials}

Here we compare the GAP approach based on molecular terms with the MSCG/FM approach based on pair potentials.
Both approaches are based on forces, and we use a training set containing 11400 dimer configurations that were randomly sampled from the atomistic trajectory and a further 4894 dimer configurations sampled at short range (with a dimer COM separation less than \SI{3.3}{\angstrom}).

\paragraph{GAP.}

The methanol dimer is completely described by the \descdimer descriptor, which for this system contains six distances (two bonded and four non-bonded).

As discussed in \cref{sec:gap:multipledescriptors}, it is advantageous to separate monomer and dimer contributions by explicitly including a \descmonomer descriptor, as the largest contribution to the mean forces comes from the monomer bond forces. This can be seen in \cref{fig:methanol:dimer:monomerprediction}, where we compare mean forces in the dimer system with the force prediction of the monomer GAP discussed in the previous section.
\begin{figure}[tbhp]
  \centering
  \input{thesis_figures/methanol/dimer/fig_monomer_for_dimer.pgf}
  \mycaption{Comparison between monomer and dimer forces.}{This graph shows that a large contribution to the mean forces in the methanol dimer is already explained by the force predictions given by the monomer term. Two-body contributions to the dimer mean forces show up as deviations from the diagonal line.}
  \label{fig:methanol:dimer:monomerprediction}
\end{figure}

In principle, GAP can automatically split the contributions associated with different descriptors (and we will make use of this for the bulk systems).
Here, we simply use the \descmonomer GAP discussed in the previous section as a core potential. On top of this core potential, we teach the \descdimer descriptor to capture the two-body contributions to the mean forces that are not described by the monomer term.

As discussed in \cref{sec:gap:insufficientsampling}, at short range the interaction between atoms and, hence, between CG sites becomes highly repulsive, which may lead to insufficient sampling and unphysical behaviour of the CG interactions. In our approach, we handle this by introducing a short-range repulsive interaction between pairs of CG sites. We generate the repulsive pair interaction from mean force calculations on a grid, where the different combinations of CG sites (\pairCC, \pairCO and \pairOO) are brought together. This is then combined with the monomer GAP discussed previously. The resulting core potentials are shown in \cref{fig:methanol:dimer:pairpotentials}.
\label{ref:methanol:corepotential}

As discussed in \cref{sec:gap:transferfunction}, we need to take into account the change in length scale between long-range and short-range interactions. To this end we apply the transfer function \cref{eq:gap:transferfunction} to all non-bonded distances in the \descdimer descriptor.%
\label{ref:methanol:transferfunction}
The hyperparameters for the \descdimer-descriptor GAP are listed in \cref{tab:methanol:gap:dimer} in the Appendix.

\paragraph{MSCG/FM pair potentials.}

For comparison, we also include the spline-based MSCG/\allowbreak{}FM approach. This is equivalent to a GAP using \descdist descriptors.
To describe the interactions between two-site methanol monomers in terms of pair potentials between sites, we need one spline interaction for the CG bond and three non-bonded splines. %
We obtained the splines for these interactions using the MSCG/FM code provided by the Voth group. As a consistency check, we also taught a GAP with the corresponding four \descdist descriptors; this leads to the same potentials.
The resulting non-bonded pair potentials are shown in \cref{fig:methanol:dimer:pairpotentials}. (The bonded pair potential is equivalent to the monomer potential and is not shown again.)
\begin{figure}[H]
  \centering
  \input{thesis_figures/methanol/dimer/pairpotentials.pgf}
  \mycaption{Methanol dimer pair potentials.}{The non-bonded interactions are described by three pair potentials for the \pairCC, \pairCO, and \pairOO interactions. Here we show the spline potentials obtained in the MSCG/FM approach (solid lines) and, for comparison, the potentials obtained in the GAP approach using one-dimensional pairwise distance descriptors (dotted lines). This shows the equivalence to the MSCG/FM approach. The highly attractive hydrogen bond between two \siteOH sites results in the much deeper well depth for the \pairOO pair potential.
  We also show the short-ranged core potentials used in the training of the dimer-descriptor GAP (dashed lines).}
  \label{fig:methanol:dimer:pairpotentials}
\end{figure}

The MSCG/FM approach is less reliant on core potentials because the splines do not have the Gaussian process's tendency to drive the energy back to zero far away from the data. Moreover, due to the much lower dimensionality of pairwise splines compared to the molecular descriptors, they are much easier to saturate with data. In all our tests, we did not run into issues due to insufficient sampling for the MSCG/FM splines. We note, though, that in practice the MSCG/FM approach often uses a ``core potential'' to ensure short-range repulsion as well (see, for example, \textcite{voth:lipidhcg}).

\subsubsection{Comparison of potentials}

While it is not feasible to visualise predictions of the potentials in the full six-dimensional space that describes the methanol CG dimer, we can display forces along cutlines through this space. Here, we consider fixed relative orientations of the methanol molecules and vary their COM separation. %
In \cref{fig:methanol:dimer:cutline} we compare the COM force predictions of dimer GAP and pair potentials with mean force reference values. This shows that GAP captures the MF with very small error, whereas the pair potentials exhibit a visible discrepancy. In \cref{fig:methanol:dimer:cutline:transfer,fig:methanol:dimer:cutline:core} we show the effects of transfer function and short-range core potential on the dimer GAP predictions.
\begin{figure}[tbhp]
  \centering
  \input{thesis_figures/methanol/dimer/fig_dimercf_cutline.pgf}
  \mycaption{Methanol dimer: forces along a cutline (1/3) --- comparison of dimer-descriptor GAP and pair potentials.}{\textit{Top:} centre-of-mass force between the molecules along the COM separation, comparing CG predictions with MF. \textit{Bottom:} deviation of CG potential predictions from MF [solid: COM force along COM separation; dash-dotted: perpendicular (shearing) component of COM force; dashed: torque-inducing force (not along a CG bond)]. This shows the core potential, how GAP captures the MF with very small error, and how the MSCG/FM pair potentials produce forces significantly different from the MF.}
  \label{fig:methanol:dimer:cutline}
\end{figure}
\begin{figure}[tbhp]
  \centering
  \input{thesis_figures/methanol/dimer/fig_dimercf_coretrans_cutline9.pgf}
  \mycaption{Methanol dimer: forces along a cutline (2/3) --- transfer function and core potential.}{This shows that, without a transfer function, there is no single length scale $\theta$ that is appropriate for the whole range of COM separations which, in turn, is responsible for the resulting ``ringing'' that can be clearly observed along the cutline. It also shows how, without a repulsive core potential, the force prediction peaks and then turns to zero (and will eventually become negative) in the regions where there is insufficient sampling.}
  \label{fig:methanol:dimer:cutline:transfer}
\end{figure}
\begin{figure}[tbhp]
  \centering
  \input{thesis_figures/methanol/dimer/fig_dimercf_coretrans_cutline22.pgf}
  \mycaption{Methanol dimer: forces along a cutline (3/3) --- drawback of core potential.}{Here we can see one drawback of using core potentials: the GAP without the core potential actually captures the dimer cutline more accurately. Though the aim of the core potential is to make the function that remains to be fitted smoother, and specifically to reduce the steep short-range repulsion, here it may have led to a somewhat more rugged function that is then slightly less well-represented by the GAP. This may be due to the repulsive \pairOO core potential extending further than the full MSCG/FM pair potential, as can be seen in \cref{fig:methanol:dimer:pairpotentials}. This issue will be investigated in more detail in future research.}
  \label{fig:methanol:dimer:cutline:core}
\end{figure}

We also compare the force prediction error of the CG potentials by measuring the deviation from the components of the MF reference on a separate test set of 1400 randomly sampled dimer configurations that were not included in the training. In \cref{fig:methanol:dimer:forceerrorvsmagnitude} we show this error as a function of the two-body contribution to the force magnitude. Note that the error for the GAP predictions is distributed symmetrically around zero. In contrast, the pair potential predictions have a much larger error that also contains a systematic component. In \cref{fig:methanol:dimer:forceerrorvsdistance} we show the force prediction error as a function of the dimer COM separation, compared to the two-body contribution to the force magnitude. The two-body force magnitude decreases with the COM distance. This shows that the relative error of the GAP forces is much smaller than that of the forces given by the pair potentials.
The root mean squared error (RMSE) for the different potentials across the whole test set is given in \cref{tab:methanol:dimer:rmse}. %

Whereas errors are usually assumed to follow a Gaussian distribution, we note that in the case of the clusters the deviation between CG potential prediction and mean force reference is more closely described by a Laplace distribution: in a logarithmic histogram of the distribution of force component deviations, the tails show a linear trend. (In bulk methanol, the errors follow a Gaussian distribution as expected.)

\begin{figure}[tbhp]
  \centering
  \input{thesis_figures/methanol/dimer/fig_dimer_error_mag.pgf}
  \mycaption{Methanol dimer: force prediction error \vs force magnitude.}{\\ Here we compare the deviation of the force predictions of CG potentials from MF on a separate test set as a function of the two-body contribution to the force component magnitude. \textit{Top:} dimer-descriptor GAP taught from dimer MF. \textit{Middle:} dimer prediction of a GAP with dimer and trimer descriptors simultaneously taught from trimer MF (\cf \cref{sec:methanol:trimer}). \textit{Bottom:} MSCG/FM pair potentials from dimer MF. This shows that the largest errors do not occur for the largest forces, but rather at small force magnitude. Moreover, whereas the error in the GAP predictions is small and symmetrically distributed around zero, the pair potential predictions contain a systematic error.}
  \label{fig:methanol:dimer:forceerrorvsmagnitude}
\end{figure}
\begin{figure}[tbhp]
  \centering
  \input{thesis_figures/methanol/dimer/fig_dimer_error_dist.pgf}
  \mycaption{Methanol dimer: force prediction error \vs COM distance.}{\\ Here we show the force prediction error as a function of COM separation (red) and, for comparison, the absolute magnitude of the mean force components (green). \textit{Top:} dimer-descriptor GAP taught from dimer MF. \textit{Middle:} dimer prediction of a GAP with dimer and trimer descriptors simultaneously taught from trimer MF (\cf \cref{sec:methanol:trimer}). \textit{Bottom:} MSCG/FM pair potentials from dimer MF.}
  \label{fig:methanol:dimer:forceerrorvsdistance}
\end{figure}

\begin{table}[htbp]
  \centering
  \begin{tabular}{llS[table-format=1.1]}
    \toprule
    potential & derived from & \textcell{RMSE / \si{\kT\per\angstrom}} \\
    \midrule
    \descmonomer GAP & monomer MF & 2.3 \\ %
    \descdimer GAP   & dimer MF   & 0.2 \\ %
    MSCG/FM pair potentials          & dimer MF   & 0.9 \\ %
    \bottomrule
  \end{tabular}
  \mycaption{Methanol dimer force prediction errors.}{Comparison of the force prediction root mean squared error (RMSE) on a separate MF test set (1400 configurations). We compare the monomer GAP, the dimer GAP, and the dimer pair potentials.}
  \label{tab:methanol:dimer:rmse}
\end{table}

\clearpage
\subsubsection{MD results}

While it may be useful to assess the accuracy of CG potentials by comparing their predictions with a mean force test set, the real test of CG potentials is their performance in MD simulations. We now consider the results of simulations using the GAP and MSCG/FM CG potentials. We compare distributions in the CG ensembles with the corresponding distributions of the all-atom simulation.

Previous research has generally focused on the RDF for molecular centres of mass and for pairs of CG sites. Structure-based coarse-graining reproduces the site--site RDFs by construction. Inadequacies of these approaches only show up when considering more complex distributions and correlations between them\cite{CGManyBodyCorrelations}. Hence, here and in the following we present a wide range of observables that can be obtained from the CG coordinates, including two-dimensional correlation plots. %

In \cref{fig:methanol:dimer:mddistances} we show the distributions of COM separation and site--site distances for all-atom, GAP-CG and MSCG/FM simulations. In \cref{fig:methanol:dimer:mdangles} we show distributions of the angle $\alpha$ between a CG bond and the hydrogen bond and of the dot product $n_1 \cdot n_2$ between the normalised directions of the two CG bonds. These are captured to a very high accuracy by the dimer GAP, whereas the MSCG/FM results show significant discrepancies. \Cref{fig:methanol:dimer:2d} further demonstrates that GAP is able to reproduce the correlation of $\alpha$ and $n_1 \cdot n_2$, which is not the case for pair potentials.

\begin{figure}[tbhp]
  \centering
  \input{thesis_figures/methanol/dimer/fig_distances.pgf}
  \mycaption{Methanol dimer: distance distributions.}{Normalised distance distributions $p(r)$ for COM distance (top left), \pairCC distance (top right), \pairCO distance (bottom left) and \pairOO distance (bottom right). We compare MD results for all-atom model, \descdimer GAP CG model, and MSCG/FM CG model. $\Delta p$ is the deviation from the reference distributions in the all-atom simulation.}
  \label{fig:methanol:dimer:mddistances}
\end{figure}
\begin{figure}[p]
  \centering
  \input{thesis_figures/methanol/dimer/fig_angles.pgf}
  \mycaption{Methanol dimer: orientational distributions.}{
  \textit{Left:} distribution of the angle $\alpha$ between CG bond and hydrogen bond/\pairOO axis (\pairCO$\cdots$OH angle).
  The peak in $p(\alpha)$ coincides with the equilibrium value of the C--O--H bond angle in the all-atom model, \ang{108.5}.
  \textit{Right:} distribution of the dot product $n_1 \cdot n_2$ between CG bond directions.
  If the two bond directions were uncorrelated, the dot product $n_1 \cdot n_2$ would be distributed uniformly.
  Only bound dimer configurations with a \pairOO distance less than \SI{3.4}{\angstrom} were included in these distributions.
  }

  \label{fig:methanol:dimer:mdangles}
\end{figure}
\begin{figure}[p]
  \centering
  \input{thesis_figures/methanol/dimer/fig_2d.pgf}
  \caption{2D correlation between $\alpha$ and $n_1 \cdot n_2$ (\cf \cref{fig:methanol:dimer:mdangles}).}
  \label{fig:methanol:dimer:2d}
\end{figure}

This demonstrates that the \descdimer GAP can perfectly reproduce many kinds of distributions beyond simple RDFs. Indeed, any discrepancies are due to insufficient sampling. While the pair potentials provide a reasonable description of the methanol dimer, they show systematic deviation from the all-atom distributions.

\clearpage%
\subsection{Trimer}
\label{sec:methanol:trimer}

\subsubsection{Atomistic structure}

The methanol trimer exhibits two distinct stable configurations.

On the one hand, there are configurations in which there are three hydrogen bonds that form an equilateral triangle, with all angles between \siteOH sites approximately \ang{60}. The molecules tend to form a planar configuration in which the bond vectors are pointing towards the centre of the triangle.

On the other hand, there are configurations with two hydrogen bonds directed towards a central molecule, forming an isosceles triangle. There, the angle between the hydrogen bonds varies between approximately \SIlist{90;120}{\degree}. Hence, the other two angles in the triangle of \siteOH sites generally lie in the range \SIrange{30}{45}{\degree}.

These two configurations can be clearly distinguished in plots of the angular distribution function. %
To analyse orientations of the bonds and differences between the two types of structure, it makes sense to consider these configurations separately. We define the following masks:
\begin{description}
  \item[``Bound configuration'':] at least two \pairOO distances are less than \SI{3.4}{\angstrom}, indicating a hydrogen bond.
  \item[``Equilateral configuration'':] bound configuration where all angles within the triangle of \siteOH sites are larger than \ang{55}; this implies that any angle can be at most \ang{70}.
  \item[``Isosceles configuration'':] bound configuration where two of the angles differ by at most \ang{5}\!\!, and the third angle is larger than \ang{70}.%
\end{description}
We show the distribution of angles within bound configurations in \cref{fig:methanol:trimer:ternary:atmask}, where we also highlight those configurations that we count as ``equilateral'' or ``isosceles''.

\begin{figure}[tbhp]
  \centering
  \input{thesis_figures/methanol/trimer/fig_atom_ternary.pgf}
  \mycaption{Methanol trimer: distribution of angles between \siteOH sites and angle masks in all-atom model.}{\textit{Left:} ternary plot of the angles in the triangle formed by \siteOH sites. Each axis represents one of the angles from \ang{0} to \ang{180}. (Only bound trimer configurations with at least two \pairOO distances less than \SI{3.4}{\angstrom} are included.) \textit{Middle:} the mask used to capture equilateral configurations. \textit{Right:} the mask used to capture isosceles configurations.}
  \label{fig:methanol:trimer:ternary:atmask}
\end{figure}

\label{ref:trimer:rst}
As in the case of the dimer, we employ half-open distance restraints between the centres of mass of the molecules to keep the cluster from dissociating. Here we apply restraints between each pair of molecules, with the restraint setting in at \SI{15}{\angstrom}. The force constant is $\SI{50}{\kcal\per\mole\per\angstrom^2} \approx \SI{84}{\kT\per\angstrom^2}$.

\subsubsection{CG potentials}

In the all-atom model used for this work, adding more molecules to the system does not change the form of each individual interaction. 
If this were true for the CG model, we could simply take the potentials obtained from dimer MF as described in the previous section and apply them to systems containing any number of molecules. This would also imply that we would obtain the same interactions whether teaching the \descdimer GAP from dimer MF or trimer MF. Unfortunately, this is not the case, as we will see in the deviation of the predicted forces from the all-atom mean forces along a dimer cutline for a dimer-descriptor GAP trained on \emph{trimer} MF, shown in \cref{fig:methanol:trimer:dimercutline}. In this case the dimer term will acquire an effective contribution due to the actual trimer interactions. %

\paragraph{GAP.}

We compare three different GAPs.
One is the \descdimer GAP trained on dimer MF that was discussed in the previous section.
Furthermore, we consider two GAPs taught from trimer MF, one using only the \descdimer descriptor and one using the \desctrimer descriptor (using the \descdimer GAP trained on dimer MF as a core potential). We will also discuss a simultaneous teaching of dimer and trimer descriptor from trimer MF.
\label{ref:simultaneoustrimer}

For teaching a \descdimer GAP from trimer MF we use the same hyperparameters as for the methanol dimer GAP (see \cref{tab:methanol:gap:dimer} in the Appendix). As the dimer descriptor does not capture the full dimensionality of the trimer system and cannot exactly represent the trimer PMF, there is an additional, systematic error beyond just the noise in mean forces. Hence, in principle, the noise parameter should be increased, corresponding to an increased amount of regularisation.
In this case, increasing the noise level hyperparameter turned out not to have any noticeable effect in practice.
The \desctrimer descriptor of the two-site methanol contains 15 distances (three bonded and 12 non-bonded).
For the trimer-descriptor GAP used in MD simulations, we use the \descdimer GAP that was presented in the previous section as the core potential. The hyperparameters for the \desctrimer descriptor are given in \cref{tab:methanol:gap:trimer} in the Appendix.
It is also possible to teach the dimer and trimer descriptors simultaneously from trimer MF. For this we use the hyperparameters for dimer and trimer descriptor given in \cref{tab:methanol:gap:dimer,tab:methanol:gap:trimer}.
In \cref{fig:methanol:trimer:dimercutline} we compare the predicted forces along a dimer cutline for the different GAP potentials with the dimer MF reference values.
Teaching only the dimer descriptor from trimer MF results in a significantly different curve for the forces along the cutline.
This shows that the dimer term picks up an effective contribution due to many-body effects. In the following we will see that neither of the potentials using only the \descdimer descriptor is able to reproduce the atomistic structure correctly.
In our two-site methanol trimer, there are genuine three-body effects that are not captured by the \descdimer descriptor alone; they can only be described by considering the molecular three-body interactions.
Teaching dimer and trimer descriptors simultaneously allows the dimer descriptor to recover the actual dimer cutline, while the trimer descriptor captures all three-body effects.
This supports our decision to separate the PMF into molecular monomer-based terms.

\begin{figure}[tbhp]
  \centering
  \input{thesis_figures/methanol/trimer/fig_trimer_cutline.pgf}
  \mycaption{Methanol trimer GAP: force prediction along dimer cutline.}{Training only a dimer-descriptor GAP from trimer MF results in a systematic deviation of the dimer force prediction from the MF reference. In contrast, training both dimer and trimer descriptors simultaneously recovers the dimer forces along the cutline very well. For comparison, we include the dimer-descriptor GAP trained from dimer MF.}
  \label{fig:methanol:trimer:dimercutline}
\end{figure}

\paragraph{MSCG/FM pair potentials.}

The methanol trimer system contains the same types of pairwise interactions as the dimer. However, we do not assume that the dimer pair potentials accurately describe the trimer system. Hence we re-calculate the splines from trimer MF. The resulting potentials are shown in \cref{fig:methanol:trimer:pairpotentials} in the Appendix.

\paragraph{Structure-based pair potentials (IBI).}
\label{ref:methanol:ibi}

For the methanol trimer, we include results for structure-based pair potentials that reproduce the site--site distance distributions.
As mentioned in \cref{ch:cg}, different approaches such as IBI and IMC would arrive at the same pair potentials; as we consider comparatively small systems, for the present work, IBI is our method of choice.

For technical reasons, it was impractical to carry out IBI directly on the trimer in vacuum with restraining potential.\footnote{The \VOTCA package used to carry out the Iterative Boltzmann Inversion was not able to calculate distance distributions for the open boundary conditions of the vacuum simulations.} Hence, we placed the trimer inside a box with side length $L=\SI{26}{\angstrom}$ and periodic boundary conditions (PBC); this simulation produced sufficient data to obtain converged RDFs for pairs of CG sites.

In a periodic box, the long-range electrostatic interactions of the partial charges lead to interactions with their mirror images. This results in forces on the atoms (and hence, mean forces on the CG sites) that are slightly different from those in a vacuum simulation with open boundary conditions (and otherwise equivalent settings). Hence, the GAP taught from vacuum MF does not, and is not expected to, exactly reproduce the all-atom distance distributions in a periodic box.
To overcome this problem, we could have recalculated MF in the atomistic periodic box simulation and trained a new GAP from these forces.
However, here we will instead present results for GAP and MSCG/FM (compared to the all-atom simulation with restraints) separately from results for IBI (compared to the all-atom simulation with PBC).
In any case, this is only a limitation with respect to overall distributions of distance variables.
When only considering configurations in which all three pair distances are within a certain cutoff (``bound states''), there is no difference between the distributions from simulations carried out in vacuum using distance restraints and those from simulations in a box with periodic boundary conditions.

The CG bond potential was obtained by Direct Boltzmann Inversion using the harmonic approximation.
The non-bonded pair potentials were refined iteratively using the \VOTCA package\cite{VOTCA} (starting from the results of the Direct Boltzmann Inversion). The three pair potentials were defined on a grid up to a cutoff of \SI{10}{\angstrom} with spacing \SI{0.1}{\angstrom}.
To improve the stability of the IBI algorithm, potential updates were smoothed twice and scaled by a factor of \num{0.2}. In each step, only one of the three non-bonded potentials was updated.

The resulting pair potentials accurately reproduce the site--site distance distributions of the all-atom model, as can be seen in \cref{fig:methanol:trimer:distancedist:box} on \mypageref{fig:methanol:trimer:distancedist:box}. There is a minor discrepancy for the \pairOO distance distribution, which is due the sharpness of the peak corresponding to the hydrogen bond. To capture this peak accurately would require a much finer grid, which would in turn require much longer simulation runs to obtain converged distributions, making the calculations more difficult.

\subsubsection{Comparison of potentials}

We show the force prediction errors of the potentials on a separate trimer MF test set in \cref{tab:methanol:trimer:rmse}. It can be clearly seen that the trimer-descriptor GAP trained on trimer MF has a significantly smaller error in the predictions (as compared to reference MF values) than the other potentials. However, while RMSE values can give a first indication as to the accuracy of a potential, the true test for the quality of a CG potential is its performance in MD simulations.

\begin{table}[b]
  \centering
  \begin{tabular}{llS[table-format=1.1]S[table-format=1.1]S[table-format=1.1]}
    \toprule
    \multirow{2}{*}{potential} & \multirow{2}{*}{derived from} & \multicolumn{3}{c}{RMSE / \si{\kT\per\angstrom}} \\
    \cmidrule(l){3-5}
    & & {total force} & {dimer part} & {trimer part} \\
    \midrule
    monomer GAP                 & monomer MF  & 3.7 & 3.5 & 1.6 \\
    \descdimer GAP              & dimer MF    & 1.7 & 0.4 & 1.6 \\
    \descdimer GAP              & trimer MF   & 1.4 & 0.8 & 1.6 \\
    dimer+trimer GAP            &             & 0.9 & 0.5 & 1.0 \\
    \desctrimer GAP             &             & 1.0 & 0.4 & 1.0 \\
    MSCG/FM pair potentials     & dimer MF    & 2.0 & 1.4 & 1.6 \\
                                & trimer MF   & 1.9 & 1.5 & 1.6 \\
    IBI	pair potentials         & trimer RDF  & 2.2 & 1.7 & 1.6 \\
    \bottomrule
  \end{tabular}
  \mycaption{Methanol trimer force prediction errors.}{Comparison of the force prediction root mean squared error (RMSE) on a separate mean force test set for different potentials. We also compare the RMSE of those contributions to the total force that are only due to two-body interactions (``dimer part'') and of those that are only due to three-body interactions (``trimer part'').
  
  Note that the simultaneous dimer+trimer GAP training results in a slightly smaller RMSE for the total force prediction than the trimer-descriptor GAP that uses the methanol dimer GAP as a core potential. On the other hand, the dimer+trimer GAP shows a slightly larger error for the dimer contributions to the total forces.
  This suggests that the split of the interactions between dimer and trimer descriptor could be somewhat improved still; this might also reduce the small discrepancy between the forces along cutlines that can be seen in \cref{fig:methanol:trimer:dimercutline} on \mypageref{fig:methanol:trimer:dimercutline}.}
  \label{tab:methanol:trimer:rmse:split}
  \label{tab:methanol:trimer:rmse}
\end{table}

\clearpage
\subsubsection{MD results}

The simulation with the IBI pair potentials was carried out in periodic boundary conditions; for comparison, we also carried out a simulation in periodic boundary conditions with the \desctrimer GAP trained on vacuum trimer MF.
All other simulations were performed using the distance restraints discussed on \mypageref{ref:trimer:rst}. We show the resulting distance distributions (COM, \pairCC, \pairCO, \pairOO) for vacuum simulations using the distance restraints in \cref{fig:methanol:trimer:distancedist:rst} and for periodic boundary conditions in \cref{fig:methanol:trimer:distancedist:box}. Though the IBI pair potentials reproduce the site--site distance distributions by construction, they show a small but systematic deviation in the COM distance distribution. In contrast, the trimer-descriptor GAP reproduces both site--site \emph{and} COM distance distribution of the all-atom model on whose forces it was trained. All other CG potentials show some deviation from the all-atom results.
Note that there is a clear bound state for all the potentials.

As we have discussed previously, obtaining the correct distance distributions does not imply that other distributions are also captured accurately. This can be seen in \cref{fig:methanol:trimer:ternary:md}, which shows ternary plots of the distributions of the angles in the triangle formed by the three \siteOH sites. We show a slice through this distribution in \cref{fig:methanol:trimer:eqslice}; this highlights that the trimer-descriptor GAP reproduces the all-atom distribution very accurately, whereas all other CG potentials do not represent the isosceles configuration correctly.

Moreover, even though the IBI pair potential captures equilateral configurations, it does not reproduce the bond orientations correctly. This can be seen in \cref{fig:methanol:trimer:pairorient}, which shows the correlation between relative orientation and \pairOO distance for pairs of molecules separately for configurations classified as ``equilateral'' or ``isosceles''.
The trimer GAP is the only CG potential that accurately reproduces the distributions of the CG variables in the all-atom model.
Note that even though the \descdimer GAP trained from trimer MF does not describe the isosceles configurations correctly, it reproduces the pair orientations in equilateral configurations much better than the dimer GAP from dimer MF or the pair potentials.

\begin{figure}[H]
  \centering
  \input{thesis_figures/methanol/trimer/fig_distances_rst.pgf}
  \mycaption{Methanol trimer: distance distributions (for simulations carried out with distance restraints).}{COM and site--site distance distributions $p(r)$, normalised over the range \SIrange{2}{16}{\angstrom}, and deviation $\Delta p(r)$ from the all-atom result. Note that the distance distributions for simulations with the trimer-descriptor GAP are indistinguishable from the all-atom distributions.}
  \label{fig:methanol:trimer:distancedist:rst}
\end{figure}
\begin{figure}[H]
  \centering
  \input{thesis_figures/methanol/trimer/fig_distances_box.pgf}
  \mycaption{Methanol trimer: distance distributions (for simulations carried out with periodic boundary conditions).}{COM and site--site distance distributions $p(r)$, normalised over the range \SIrange{2}{16}{\angstrom}, and deviation $\Delta p(r)$ from the all-atom result. Note that the IBI pair potentials exactly reproduce the site--site distance distributions (this is by construction), but nonetheless exhibit a systematic discrepancy in the COM distance distribution.
  
  \protect\footnotemark[1]{The \desctrimer GAP was trained on trimer MF obtained from simulations with open boundary conditions and, for this reason, does not exactly reproduce the distance distributions for periodic boundary conditions}.
  }
  \label{fig:methanol:trimer:distancedist:box}
\end{figure}

\begin{figure}[tbhp]
  \centering
  \input{thesis_figures/methanol/trimer/fig_md_ternary.pgf}
  \mycaption{Methanol trimer: distribution of angles between \siteOH sites.}{Only bound configurations (at least two \pairOO distances less than \SI{3.4}{\angstrom}) are included. The color scale is equivalent for all plots. The scale has been capped such that the secondary peaks for all-atom and \desctrimer GAP simulations are visible; this means that the central peaks of the other plots saturate the scale. %
  }
  \label{fig:methanol:trimer:ternary:md}
\end{figure}
\begin{figure}[tbhp]
  \centering
  \input{thesis_figures/methanol/trimer/fig_eqcut.pgf}
  \mycaption{Methanol trimer: slice through the angle distributions}{shown in \cref{fig:methanol:trimer:ternary:md}. We show the distribution $p(\alpha)$ for one angle, where the other two angles $\beta$, $\gamma$ are approximately equal ($|\beta-\gamma|<\ang{5}$). The peak at \ang{60} corresponds to the equilateral configurations. The second, broader peak around \ang{120} corresponds to isosceles configurations. Only the trimer-descriptor GAP reproduces the distribution of the atomistic model over the whole range of bond angles.}
  \label{fig:methanol:trimer:eqslice}
\end{figure}

\begin{figure}[tbhp]
  \centering
  \input{thesis_figures/methanol/trimer/fig_pairorient.pgf}
  \mycaption{Methanol trimer: correlation of bond orientations.}{We measure the relative bond orientation between methanol molecules by the scalar product $n_1 \cdot n_2$ of the vectors which label the directions of the CG bonds. We show the distributions of relative bond orientation against \pairOO distance. These are plotted separately for equilateral configurations (top row) and isosceles configurations (bottom row). In the all-atom equilateral configuration, the \siteOH sites point towards the centre of the triangle, hence they form angles of \ang{120} with each other, corresponding to a dot product of \num{-0.5}. This is reproduced only by the trimer-descriptor GAP.}
  \label{fig:methanol:trimer:pairorient}
\end{figure}

\clearpage

\subsection{Tetramer}
\label{sec:methanol:tetramer}

\subsubsection{Atomistic structure}

As in the cases of dimer and trimer, we employ half-open distance restraints between the centres of mass of the molecules to keep the cluster from dissociating. There are six restraints between all pairs of molecules, with the restraint setting in at \SI{20}{\angstrom}, and the force constant is $\SI{50}{\kcal\per\mole\per\angstrom^2} \approx \SI{84}{\kT\per\angstrom^2}$. For IBI, we place the tetramer in a periodic cubic box with side length $L=\SI{26}{\angstrom}$.

In a system of four methanol molecules, configurations can contain tetramers, trimers, dimers, and free molecules.
However, clusters of four methanol molecules predominantly form a square structure in which there are four hydrogen bonds linking all four molecules into a planar system. This can be seen in the ADF (\cref{fig:methanol:tetramer:adf} on \mypageref{fig:methanol:tetramer:adf}), where there are clear peaks at \ang{90} and \SI{2.7}{\angstrom} (corresponding to the triplet formed by a corner of the square and its two hydrogen-bonded nearest neighbours) and at \ang{45} and $\sqrt{2} \times \SI{2.7}{\angstrom} \approx \SI{3.8}{\angstrom}$ (corresponding to the triplet formed with one hydrogen-bonded nearest neighbour and the molecule diagonally across the square).
The presence of other types of configurations is indicated by the smaller, secondary peak in the ADF at \ang{60} and \SI{2.7}{\angstrom} that corresponds to configurations containing equilateral trimers.

\subsubsection{CG potentials}

\paragraph{GAP.}

As discussed in \cref{sec:gap:generalx}, a descriptor based on quadruplets of molecules would not be practicable. We already saw that the \descdimer descriptor is not sufficient to reproduce the methanol trimer distributions. Hence, we only consider GAPs that include the \desctrimer descriptor.

We train a trimer-descriptor GAP from tetramer MF using the same hyperparameters as for the methanol trimer (\cref{tab:methanol:gap:trimer} in the Appendix); this includes the dimer core potential. The difference is only in the training data. Here, we use 12684 randomly sampled tetramer configurations.
We note that it is possible to combine configurations from different systems in a single training set. Training a GAP from such a mixed set results in a potential that describes ``average'' molecular free energy contributions that are an approximation between the different monomer, dimer and trimer terms that best describe the PMF across systems.
Here we include a GAP trained on a mixed set of trimer and tetramer MF that contained 12684 tetramer configurations (corresponding to 35508 triplets within the cutoff of the \desctrimer descriptor) and 33380 trimer configurations.
For comparison, we include the \desctrimer GAP trained on trimer MF.

\paragraph{Pair potentials.}

As for the methanol trimer, we also present results for pair potentials obtained both by force-matching (using the same training data as for GAP) and by IBI (in a periodic box). The pair potentials are shown in \cref{fig:methanol:tetramer:pairpotentials} in the Appendix.

\subsubsection{Comparison of potentials}
We show the force prediction performance of the different potentials, measured by the root mean squared error on a separate tetramer test set, in \cref{tab:methanol:tetramer:rmse}.

\begin{table}[htbp]
  \centering
  \begin{tabular}{llS[table-format=1.1]}
    \toprule
    potential & derived from & \textcell{RMSE / \si{\kT\per\angstrom}} \\
    \midrule
    monomer GAP & monomer MF      & 4.0 \\
    dimer GAP   & dimer MF        & 2.1 \\
    trimer GAP  & trimer MF       & 1.5 \\
    trimer GAP  & trimer+tetramer MF & 1.3 \\
    trimer GAP  & tetramer MF     & 1.2 \\
    MSCG/FM     & tetramer MF     & 2.3 \\
    IBI         & tetramer RDF    & 2.6 \\
    \bottomrule
  \end{tabular}
  \mycaption{Methanol tetramer force prediction errors.}{Comparison of the force prediction root mean squared error (RMSE) on a separate mean force test set for different potentials.}
  \label{tab:methanol:tetramer:rmse}
\end{table}

\subsubsection{MD results}

As in the case of the methanol trimer, we separately show the distance distributions for vacuum/restrained simulations in \cref{fig:methanol:tetramer:distances:rst} and for simulations using periodic boundary conditions in \cref{fig:methanol:tetramer:distances:box}.
The angular distribution functions shown in \cref{fig:methanol:tetramer:adf} clearly demonstrate that reproducing the pair distances does not, necessarily, imply a good reproduction of other measures of the structure.

We calculate the number of hydrogen bonds within the CG simulations based on an \pairOO cutoff of \SI{3.4}{\angstrom}. The fractional occupation of states with a given number of hydrogen bonds is shown in \cref{tab:methanol:tetramer:fractions} and visualised in \cref{fig:methanol:tetramer:hbonds}.
Configurations with at least four hydrogen bonds describe a single bound tetramer. For these configurations, we show the distribution of angles and solid angles in \cref{fig:methanol:tetramer:angles}.

\begin{figure}[H]
  \centering
  \input{thesis_figures/methanol/tetramer/fig_distances_rst.pgf}
  \mycaption{Methanol tetramer: distance distributions (for simulations carried out with distance restraints).}{COM and site--site distance distributions.}
  \label{fig:methanol:tetramer:distances:rst}
\end{figure}
\begin{figure}[H]
  \centering
  \input{thesis_figures/methanol/tetramer/fig_distances_box.pgf}
  \mycaption{Methanol tetramer: distance distributions (for simulations carried out with periodic boundary conditions).}{COM and site--site distance distributions. By construction, the site--site distance distributions obtained for the IBI pair potentials match the all-atom distributions perfectly; the two lines are nearly indistinguishable. However, there is systematic discrepancy in the distribution of COM distances.}
  \label{fig:methanol:tetramer:distances:box}
\end{figure}
\begin{figure}[H]
  \centering
  \input{thesis_figures/methanol/tetramer/fig_adf.pgf}
  \mycaption{Methanol tetramer: Angular Distribution Function (ADF) for \siteOH sites.}{The ADF (left-hand plot of each pair) is determined by considering all triplets of \siteOH sites where the distance $r_\text{c1}$ between a central site and its first neighbour is less than \SI{3.4}{\angstrom}. Here we show the 2D distribution of the angle $\theta_\text{c12}$ at the central site against the distance $r_\text{c2}$ between the central site and its second neighbour. We only included configurations in the ADF calculation in which all pairwise COM distances were less than \SI{10}{\angstrom}. For the all-atom model, there is no discernible difference between the results for simulations using periodic boundary conditions and for simulations with distance restraints and hence we do not consider them separately in this figure. The right-hand plot of each pair shows the deviation from the all-atom result.}
  \label{fig:methanol:tetramer:adf}
\end{figure}

\begin{table}[H]
  \centering
  \begin{tabular}{llS[table-format=2.1]S[table-format=2.1]S[table-format=2.1]S[table-format=2.1]S[table-format=2.1]S[table-format=2.1]S[table-format=2.1]}
    \toprule
    \multirow{2}{*}{potential} & \multirow{2}{*}{derived from} & \multicolumn{7}{c}{number of hydrogen bonds / \%} \\
    \cmidrule(l){3-9}
    & & 6 & 5 & 4 & 3 & 2 & 1 & 0 \\
    \midrule
    \multicolumn{2}{l}{all-atom (restraints)}   &  0.0 & 10.0 & 27.6 & 21.8 & 21.9 & 13.8 &  4.9 \\
    \multicolumn{2}{l}{all-atom (periodic box)} &  0.0 & 11.1 & 30.8 & 23.6 & 18.8 & 11.1 &  4.5 \\
    trimer GAP & trimer MF                      &  9.6 & 22.4 & 24.8 & 16.9 & 15.7 &  8.1 &  2.5 \\
    trimer GAP & trimer+tetramer MF                &  2.2 & 12.9 & 22.9 & 22.3 & 21.7 & 12.9 &  5.0 \\
    trimer GAP & tetramer MF                    &  0.2 &  7.4 & 17.6 & 20.1 & 22.4 & 22.4 & 10.0 \\
    MSCG/FM & tetramer MF                       & 23.9 & 29.0 & 16.3 & 12.5 &  6.6 &  6.8 &  4.8 \\
    IBI & tetramer RDF                          &  4.4 & 18.5 & 20.4 & 18.6 & 13.5 & 12.9 & 11.7 \\
    \bottomrule
  \end{tabular}
  \mycaption{Methanol tetramer: relative frequencies of number of hydrogen bonds.}{We measure the number of hydrogen bonds in each configuration by counting the number of \pairOO distances that are less than \SI{3.4}{\angstrom}.}
  \label{tab:methanol:tetramer:fractions}
\end{table}
\begin{figure}[H]
  \centering
  \input{thesis_figures/methanol/tetramer/fig_hbonds_pie.pgf}
  \mycaption{Methanol tetramer: relative frequencies of number of hydrogen bonds.}{We show the fractional occupation of states with different numbers of hydrogen bonds, measured by the number of \pairOO distances less than \SI{3.4}{\angstrom}. This is a visualisation of the values in \cref{tab:methanol:tetramer:fractions}.}
  \label{fig:methanol:tetramer:hbonds}
\end{figure}
\begin{figure}[tbp]
  \centering
  \input{thesis_figures/methanol/tetramer/fig_angles.pgf}
  \mycaption{Methanol tetramer: angle distributions within bound configurations.}{Only configurations with at least four \pairOO distances less than \SI{3.4}{\angstrom} were included, corresponding to all four molecules linked together by hydrogen bonds. \textit{Left:} distribution of angles between triplets of \siteOH sites. \textit{Right:} distribution of solid angles from each \siteOH site to the others. A planar square structure corresponds to vanishing solid angles and two peaks at \ang{45} and \ang{90} degrees for the angles. A regular tetrahedron has solid angles of approximately \SI{0.55}{\steradian}, and a peak at \ang{60} for the angles. Whereas the \desctrimer GAP reproduces the all-atom distribution to a large extent, both structure- and force-based pair potentials fail to describe the planar structure in the all-atom model correctly and tend to form tetrahedrons instead.}
  \label{fig:methanol:tetramer:angles}
\end{figure}

\clearpage
Unlike the cases of the methanol dimer and trimer, our GAP-CG simulations do not exactly reproduce the behaviour of the all-atom tetramer model.
This implies that, in our two-site CG description, the PMF of the methanol tetramer cannot be partitioned exactly into monomer, dimer, and trimer terms. Hence, the GAP approach is no longer exact and, instead, becomes an approximation. This approximation might be improved by increasing the size of the training set.

However, even though the trimer-descriptor GAP does not reproduce the distance distributions as well as the IBI pair potentials, it can correctly reproduce the \emph{structure} of four methanol molecules that form a planar square using four hydrogen bonds. The pair potentials completely fail to reproduce this structure, as is clearly demonstrated by the angular distribution functions.
Overall, the \desctrimer-descriptor GAP still provides a much more accurate representation of the distributions of the methanol tetramer as a whole than the other potentials.

This is true even for the \desctrimer GAP trained on \emph{trimer} MF. However, it puts too much weight on trimer configurations within the tetramer system and does not describe the square structure as well as the GAP trained on tetramer configurations. The reverse is true as well, that is, the GAP trained on tetramer MF does not describe the trimer system as well as the GAP trained on trimer MF.
We note that, by combining training configurations from both tetramer and trimer systems, we can obtain a GAP that interpolates between both systems and finds the best separation into (effective) dimer and trimer interactions that gives reasonable results both for trimer and tetramer systems.

Generally, the best results will be achieved by training the GAP from data of the system in which it will be used.
Here, we simply reused the vacuum \descdimer GAP as a core potential.
In the general case, there will be effective contributions to the dimer/trimer interactions due to environmental influences (four- and higher-body effects such as contributions to the structure due to dense packing). Hence, for the bulk systems considered in the remainder of this thesis, we will not use a monomer or dimer core potential; instead, we will teach monomer, dimer, and/or trimer descriptors simultaneously.

\clearpage
\section{Bulk Simulations}
\label{sec:gapcg:bulk}

Whereas small clusters are dominated by direct interactions, in bulk systems environmental effects play a significant role. This was already discussed in relation to Boltzmann Inversion (see \cref{sec:cg:dbi,sec:cg:ibi}), where an iterative procedure became necessary for an accurate reproduction of all-atom distributions due to correlations between different degrees of freedom. A large contribution to the overall structure simply comes from the dense packing of the molecules, determined by the balance between the external pressure and the avoidance of atomic overlap due to the strong core repulsion of atoms.
This allows even a potential that only describes short-range repulsion of CG sites (``core-only potential'') to already reproduce the all-atom structure to a large extent.
The predominance of packing effects has several implications.

In structure-based approaches, the site--site pair potentials only need to provide a small adjustment to reproduce distance distributions, and this allows for a much better reproduction of orientational distributions in the bulk than in the clusters.

In force-based approaches such as GAP, the mean forces that we aim to reproduce are different from those in vacuum or clusters. Hence, we cannot simply reuse vacuum or cluster potentials.\footnote{In principle, we could use the cluster GAP as a core potential to describe the direct contribution to the effective forces in the bulk system, and teach the difference. However, this may not always improve the training, and in the following we only keep the repulsive short-range core potential for the pairwise non-bonded interactions between CG sites.}
The molecular contributions (dimer and trimer terms) now have to take into account the effective, environment-mediated interactions. This is demonstrated by the difference between vacuum and bulk mean forces along a cutline (shown in \cref{fig:methanol:bulk:cutline:mf} on \mypageref{fig:methanol:bulk:cutline:mf}).
However, even though the molecular terms will be different between cluster and bulk, we can still consider a decomposition of the PMF into monomer, dimer, and trimer contributions.

To demonstrate the general applicability of the GAP approach to different systems, here we consider bulk models for methanol as well as for benzene.
Whereas small clusters have very short decorrelation times and can be considered equilibrated almost instantly, in bulk systems careful equilibration becomes crucial. Moreover, to ensure there are no spurious effects due to interactions with the periodic mirror images of the system, we need to check that the properties of interest in the system are converged with respect to the box size. In \cref{sec:gapcg:bulk:equilibration} we outline our equilibration protocol and discuss convergence with respect to system size.
We then show in \cref{sec:gapcg:bulk:locality} that, indeed, the mean forces only depend on a local environment, which is a necessary condition for a successful decomposition of the PMF into a cluster expansion of molecular terms with a finite cutoff.
We discuss the results of our CG simulations for methanol in \cref{sec:gapcg:bulk:methanol} and for benzene in \cref{sec:gapcg:bulk:benzene}.

\subsection{Simulation Protocol}
\label{sec:gapcg:bulk:equilibration}

In all atomistic simulations of the bulk systems, we use periodic boundary conditions, and employ a force field cutoff of \SI{9}{\angstrom}.
The systems were equilibrated using the protocol given in \cref{tab:bulk:equilibrationprotocol}. The trajectory root mean squared deviation from the initial structure, calculated using the minimum image convention, reached a plateau after \SI{0.5}{\ns} for methanol and after \SI{1}{\ns} for benzene, indicating that our equilibration times are sufficiently long. %

\begin{table}[htbp]
  \centering
  \begin{tabular}{lcS[table-format=7]>{(}c<{)}}
    \toprule
    & type & \multicolumn{2}{c}{steps (time)} \\
    \midrule
    energy minimisation & steepest descent & 100 \\
    & conjugate gradient & 9900 \\
    \specialcell[l]{temperature ramp \\ \qquad from \SI{0}{\kelvin} to $T=\SI{300}{\kelvin}$} & $NVT$ & 50000 & \SI{25}{\pico\second} \\
    equilibration at $T=\SI{300}{\kelvin}$ & $NVT$ & 50000 & \SI{25}{\pico\second} \\
    pressure equilibration & $NPT$ & 3000000 & \SI{1.5}{\nano\second} \\
    determine average volume & $NPT$ & 2000000 & \SI{1.0}{\nano\second} \\
    equilibration at averaged volume & $NVT$ & 2000000 & \SI{1.0}{\nano\second} \\
    \bottomrule
  \end{tabular}
  \caption{Equilibration protocol for bulk systems.}
  \label{tab:bulk:equilibrationprotocol}
\end{table}

\subsubsection{System size convergence}
\label{ref:systemsize:convergence}

\newcommand*{\systemboxmol}[1]{\ensuremath{#1\!\times\!#1\!\times\!#1}}

For bulk systems, it is important to check that our model is converged with respect to the system size. 

\begin{figure}[tbhp]
  \centering
  \begin{subfigure}[b]{0.48\textwidth}
    \includegraphics[width=\textwidth]{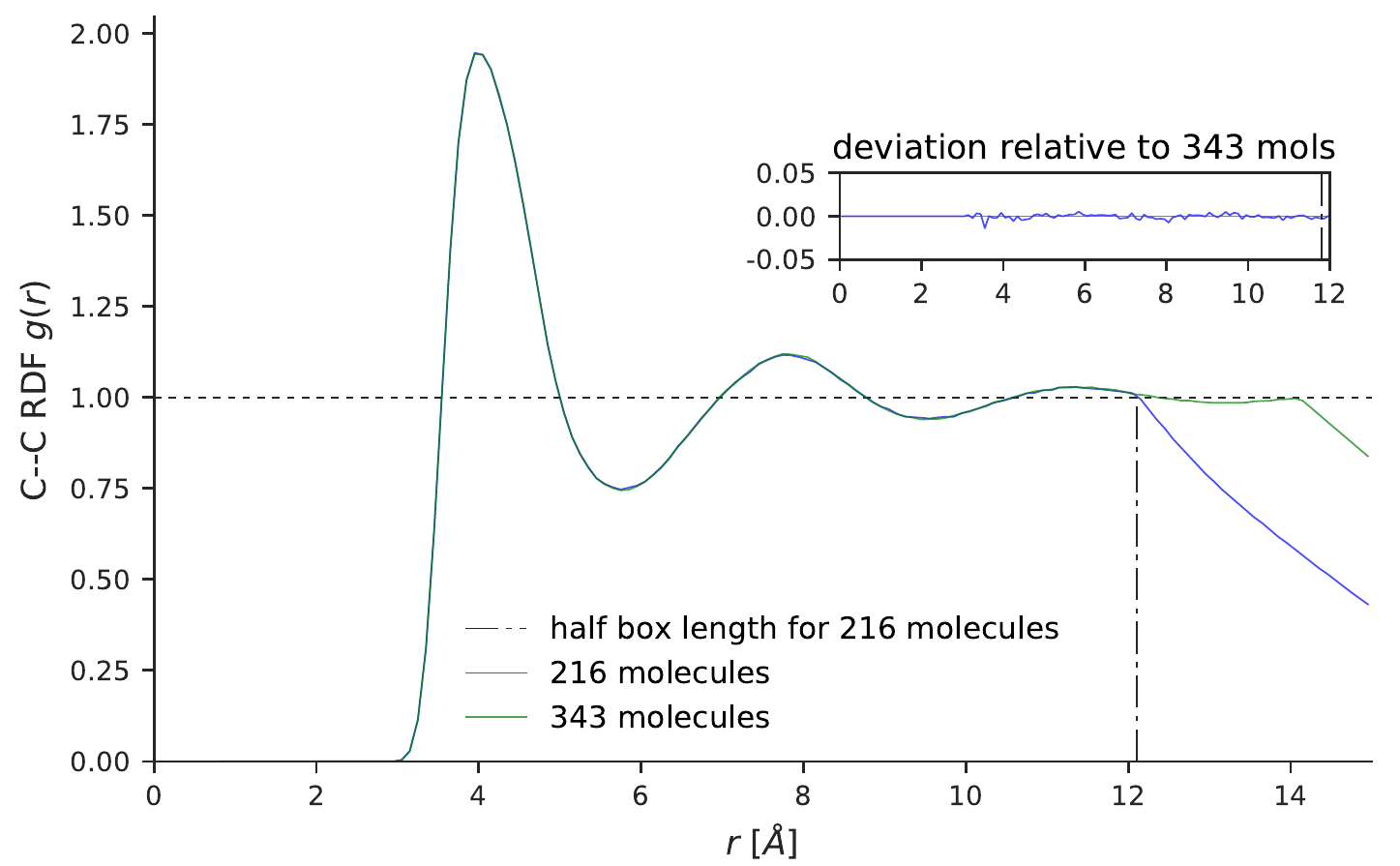}%
    \caption{Methanol}
    \label{fig:methanol:systemsize}
  \end{subfigure}
  \begin{subfigure}[b]{0.48\textwidth}
    \includegraphics[width=\textwidth]{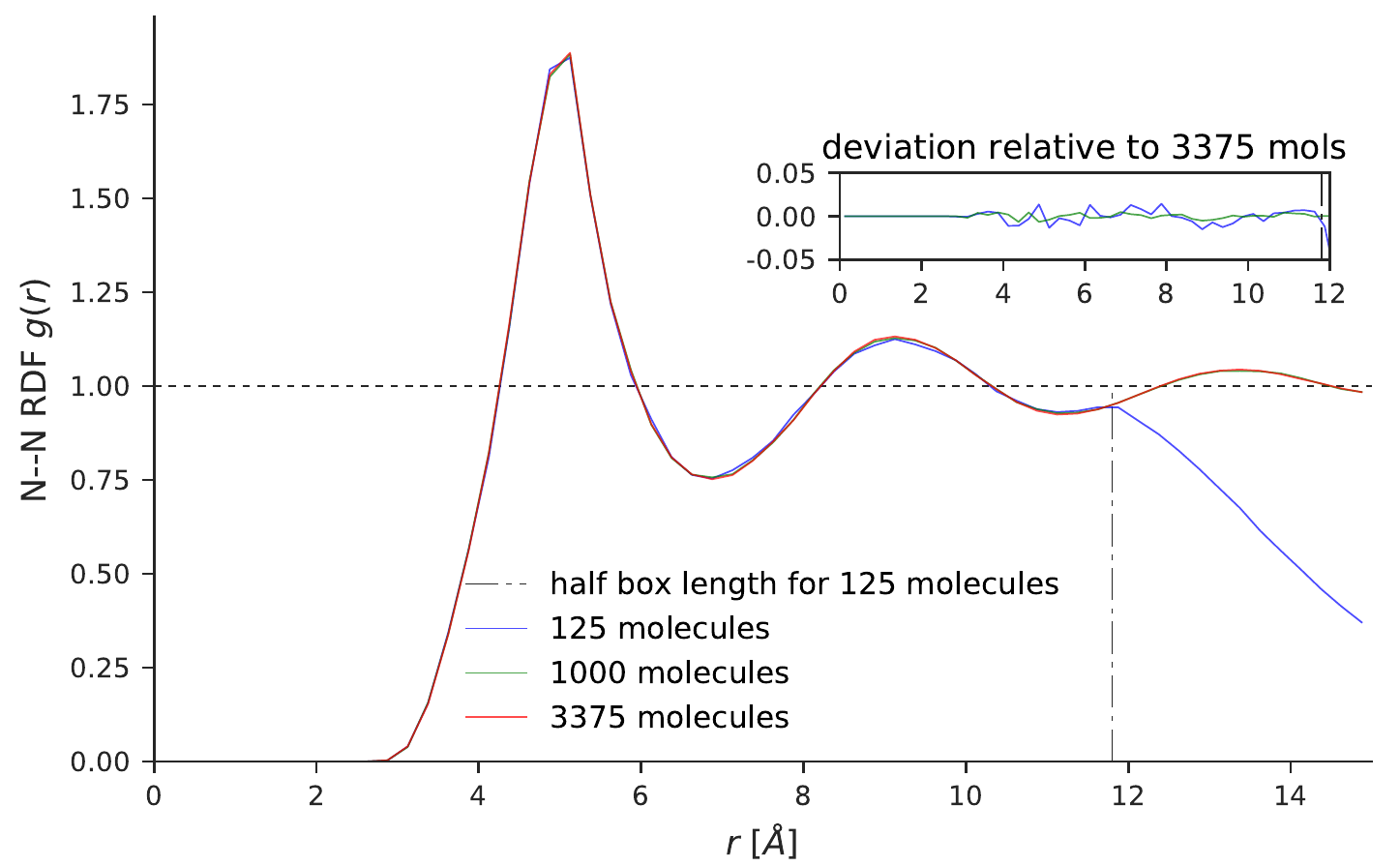}
    \caption{Imidazole}
    \label{fig:imidazole:systemsize}
  \end{subfigure}
  \mycaption{Convergence with respect to system size.}{
    \textit{Left:} bulk methanol RDF between carbon atoms, for 216 and 343 molecules.
    \textit{Right:} bulk imidazole RDF between nitrogen atoms, for 125, 1000, and 3375 molecules.}
\end{figure}

For methanol, we considered initial configurations of $\systemboxmol{6}=216$ and $\systemboxmol{7}=343$ molecules. Following the equilibration procedure described in \cref{tab:bulk:equilibrationprotocol}, this resulted in final box lengths of $L=\text{\SIlist{24.205;28.226}{\angstrom}}$, respectively. In each case, we carried out a \SI{10}{\nano\second} production run in the $NVT$ ensemble. We compare RDFs from these final runs in \cref{fig:methanol:systemsize} (see following page).
The difference between the RDFs is negligible. Hence, in the following we only consider the system with \num{216} methanol molecules.

We did not carry out a separate system size test for benzene. However, in an earlier investigation we considered the convergence with respect to system size for imidazole, a planar molecule of similar size and structure to benzene. For the imidazole system, we considered initial configurations of \systemboxmol{5}, \systemboxmol{10} and \systemboxmol{15} molecules. As can be seen in \cref{fig:imidazole:systemsize}, even for $N=125$ the results are sufficiently converged. As we used the same force field, we would expect very similar results for benzene. In the following, we use $N=216$ for the benzene bulk simulations, corresponding to a starting configuration of $6 \times 6 \times 6$ molecules in a box with side length $L=\SI{32.0205}{\angstrom}$.

\subsection{Locality Test}
\label{sec:gapcg:bulk:locality}

We will show that the PMF is indeed a \emph{local} function, in the sense that it can be written as a sum of local molecular contributions. %
We want to know to what extent it is possible to describe the (full) interactions of the system by only considering contributions associated with local environments. This is a critical requirement for our approach as any practicable CG interaction potential requires a cutoff.

To this end we consider a single molecule, referred to as the ``central molecule''; its local or inner environment, given by all molecules within a cutoff radius from the central molecule (measured in terms of COM distances between the molecules), also called the ``neighbourhood''; and the outer environment given by all other molecules.

For the locality test in methanol, we compare two cutoffs, \SI{6.0}{\angstrom} and \SI{9.2}{\angstrom}, corresponding to the first and second minima of the COM RDF, respectively.
To take into account statistical uncertainty, we repeat the procedure discussed below for 10 different starting configurations. 
In these configurations, the smaller cutoff includes \numrange{12}{17} molecules (average \num{14.0}); the larger cutoff includes \numrange{44}{55} molecules (average \num{50.5}).

In the first stage, the central molecule and its inner environment are constrained on their CG sites, with all other molecules moving freely, and we run MD for \SI{2}{\ns} to sample different configurations of the outer environment. We take snapshots from these constrained MD runs every \SI{125}{\ps}, corresponding to \num{16} different outer environments per inner environment. In the second stage, we constrain the CG sites of all molecules and run MD for \SI{100}{\ps} to calculate the mean forces on the CG sites due to the atomic fluctuations.

We now consider the mean forces on the CG sites of the central molecule for different environments. We denote these mean forces as $F^\text{MF}_k(\gamma,\omega)$, where $k$ indexes the six mean force components on the two sites of the central molecule, $\gamma$ identifies the configuration of the inner environment, and $\omega$ identifies the configuration of the outer environment (for a given $\gamma$). For each inner environment, there is some variation in the mean forces due to the changes in the outer environment. This can be quantified by the deviation between the individual $F^\text{MF}_k(\gamma,\omega)$ and the average value over the outer environments for a given inner environment. In \cref{fig:methanol:locality} (on the following page) it can be seen that this mean force deviation is generally smaller for a larger cutoff.

Note that this locality test does not consider whether coarse-grained particles and forces are a good approximation to the atomistic interactions; it only investigates whether the CG forces due to a local neighbourhood can be a good approximation to the ``converged'' CG forces calculated in the full system. Thus we are addressing the question ``How many neighbours do we have to include to appropriately describe the system?''

The spread we measure in a locality test for a given cutoff describes the maximum accuracy we can expect to obtain when separating the PMF into a function of local neighbourhoods based on this cutoff.

\begin{figure}[tbhp]
  \centering
  \input{thesis_figures/methanol/bulk/fig_localisation.pgf}
  \mycaption{Locality test of the mean forces for methanol.}{Distribution of the mean force components (\numrange{1}{6}) of the central molecule for different configurations beyond the neighbour shell, for two different sizes of the neighbour shell. The reference values are given by the average over outer environments for the larger \SI{9.2}{\angstrom} cutoff. Horizontal bars indicate the deviation of the \emph{average} over outer environments for the smaller \SI{6.0}{\angstrom} cutoff. Each plot corresponds to one of the ten configurations of the inner environment.}
  \label{fig:methanol:locality}
\end{figure}

\clearpage
\subsection{Methanol}
\label{sec:gapcg:bulk:methanol}

For the methanol monomer, the bond length distribution in bulk shows no significant difference from that of the vacuum monomer. Hence, we expect the monomer term for the bulk to be similar to the \descmonomer GAP trained on vacuum grid MF.
[Based on the difference in bond length average and the linear fit force constant, we would expect a deviation of about \SI{0.6}{\kT\per\angstrom}.] %

While the monomer terms for methanol are very similar, intermolecular interactions in bulk are significantly different from the clusters due to the environmental contributions to the mean forces. This can be seen in the MF cutlines, which we compare for a dimer in bulk and for a dimer in vacuum in \cref{fig:methanol:bulk:cutline:mf}. Note that the bulk cutline goes to zero much faster than the vacuum cutline; this motivates the use of much smaller cutoffs for the bulk systems.
\begin{figure}[tbhp]
  \centering
  \input{thesis_figures/methanol/bulk/fig_cutline_dimerbulk.pgf}
  \mycaption{Comparison of COM force between bulk methanol and vacuum dimer.}{``Oriented'' refers to a fixed relative orientation of two methanol molecules at a range of COM distances. ``COM'' refers to only the COM distance being fixed, while averaging over the orientations of the molecules.}
  \label{fig:methanol:bulk:cutline:mf}
\end{figure}

\subsubsection{CG Potentials}

\paragraph{GAP.}

Because the monomer terms are very similar for bulk and for vacuum, we could use the vacuum monomer potential as a core potential for the bulk GAP potentials. When teaching just monomer and dimer terms simultaneously, there is indeed a slight systematic difference between the monomer term derived from bulk and the vacuum monomer. However, when teaching monomer, dimer and trimer terms simultaneously, the bulk-induced monomer term exactly corresponds to the vacuum monomer MF.

It would also be possible to use a GAP from the cluster expansion as a core potential; this would describe the \emph{direct} contribution to the mean force, and we could teach a \descdimer and/or \desctrimer descriptor on top of this core potential to capture the environmental contribution to the mean forces.%

However, it may not always be feasible to carry out the cluster expansion, and here we show that we can learn monomer, dimer and trimer terms directly from force observations in the bulk system.
We use a training set of 30 configurations of the CG coordinates. Due to the large number of collective variables, we only collected MF in runs of approximately \SI{100}{\ps}. As these are relatively short MD runs, the error in the mean forces is larger, and we set $\sigma_f = \SI{5}{\milli\eV\per\angstrom} \approx \SI{0.2}{\kT\per\angstrom}$ instead of $\SI{0.5}{\meV\per\angstrom}$.

We consider two GAP potentials for bulk methanol. In the first one, we teach monomer and dimer descriptors simultaneously; we refer to this as ``dimer GAP''. In the second one, we teach monomer, dimer, and trimer descriptors simultaneously; we refer to this as ``trimer GAP''. In both cases we do not make use of cluster potentials as a core potential. However, we do keep the short-range repulsive pair potentials discussed in \cref{ref:methanol:corepotential} on \mypageref{ref:methanol:corepotential}.
Compared to the methanol cluster GAPs, we reduce the \descdimer cutoff from \SI{9}{\angstrom} to \SI{6}{\angstrom}, and the \desctrimer cutoff from \SI{6}{\angstrom} to \SI{4}{\angstrom}, corresponding to the second and first minima in the all-atom COM RDF, respectively.
For all other hyperparameters we use the same values as for the methanol cluster GAPs (see \cref{tab:methanol:gap:dimer,tab:methanol:gap:trimer} in the Appendix).

\paragraph{Pair potentials.}

As in the case of the clusters, we consider both a force-matching and a structure-matching set of pair potentials.
For the force-matching pair potentials, it turns out that there is no significant difference between a \SI{6}{\angstrom} and a \SI{9}{\angstrom} cutoff.
Our force-matching pair potentials are equivalent to those derived by Izvekov and Voth\cite{voth:mscg2}, and we obtain the same COM and site--site RDFs in our simulations.

The structure-matching pair potentials are, again, obtained by Iterative Boltzmann Inversion using the \VOTCA package, using the same settings as described in \cref{ref:methanol:ibi} on \mypageref{ref:methanol:ibi}.

\paragraph{Core-only potential.}
To demonstrate the significant contribution to the overall structure that is solely due to the short-range repulsion which is needed to prevent overlap of atoms, we considered two different ``core-only'' potentials. The MF core potential is the one used to provide the short-ranged repulsion for non-bonded interactions in the methanol clusters (\cf \cref{ref:methanol:corepotential}). The DBI core potential is based on a spline fit to the Direct Boltzmann Inversion of the short-range part of the all-atom RDFs. Beyond their respective short-range cutoffs, the core potentials are exactly zero.

\myvspace
The pair potentials are shown in \cref{fig:methanol:bulk:pairpot} in the Appendix. We compare the COM force predictions along a cutline in \cref{fig:methanol:bulk:cutline}.
In \cref{tab:methanol:bulk:rmse} we compare the RMSE of the predictions of the various potentials on a test set of 10 configurations.

\begin{figure}[H]
  \centering
  \input{thesis_figures/methanol/bulk/fig_bulkcf_cutline0.pgf}
  \mycaption{Bulk methanol: forces along a dimer cutline.}{Comparison of CG force predictions with bulk MF. For a more detailed explanation, see \cref{fig:methanol:dimer:cutline} on \mypageref{fig:methanol:dimer:cutline}.}
  \label{fig:methanol:bulk:cutline}
\end{figure}
\begin{table}[H]
  \centering
  \begin{tabular}{llS[table-format=1.1]}
    \toprule
    potential & derived from & \textcell{RMSE / \si{\kT\per\angstrom}} \\
    \midrule
    \descmonomer GAP & monomer MF & 6.1 \\
    \descdimer GAP   & dimer MF   & 3.3 \\
    \descdimer GAP   & trimer MF  & 3.2 \\
    \desctrimer GAP  & trimer MF  & 6.6 \\
    IBI pair potential & bulk RDF & 3.7 \\
    MSCG/FM pair potential   & bulk MF    & 3.5 \\
    \descdimer GAP   & bulk MF    & 2.7 \\
    \desctrimer GAP  & bulk MF    & 2.6 \\
    \bottomrule
  \end{tabular}
  \mycaption{Bulk methanol force prediction errors.}{Comparison of the force prediction root mean squared error (RMSE) with respect to MF in a separate test set of 10 bulk methanol configurations for different potentials.}
  \label{tab:methanol:bulk:rmse}
\end{table}

\subsubsection{MD results}

\begin{figure}[H]
  \centering
  \input{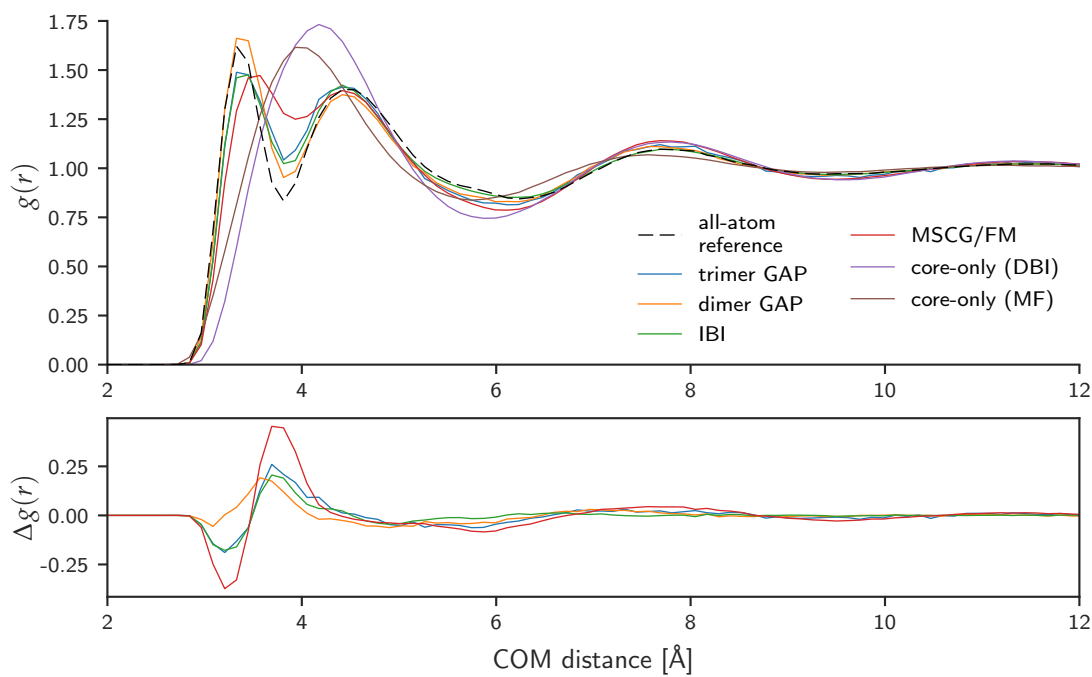}
  \caption{Bulk methanol: COM RDF for different potentials.}
  \label{fig:methanol:bulk:comrdf}
\end{figure}

We show the COM and site--site RDFs for simulations using the different potentials in \cref{fig:methanol:bulk:comrdf,fig:methanol:bulk:siterdfcc,fig:methanol:bulk:siterdfco,fig:methanol:bulk:siterdfoo}. We show the angular distribution function of the centres of mass of the molecules in \cref{fig:methanol:bulk:adf}.
In \cref{fig:methanol:bulk:orientdot,fig:methanol:bulk:orientdih} we show distributions of the relative orientation between pairs of methanol molecules as a function of their distance. We consider both the dihedral angle formed by the two CG bonds as well as the dot product of the vectors indicating their direction.

The core-only potentials already reproduce the all-atom RDFs to a large extent, with the exception of the sharp first peak in the \pairOO and COM RDFs that is due to the hydrogen bonds between \siteOH sites.
Clearly, the force-matching pair potentials do not reproduce the structure as well as the IBI pair potentials.
Here, the \descdimer GAP shows some overstructuring compared to the IBI pair potentials; this is especially prominent in the peak at \ang{60} and \SI{3}{\angstrom} in the ADF. However, when including molecular trimer interactions in the GAP, it can reproduce the all-atom orientational and angular distributions better than the IBI pair potential, despite being trained on forces.

Overall, the results in this section demonstrate that we can capture the full many-body Potential of Mean Force of the bulk methanol system to a high degree of accuracy using GAP, the resulting potential being able to reproduce the structure even better than structure-matching pair potentials.

\begin{figure}[H]
  \centering
  \input{thesis_figures/methanol/bulk/fig-bulkmd-siterdf-cc.pgf}
  \caption{Bulk methanol: \pairCC RDF for different potentials.}
  \label{fig:methanol:bulk:siterdfcc}
\end{figure}
\begin{figure}[H]
  \centering
  \input{thesis_figures/methanol/bulk/fig-bulkmd-siterdf-co.pgf}
  \caption{Bulk methanol: \pairCO RDF for different potentials.}
  \label{fig:methanol:bulk:siterdfco}
\end{figure}
\begin{figure}[H]
  \centering
  \input{thesis_figures/methanol/bulk/fig-bulkmd-siterdf-oo.pgf}
  \caption{Bulk methanol: \pairOO RDF for different potentials.}
  \label{fig:methanol:bulk:siterdfoo}
\end{figure}

\begin{figure}[p]
  \centering
  \input{thesis_figures/methanol/bulk/fig_adf.pgf}
  \mycaption{Bulk methanol: Angular Distribution Function}{of molecule COMs for different potentials. For a more detailed explanation, see \cref{fig:methanol:tetramer:adf} on \mypageref{fig:methanol:tetramer:adf}.}
  \label{fig:methanol:bulk:adf}
\end{figure}
\clearpage

\begin{figure}[H]
  \centering
  \input{thesis_figures/methanol/bulk/fig_orient_dot.pgf}
  \mycaption{Bulk methanol: orientational distributions (1/2).}{This shows the distribution of the dot product between the normalised vectors that correspond to the directions of the CG bonds as a function of the COM separation between pairs of methanol molecules. \textit{Left:} 2D distribution. \textit{Right:} horizontal slices through this distribution. The distributions have been normalised along the distance axis according to the RDF convention.}
  \label{fig:methanol:bulk:orientdot}
\end{figure}
\begin{figure}[H]
  \centering
  \input{thesis_figures/methanol/bulk/fig_orient_dih.pgf}
  \mycaption{Bulk methanol: orientational distributions (2/2).}{This shows the distribution of the dihedral angle between the CG bonds as a function of the COM separation between pairs of methanol molecules. \textit{Left:} 2D distribution. \textit{Right:} horizontal slices through this distribution. The distributions have been normalised along the distance axis according to the RDF convention.}
  \label{fig:methanol:bulk:orientdih}
\end{figure}

\clearpage
\subsection{Benzene}
\label{sec:gapcg:bulk:benzene}

We now consider bulk benzene, to demonstrate that the GAP approach can be easily applied to different systems. We described the molecular structure of benzene in \cref{ref:benzene:structure} on \mypageref{ref:benzene:structure}. In \cref{ref:systemsize:convergence} we discussed the convergence with respect to the size of the periodic box for imidazole, a molecule very similar to benzene, and hence chose to simulate a system with \num{216} benzene molecules in a periodic box. Following the equilibration protocol (\cref{tab:bulk:equilibrationprotocol}), we obtain a final box length of $L=\SI{32.0205}{\angstrom}$.

\subsubsection{CG Potentials}

For benzene, we derive a core potential by fitting pair potentials to the bulk mean forces using a short-range cutoff of \SI{4}{\angstrom}. The resulting potential is shown in \cref{fig:benzene:bulk:pairpot} in the Appendix.
Note that the exact form of the core potential does not matter %
 as long as it prevents overlap of CG sites and does not extend significantly into regions that are considerably populated.

\paragraph{GAP.}
In the case of benzene we do not go beyond dimer interactions, as we will see that this already describes the full interactions sufficiently well. Moreover, the \desctrimer descriptor would be too computationally expensive for bulk benzene.
We teach both \descmonomer and \descdimer simultaneously.
The hyperparameters used in this case are similar to those used for methanol; the main difference is the covariance cutoff, which is now \SI{8}{\angstrom}, corresponding to the first minimum of the benzene COM RDF. The hyperparameters for the GAP used in MD simulations are given in \cref{tab:benzene:bulkgap:dimer} in the Appendix. We will discuss the influence of different hyperparameters and how changing them affects the force predictions in \cref{sec:gapcg:factors}.

\paragraph{Pair potentials.}

We generate pair potentials using both the force-based MSCG/FM approach and the structure-based IBI approach, as we did for methanol. The resulting pair potentials are shown together with the core-only potential in \cref{fig:benzene:bulk:pairpot} in the Appendix.

\clearpage
\subsubsection{MD results}

We show the COM and site--site RDFs in \cref{fig:benzene:bulk:rdf} and the COM ADF in \cref{fig:benzene:bulk:comadf}. We consider the orientational distributions of pairs of molecules in \cref{fig:benzene:bulk:pairangle,fig:benzene:bulk:planecomangle}.

The MD results demonstrate that the GAP can accurately capture the oscillations in the benzene pair orientations, even beyond the interaction cutoff of \SI{8}{\angstrom}. In contrast, the pair potentials only manage to describe the orientation up to about \SI{6}{\angstrom}, even though their cutoff is \SI{10}{\angstrom}.

\begin{figure}[bthp]
  \centering
  \input{thesis_figures/benzene/bulk/fig_rdf.pgf}
  \mycaption{Bulk benzene: Radial Distribution Functions}{for COM and CG sites.}
  \label{fig:benzene:bulk:rdf}
\end{figure}

\begin{figure}[H]
  \centering
  \input{thesis_figures/benzene/bulk/fig_orient_dot.pgf}
  \mycaption{Bulk benzene: orientational distributions (1/2).}{This shows the distribution of plane orientations for pairs of benzene molecules as a function of their COM separation. The relative plane orientation is measured by the dot product $\abs{n_1 \cdot n_2}$ between plane normal vectors; due to the symmetry of benzene, only the absolute value matters. A value of \num{1} corresponds to parallel orientation; a value of \num{0} corresponds to perpendicular orientation.
  \textit{Left:} 2D distribution. \textit{Right:} horizontal slices through this distribution.

  This is an average over all pairs of molecules in the simulation box.
  The distributions have been normalised according to the RDF convention along the distance axis.
  Uniformly random orientations would lead to a uniform profile of the dot product, corresponding to all lines in the right-hand plot lying on top of each other.

  Note that the GAP model describes the oscillations in the distribution as well as the all-atom model over the whole range of COM separations. This can still be seen at a COM distance of \SI{10}{\angstrom}, even though the (COM-based) GAP cutoff used in this model is only \SI{8}{\angstrom}. In contrast, the IBI pair potential that reproduces the site--site RDF perfectly has a cutoff of \SI{10}{\angstrom}, but it does not describe the oscillations in the orientational distribution beyond the first peak at \SI{6}{\angstrom}.
  This demonstrates that GAP based on the many-body dimer descriptor can reproduce the all-atom distribution much better than the pairwise spline-based potential.}
  \label{fig:benzene:bulk:pairangle}
\end{figure}

\begin{figure}[H]
  \centering
  \input{thesis_figures/benzene/bulk/fig_orient_d2.pgf}
  \mycaption{Bulk benzene: orientational distributions (2/2).}{Here we consider the distribution of the absolute value of the dot product between the plane normal vector $n_\text{plane}$ of one molecule and the normalised vector along the COM separation axis $n_\text{COM}$ to another molecule, as a function of their COM separation. This figure is otherwise analogous to \cref{fig:benzene:bulk:pairangle}.}
  \label{fig:benzene:bulk:planecomangle}
\end{figure}

\begin{figure}[p]
  \centering
  \input{thesis_figures/benzene/bulk/fig_adf.pgf}
  \mycaption{Bulk benzene: Angular Distribution Function (ADF)}{of the centres of mass of the benzene molecules, for the all-atom model and the different CG potentials. The ADF (left-hand plot of each pair) is determined by considering all benzene triplets in which the COM distance $r_\text{c1}$ between a central molecule and its first neighbour is less than \SI{7.5}{\angstrom}. Here we show the 2D distribution of the angle $\theta_\text{c12}$ between the central molecule and its two neighbours against the distance $r_\text{c2}$ between the central molecule and its second neighbour.

  Even though the error is large, the core-only potential already reproduces the all-atom ADF qualitatively. Both IBI and MSCG/FM pair potentials show some systematic, patterned deviation from the all-atom ADF. The results for the dimer-descriptor GAP contain more noise, as the GAP-CG MD in this case was very expensive and we only obtained limited sampling. However, the GAP results show much less \emph{systematic} deviation than the pair potentials.}
  \label{fig:benzene:bulk:comadf}
\end{figure}

\clearpage

\subsection*{GAP can recover higher symmetry than descriptor}
\label{sec:benzene:sixfold}

Another demonstration of the power of the GAP approach and the flexibility of the potentials it can learn is the recovery of higher symmetry than that prescribed by the descriptor.

Benzene has six-fold symmetry; rotating a molecule by an integer multiple of \ang{60} around the plane normal axis brings it into a state completely equivalent to the previous one. However, our three-site CG description of the benzene molecule only has three-fold symmetry: one would have to rotate a molecule by an integer multiple of \ang{120} to return it to an equivalent state. Despite this limitation, the GAP successfully learns that the three-site description is of a molecule with six-fold symmetry. This can be seen in the oscillation of the potential landscape of the dimer descriptor as a function of the in-plane rotation angles of both molecules, shown in \cref{fig:benzene:sixfold}.
\begin{figure}[tbhp]
  \centering
  \input{thesis_figures/benzene/bulk/fig_gap_rot.pgf}
  \mycaption{GAP-predicted PMF as function of rotation around plane normal axes}{for a dimer descriptor trained on bulk benzene MF. This reproduces the six-fold rotational symmetry of the all-atom description, even though the CG description only prescribes three-fold rotational symmetry: rotating either molecule by \ang{30} around its plane normal axis leads to an (approximately) equivalent point in the GAP-predicted free energy surface.}
  \label{fig:benzene:sixfold}
\end{figure}

\clearpage
\section{Influence of Hyperparameters, Sparsification, and Training Data}
\label{sec:gapcg:factors}

So far, we have simply stated the values for hyperparameters and other settings used in generating GAP-CG potentials. Here, we discuss the influence of the hyperparameters (\cref{sec:gapcg:hyper}), the number and selection of sparse points (\cref{sec:gapcg:sparsification}), and the size and noisiness of the training set. This is applied to the bulk benzene model presented in the previous section.

\subsection{Hyperparameters}
\label{sec:gapcg:hyper}

The squared exponential kernel discussed in \cref{sec:gp:kernel} has several hyperparameters that influence the behaviour of the Gaussian process and hence the prediction performance (\cf \mypageref{sec:gp:hyperparameters}). The covariance cutoff, length scale and energy scale are properties of the \emph{system} we want to model and are independent of the training data and its noise.

The choice of the covariance cutoff can be based on minima of the RDF; this minimises the number of interactions at the edge of the cutoff, which might be affected by cutoff effects. Whether only the first or also the second neighbour shell (or beyond) needs to be included is directly related to the locality of the mean forces, and we discussed this for bulk methanol in \cref{sec:gapcg:bulk:locality}.

The length scale $\theta$ of the tail of non-bonded interactions at longer range is on the order of \SIrange{1}{2}{\angstrom}; for the more rapidly changing part of the interaction at short range or for stiffer bonded interactions, a length scale of around \SI{0.3}{\angstrom} is more appropriate. This variation in length scale necessitates the use of a ``transfer function'' to be able to describe both regimes at once (\cf \cref{sec:gap:transferfunction}). We demonstrated this for the methanol dimer in \cref{ref:methanol:transferfunction}. %
In \cref{fig:gapcg:hyper:thetatransfer} in the Appendix we show the effect of varying the length scales and the parameters of the transfer function.

The hyperparameter $\sigma_f$ for the noise level in the force observations depends on the training data. When training from mean forces, we can calculate the error within the mean force calculations. In our case, the mean force is the mean of all constraint force samples, and we can measure the error based on the standard deviations of the constraint forces. %
We will discuss the much larger noise that is present when teaching from instantaneous collective forces in \cref{sec:gapcg:rawmean}.

It should be noted that another source of noise from the point of view of the Gaussian process regression is a mismatch between the true function that we want to infer --- in this case the PMF as a function of all $3N$ CG coordinates --- and what is representable by the descriptors used in the GAP model. If the descriptors cannot represent the true shape of the PMF, the noise level hyperparameter would need to be increased. Fortunately, our work suggests that the molecular descriptors based on monomers, dimers, and trimers can already represent the PMF to a high degree.

The energy scale $\delta$ is hard to determine when we are only teaching from forces. In the simplest case, it is only the ratio $\delta / \sigma_f$ that matters.
In \cref{fig:gapcg:hyper:deltasigma} we show the force prediction error (relative to reference mean forces for a separate test set of 10 CG configurations that were not part of the training set) as a function of the energy scales for monomer and dimer descriptor, $\delta_\text{mono}$ and $\delta_\text{dim}$, for different values of $\sigma_f$.
\begin{figure}[tbhp]
  \centering
  \input{thesis_figures/benzene/bulk/fig_hyper_delta_sigma.pgf}
  \mycaption{Influence of hyperparameters on GAP prediction: energy and noise scales.}{Force RMSE in \si{\kT\per\angstrom} in test set and training set for different combinations of the monomer and dimer energy scales ($\delta_\text{mono}$ and $\delta_\text{dim}$, respectively), for different values of the noise level hyperparameter $\sigma_f$. All other settings as given in \cref{tab:benzene:bulkgap:dimer}. A large discrepancy between test set error and training set error indicates overfitting.}
  \label{fig:gapcg:hyper:deltasigma}
\end{figure}

Finally, an important point to appreciate is that the more training data is available, the less is the influence of the prior as determined by the covariance function and its hyperparameters. Hence, with a sufficient number of training configurations, GAP predictions are robust with respect to changes in the hyperparameters.

\subsection{Sparsification}
\label{sec:gapcg:sparsification}

Sparsification is a further approximation in the GAP approach to reduce the memory and computing requirements. The number of sparse points should be high enough to be able to represent the interaction potential accurately.
In the limit of the number of sparse points being equal to the number of training points, there is no distinction between different methods for selecting the sparse points. However, in practice, the number of sparse points needs to be significantly smaller than the number of training points. In \cref{fig:gapcg:sparse} we compare the force prediction error for the different selection algorithms introduced in \cref{sec:gap:sparsepoints} as a function of the number of sparse points.

The sparse points should evenly cover that part of the descriptor space that is actually explored by the system.
A uniform selection of sparse points is the most favourable choice for low-dimensional systems and when a sufficiently large number of sparse points can be used.
It avoids selecting sparse points that are close to each other, which would give rise to a high condition number of the covariance matrix.
Interestingly, uniform selection still appears to perform best for the relatively high-dimensional benzene dimer descriptor (15 dimensions).
For more drastic sparsification (smaller number of sparse points or higher descriptor dimension), we would expect CUR decomposition to outperform uniform selection.

\begin{figure}[H]
  \centering
  \input{thesis_figures/benzene/bulk/fig_hyper_sparse.pgf}
  \mycaption{Influence of number and selection method of sparse points on GAP prediction.}{Force RMSE as a function of the number of sparse points, for different selection methods. All other settings as given in \cref{tab:benzene:bulkgap:dimer}. Covariance-based sparse point selection was too slow to be feasible for higher numbers of sparse points. Note that adding sparse points never reduces prediction performance, but yields diminishing returns.}
  \label{fig:gapcg:sparse}
\end{figure}

\subsection{Noisy Training Data}
\label{sec:gapcg:rawmean}

Throughout the present work, we train GAPs from mean forces (MF), as in this case the training data has very little noise (if sufficient sampling has been used). This allows us to focus on the task of reproducing the high-dimensional PMF from gradient observations. This approach amounts to a separation between \emph{averaging} instantaneous forces to obtain the mean forces and \emph{integrating} the mean forces to obtain the PMF.
However, in practical applications it can be difficult to obtain mean forces with small errors due to the cost of calculations.
Recovering a function from \emph{noisy} observations is obviously more complicated than from data with no or only little noise. However, we showed that GAP can also be trained on samples of the instantaneous collective force (ICF), rather than the mean force, and thus carry out the averaging within the Gaussian process regression. %
The results shown in \cref{fig:gapcg:rawteach:converge} (see following page) demonstrate that with increasing volume of training data the ICF-trained GAP predictions converge towards those of the MF-trained GAP.
\begin{figure}[tbhp]
  \centering
  \input{thesis_figures/benzene/bulk/fig_verynoisy_sigma.pgf}
  \mycaption{Influence of number of training configurations and noise scale on GAP force prediction in the case of noisy training data (ICF).}{We show the force prediction RMSE of GAPs taught from ICF, for different sizes of the training set and for a range of values of the noise level hyperparameter $\sigma_f$. All other settings as given in \cref{tab:benzene:bulkgap:dimer}. For comparison, we include the results for GAP trained on converged MF as before (60 configurations). This demonstrates that, with sufficient training data, teaching from ICF converges to the teaching from MF.
  
  Importantly, this figure suggests choosing a very high noise level for a small and noisy training set. With an increasing number of training points, the RMSE curve flattens out. This means that the precise value of the noise level matters less. Note that a flat curve implies that the minimum is not stable and, hence, it does not make sense to ``optimise'' the noise level parameter; this would risk overfitting of the training data. As the prediction performance is stable with respect to the noise level for larger training sets, we can simply choose a sufficiently high noise level.}
  \label{fig:gapcg:rawteach:converge}
\end{figure}

In \cref{sec:methanol:monomer} we derived the appropriate noise level for teaching from ICF from the training data itself. In principle, this is also possible in higher dimensions. Note that we cannot simply take the \emph{global} standard deviation of force components. Instead, we have to transform the force observations --- that is, observations of gradients of the PMF with respect to positions of CG sites --- into gradients of the local free energy contributions in the space of the descriptors that we want to train. We then need to construct a grid in the descriptor space and calculate the standard deviation of the gradients \emph{in each bin}. The resulting value for the anticipated noise of the gradients in descriptor space would then have to be transformed back into Cartesian space. This is, however, beyond the scope of this thesis.

Importantly, it has been shown in the case of few dimensions that a reconstruction of the free energy surface from ICF using Gaussian processes is a more efficient use of computational resources (in terms of time-to-solution) than collecting less noisy mean forces for a smaller amount of configurations (\eg on a grid)\cite{LetifsPaper,stecher2014freeenergygp}.
In the present work, we show that this is also true in the high-dimensional case of generating a coarse-graining GAP. In \cref{tab:gapcg:rawteach:compare} we compare the predictive power of averaging within the GAP and external to the GAP based on a constant ``total information content'' of the training set. We consider ICF, constraint force samples, and the average over constraint force samples (\ie the mean force). In each case, the training set is based on 2000 samples. For ICF, these are samples for 2000 different configurations. Of these CG configurations, 20 were used to constrain the CG sites in Constrained MD runs of \SI{100}{\ps}, during which we collected 100 constraint force samples per configuration. This results in a training set of 2000 frames. We separately calculate the average over these 100 samples, resulting in (comparatively noisy) mean forces for 20 configurations. For comparison, we also include the results for a GAP trained on the converged mean forces obtained at the end of the Constrained MD runs (corresponding to an average over \num{2e5} constraint force samples). The force prediction error depends on the value of the noise level hyperparameter. Teaching from an average of constraint force samples with a smaller noise level hyperparameter is equivalent to teaching from the samples directly with a higher noise level hyperparameter.\footnote{The error of the average over \num{100} samples is smaller than the noise in the individual samples by a factor of \num{10}.} Here, training the GAP from ICF performs significantly better than training from an equivalent number of constraint force samples. Mean forces need to be converged to a much higher accuracy to provide better results.

\begin{table}[h]
  \centering
  \sisetup{
    table-format = 1.2,
  }
  \begin{tabular}{r@{}l@{}lSSSSSS}
    \toprule
    & & & \multicolumn{6}{c}{$\sigma_f$ / \si{\kT\per\angstrom}} \\
    \cmidrule(l){4-9}
    & & & {\num{0.04}} &  {\num{0.4}} &   {\num{4}} &   {\num[tight-spacing = true]{4e1}} &  {\num[tight-spacing = true]{8e1}} & {\num[tight-spacing = true]{4e2}} \\
    \midrule
    \multicolumn{3}{l}{training set} & \multicolumn{6}{c}{RMSE / \si{\kT\per\angstrom}} \\
    \midrule
     (&a&) $2000\times$ ICF                 &       &  2.45 &  2.44 &  2.44 &  2.46 &  2.69 \\
     (&b&) $20\times$ $100\times$ $L$       &       &  2.77 &  2.75 &  2.68 &  2.67 &  2.78 \\
     (&c&) $20\times$ $\average{L}_{100}$   &  2.77 &  2.75 &  2.68 &  2.78 &       &       \\
     (&d&) $20\times$ MF                    &  2.35 &  2.35 &  2.38 &       &       &       \\
    \bottomrule
  \end{tabular}
  \mycaption{Comparison of GAP performance for different sources of training data.}{Force prediction RMSE for different values of the noise level hyperparameter $\sigma_f$, for GAP trained on (a) ICF for 2000 different configurations, (b) 2000 samples of the constraint force $L$ [20 different CG configurations with 100 samples each], (c) noisy mean forces from the average $\average{L}_{100}$ over the 100 constraint force samples for each of the 20 CG configurations and, for reference/comparison, (d) converged mean forces (MF) corresponding to an average over \num{2e5} constraint force samples.}
  \label{tab:gapcg:rawteach:compare}
\end{table}

\clearpage

\section{Summary}

In this chapter we applied the GAP-CG approach to small clusters of methanol and to bulk systems of methanol and benzene. We showed for dimers and trimers that the ensembles of the GAP-CG MD simulations are indistinguishable from the ensembles of the coarse-grained coordinates in the atomistic ensemble. For larger systems we did not always achieve a perfect reproduction of the atomistic distributions, but GAP-CG still provided much more accurate results than pair potentials. Even though structure-matching approaches such as IBI reproduce pairwise distance distributions/RDFs by construction, we demonstrated that this does not, in general, allow them to capture more complex properties such as ADFs or correlations involving the orientation of molecules.

We demonstrated that in bulk simulations of the liquid phase a large contribution to the structure comes from environmentally-mediated contributions to the effective interactions due to the relatively close packing: simulations using only a very short-ranged repulsive core potential already reproduce the all-atom structure to a large extent.
Hence, in bulk models the lack of flexibility of pair potentials is not as prominent.

Force-based approaches to coarse-graining do not necessarily provide a good reproduction of structural properties. This is indeed the case when only considering pairwise interactions between sites.
Therefore, accurately reproducing the structure using a force-based approach is a powerful demonstration of having achieved a consistent CG simulation.
Whilst our GAP-CG approach is based on reproducing the mean forces, we can nevertheless recover structural properties of a CG system just as well, if not better than, the more common structure-based approaches that rely on pair potentials.\footnote{In the limit of arbitrarily flexible many-body interactions, both force-based and structure-based approaches will capture the PMF exactly\cite{RudzinskiNoidEntropyForcesStructure}.} Importantly, we showed that the GAP-CG potentials capture higher-order correlations very well, which pair potentials fail to do.
This suggests that GAP-CG provides a bridge between force-based approaches (such as MSCG/FM) that aim to obtain correct thermodynamic behaviour but do not necessarily recover the structure and structure-based approaches (such as IBI) that obtain correct RDFs but do not reproduce thermodynamic properties.

Accurately reproducing distributions of the all-atom ensemble in the GAP-CG models implies that we recovered good approximations to the respective PMFs.
This supports our assumption of separating the PMF into monomer, dimer, and trimer contributions and suggests that GAP will be generally applicable to a wide range of systems.

For the cases presented here, the GAP-CG simulations were slower than even the all-atom simulations. Each GAP force evaluation is significantly more expensive than the evaluation of a pairwise spline interaction, and in the systems we studied in this chapter there was not a sufficient reduction in the number of interactions compared to the all-atom model to make up for the increased computational cost of each interaction.

GAP-CG will only be faster than all-atom simulations when the number of CG interactions is much smaller than the number of atomic interactions. This will be the case for solvated systems with a large number of solvent molecules when we are only interested in the behaviour of the solute. In the next chapter, we will see for the example of benzene solvated by water that the accuracy of GAP allows us to generate a solvent-free CG model that reproduces the properties of the all-atom model significantly more accurately than alternative CG approaches while also providing a significant speedup over all-atom simulations.

\chapter{Practical Application: Solvent-Free Coarse-Graining}
\label{ch:waterbenzene}

In the previous chapter, we showed that we can recover the many-body PMF to high accuracy by using GAP to learn local contributions to the free energy from mean or instantaneous collective forces. While this demonstrated the \emph{accuracy} of the GAP-CG approach, for the systems discussed so far GAP would not be useful in practice, as it is actually \emph{slower} than simulations of the underlying all-atom model: each calculation of a GAP interaction between two molecules is significantly more expensive than the cost of computing the sum of the corresponding molecular mechanics interactions in the underlying all-atom model.

GAP for coarse-graining will only be of practical benefit when it is significantly faster than the underlying all-atom MD. This can be achieved when the CG model has far fewer sites and hence interactions that need to be calculated than the all-atom model on which it is based. A prime example where this is realised are solvent-free CG models for systems which contain a large volume of solvent molecules. This is common in biomolecular simulations where we need to include all the water molecules that make up the bulk of every cell but are more interested in the behaviour of the solutes, such as nucleic acids, proteins or lipids. By discarding all the water molecules we can obtain significant computational gains using GAP models, while the molecular descriptors can represent the effective interactions between solutes, mediated by the solvent molecules, much more accurately than the commonly used site-based potentials.

As an example, we here consider benzene solvated by water. Benzene is hydrophobic, which leads to interesting solvation effects. We describe the all-atom model and discuss the hydrophobic effect in \cref{sec:waterbenzene:allatom}.
As discussed in \cref{sec:gapcg:bulk}, many of the structural properties of bulk systems arise purely due to packing and other environmental effects.
In \cref{sec:waterbenzene:comparison} we highlight the influence of the surrounding environment on pair interactions by comparing structural properties and the COM PMF for pairs of benzene molecules in bulk, in gas phase, and dissolved in water at two different concentrations.
In a solvent-free CG model, the structural effects that arise due to the removed solvent molecules need to be introduced by the CG interactions. In \cref{sec:waterbenzene:potentials} we discuss different CG potentials for a solvent-free CG description of water-solvated benzene. In \cref{sec:waterbenzene:mdresults} we present results from MD simulations using the CG potentials. We will see that pair potentials cannot satisfactorily account for the structural effects due to the removed water molecules, even if the potentials are constructed to reproduce the solute--solute RDF of the all-atom model. In contrast, we will see that the GAP approach reproduces structural properties much more accurately. A further demonstration of the accuracy of the GAP approach is the reproduction of the COM PMF, which we discuss in \cref{sec:waterbenzene:compmf}.
Finally, in \cref{sec:waterbenzene:speed} we compare the computational cost of the different approaches. %

\section{Water-Solvated Benzene}

\subsection{All-Atom Description}

\label{sec:waterbenzene:allatom}

The structure of molecular benzene was described in \cref{sec:gapcg:systems}. Here we study benzene solvated by water.
To describe the structure and interactions of the water molecules in the all-atom simulations we use the TIP3P three-site water model\cite{TIP3P}.

We consider two different systems of water-solvated benzene: a benzene dimer solvated by 874 water molecules and an octamer of eight benzene molecules solvated by 868 water molecules. In each case, we use a cubic simulation box with periodic boundary conditions and follow the equilibration protocol given in \cref{tab:bulk:equilibrationprotocol}. This results in box lengths of $L_2=\SI{29.547}{\angstrom}$ and $L_8=\SI{29.804}{\angstrom}$, respectively, where the subscripts indicate the number of benzene molecules in the system. Hence, the benzene concentrations are $c_2 = \SI{129}{\mole\per\m^3} \approx \SI{0.13}{\M}$ and $c_8 = \SI{502}{\mole\per\m^3} \approx \SI{0.5}{\M}$.

The benzene molecules are hydrophobic, which leads to an additional, effective attraction between the solvated molecules compared to bulk benzene.
This is illustrated in \cref{fig:waterbenzene:waterdist}, where we show the distribution of water molecules between two benzene molecules at a range of COM distances, juxtaposed with the COM PMF and the mean force.
\begin{figure}[tbhp]
  \centering
  \input{thesis_figures/benzene/water/fig_water_dimer_waterdist.pgf}
  \mycaption{Benzene dimer in water: distribution of water molecules.}{\\ For a range of COM distances between the benzene molecules, the distribution of the water molecules is shown in blue, measured in terms of axial and radial distance of the oxygen atom from the line connecting the benzene COMs. Each box corresponds to a radial view of a cylinder with radius \SI{2.5}{\angstrom}; the black circles indicate the positions of the benzene COMs.

  Up to a COM separation of \SI{7}{\angstrom} the water molecules only encroach the edges of the cylinder between the two benzene molecules. Beyond \SI{8}{\angstrom} there is always a water molecule between them. The ``sharing'' of the hydrophobic pocket with no water between two benzene molecules effectively gives them an additional attraction. This is reflected by the behaviour of mean force (MF) and PMF for the benzene COM separation.}
  \label{fig:waterbenzene:waterdist}
\end{figure}

\FloatBarrier
\subsection{Benzene Dimer in Different Environments}
\label{sec:waterbenzene:comparison}
\label{ref:benzene:comparison}

The interactions between molecules depend significantly on their environment. We demonstrate this at the example of a benzene dimer in gas phase, in bulk benzene, and in bulk water.

In each environment we carried out mean force calculations in which we constrain the centres of mass of two benzene molecules at a range of fixed separations; details are given in \cref{appendix:benzenemeanforces}. The resulting COM PMFs are shown together with the RDFs in \cref{fig:benzene:comparison:pmf}.

\begin{figure}[tbhp]
  \centering
  \input{thesis_figures/benzene/fig_com_dimer.pgf}
  \mycaption{Centre-of-mass RDF and PMF for benzene dimers in different environments.}{We compare the RDF (top) and the PMF (bottom) for bulk benzene, a benzene dimer in gas phase ($V_m=\SI{6.6}{\dm^3\per\mole}$), a benzene dimer solvated by water ($c_2 \approx \SI{0.13}{\M}$), and water-solvated benzene at a higher concentration of $c_8 \approx \SI{0.5}{\M}$. The steeper slope in the PMF for the dimer in water corresponds to a stronger mean force due to the hydrophobic attraction.}
  \label{fig:benzene:comparison:comdist}
  \label{fig:benzene:comparison:pmf}
  \label{fig:waterbenzene:comparisonrdf}
\end{figure}

We also compare the orientational and distance distributions of a pair of benzene molecules in the different environments in \cref{fig:benzene:comparison:orient1,fig:benzene:comparison:orient2} on the following pages.
All simulations were carried out in a cubic simulation box with periodic boundary conditions. The bulk benzene results are from \cref{sec:gapcg:bulk:benzene}. The gas phase results are for a dimer in a periodic box with $L=\SI{28}{\angstrom}$, corresponding to a molar volume of $V_m=\SI{6.6}{\dm^3\per\mole}$. The results for water-solvated benzene are for the systems introduced at the beginning of this section.

It can be seen that structural distributions of a benzene dimer in water are very different from those in gas phase. In fact, the distributions of the water-solvated benzene are more similar to those in bulk benzene. The RDF shows an oscillatory structure with maxima where an integer number of water molecules neatly fit in the space between the benzene molecules. We also note that the distributions for water-solvated benzene dimer and octamer closely resemble each other; this suggests that the benzene concentrations are too small for (molecular) three-body effects to play a significant role.

In the remainder of this chapter we will see that pair potentials between CG sites do not allow enough flexibility to represent the differences between effective interactions in the different environments, whereas these differences can be captured by the molecular interaction terms available in the GAP approach.

\begin{figure}[H]
  \centering
  \input{thesis_figures/benzene/comparison/fig_orient_dot.pgf}
  \mycaption{Orientational distributions of benzene pairs in different environments (1/2).}{We show the distribution of plane orientations, measured by the absolute value of the dot product between the plane normal vectors, $\abs{n_1 \cdot n_2}$, as a function of the COM separation. \textit{Left:} 2D distribution. \textit{Right:} horizontal slices through this distribution; each line corresponds to a certain range of dot product values. The distributions have been normalised according to the RDF convention along the distance axis; the sum over orientations corresponds to the COM RDF. Uniformly random orientations would lead to a uniform profile of the dot product, corresponding to all lines lying on top of each other.

  Note that the pair structures of the two water-solvated systems are very similar to each other, and noticeably different from bulk benzene. However, the differences between a pair of benzene molecules in bulk water and in bulk benzene are small compared to the much larger difference to the gas phase dimer.}
  \label{fig:benzene:comparison:orient1}
\end{figure}
\begin{figure}[H]
  \centering
  \input{thesis_figures/benzene/comparison/fig_orient_d2.pgf}
  \mycaption{Orientational distributions of benzene pairs in different environments (2/2).}{This figure is similar to \cref{fig:benzene:comparison:orient1}; here, we measure the orientation using the absolute value of the dot product between the plane normal vector $n_\text{plane}$ of one benzene molecule and the normal vector $n_\text{COM}$ in the direction of the COM separation. Again, there is a strong similarity between the pair structures of water-solvated benzene at different concentrations, clearly distinct from the bulk benzene structure, and the gas phase structure is very different from the structures in bulk.}
  \label{fig:benzene:comparison:orient2}
\end{figure}

\clearpage
\subsection{CG Potentials}
\label{sec:waterbenzene:potentials}

As before, we compare force-based and structure-based approaches to coarse-graining.
For the force-based approaches, we collected mean forces for 17000 different CG configurations of the water-solvated benzene dimer in constrained MD runs of \SI{100}{\ps}. From these mean force observations we trained a GAP with the molecular dimer descriptor and generated a force-matching pair potential using the MSCG/FM method. Structure-based approaches are represented by the pair potential obtained by the IBI approach for the water-solvated benzene octamer. For comparison, we also include a ``traditional'' implicit solvent model of the solvated benzene dimer in which the solute is described in atomistic detail, based on the Generalised Born model for solvation effects\cite{GeneralizedBorn}.

\subsubsection{GAP} The hyperparameters we used were similar to bulk benzene, \cf \cref{tab:benzene:bulkgap:dimer}.
On a separate test set of mean forces for 3000 CG configurations, this potential has an RMSE of \SI{0.7}{\kT\per\angstrom}. %

Based on a one-site CG model for benzene in water, \textcite{BenzeneWaterCGonesite} suggests that potentials based on infinite dilution can transfer well to larger concentrations. Motivated by this, we take the dimer GAP trained entirely on dimer configurations and also apply it at higher concentration in the system of eight benzene molecules in water.

\subsubsection{Pair potentials}

The CG description of benzene used in this work has only one site type and, hence, there is just a single non-bonded pair potential.
This makes it easy to provide the pair-potential-based approaches with sufficient data to ensure that we have obtained the best effective pair potentials, according to the respective approach.
However, it also means that it is difficult for the pair potentials to approximate the true interactions as given by the PMF.

\paragraph{Force-based approach.}

For the water-solvated benzene dimer, we construct a force-matching pair potential using the MSCG/FM method.
The pair potential has an RMSE of \SI{1.2}{\kT\per\angstrom} on the mean force test set. %
We will see that the force-matching pair potential fails to accurately describe the distributions of the all-atom dimer model. We do not expect a force-matching pair potential to perform significantly better for the water-solvated benzene octamer, and for this reason we do not consider the force-matching pair potential again when discussing benzene at a higher concentration in water.

\paragraph{Structure-based approach.}

For the water-solvated benzene octamer we determine a pair potential that reproduces the site--site RDF using the IBI approach as discussed previously. On the test set of mean forces in the solvated dimer system, the IBI pair potential shows an RMSE of \SI{1.4}{\kT\per\angstrom}. %
We do not include a structure-matching pair potential for the water-solvated benzene dimer, as it would have been very computationally expensive to generate sufficiently converged RDFs at lower concentration.

\myvspace
The resulting pair potentials and the pair forces are shown in \cref{fig:waterbenzene:pairpot} in the Appendix.
We compare the force predictions of the CG potentials along cutlines through the configuration space of the benzene dimer in \cref{fig:waterbenzene:cutline} on the following page. This shows that the dimer GAP reproduces the all-atom mean forces significantly more accurately than either of the pair potentials, though we note that in this case there are some visible discrepancies between the all-atom reference and the GAP predictions.
\begin{figure}[tbhp]
  \centering
  \input{thesis_figures/benzene/water/fig_watcutline.pgf}
  \mycaption{Water-solvated benzene dimer: COM force along cutlines.}{\\ This shows the mean force along the COM separation for six different relative orientations of two benzene molecules solvated by water as a function of the COM distance, compared to the force predictions by the dimer GAP potential and the force- and structure-matching pair potentials. Though none of the potentials reproduce the mean forces exactly, the dimer GAP is the only one that qualitatively captures all the different orientations.}
  \label{fig:waterbenzene:cutline}
\end{figure}

\subsubsection{Atomistic implicit solvent model}
For comparison, we include results from simulations using a ``traditional'' implicit solvent model, in which the solute molecules are modelled with atomic detail. In such models, the solvation effect is described implicitly by changing the force field to account for electrostatic effects that are due to the bulk solvent. As we will see, this does not reproduce structural effects that are due to the absent solvent molecules. We use the Generalised Born solvation model implemented in the \AMBER code\cite{AMBER:GBSA1,AMBER:GBSA2}. This type of model requires vacuum boundary conditions; hence, we only applied it to the water-solvated benzene dimer.

\FloatBarrier
\subsection{MD Results}
\label{sec:waterbenzene:mdresults}

\subsubsection{Water-solvated benzene dimer}

We compare the all-atom model with the dimer-descriptor GAP, the force-matching pair potential and the atomistic implicit solvent model. As was the case for the methanol clusters, we impose a half-open harmonic restraint on the dimer COM separation with a force constant of $\SI{50}{\kcal\per\mole\per\angstrom^2} \approx \SI{84}{\kT\per\angstrom^2}$, beginning at a separation of \SI{12}{\angstrom}. The resulting distance distributions are shown in \cref{fig:waterbenzene:dimerdistance}, and results for orientational distributions are shown in \cref{fig:waterbenzene:dimerorient1,fig:waterbenzene:dimerorient2}. It can be seen that the atomistic implicit solvent model is a better approximation to the dimer solvated by explicit water than a pure gas phase simulation, and marginally better than the force-matching pair potential. However, both the atomistic implicit solvent model and the pair potential fail to reproduce RDF and orientational distributions correctly. In contrast, the dimer GAP model reproduces the RDF and even the orientational distributions. Whereas the distribution of $n_\text{plane} \cdot n_\text{COM}$ is reproduced almost perfectly, there is a clearly visible discrepancy in the distribution of $n_1 \cdot n_2$.

\subsubsection{Water-solvated benzene octamer}

We compare the all-atom model with the dimer-descriptor GAP trained on dimer MF and the structure-matching pair potential based on reproducing the site--site RDF. The COM and site--site RDFs are shown in \cref{fig:waterbenzene:8rdf}, and the COM ADF is shown in \cref{fig:waterbenzene:8comadf}. The orientational distributions are shown in \cref{fig:waterbenzene:8orient1,fig:waterbenzene:8orient2}. Whereas the structure-matching IBI pair potential reproduces the orientational distributions somewhat better than the force-matching pair potential, it does not reproduce the oscillations of the COM RDF correctly and provides less accurate results than the dimer GAP. Moreover, the discrepancy between the dimer GAP ADF and the all-atom ADF is on the same scale as the difference between the all-atom RDFs for different concentrations; hence, a dimer GAP trained on configurations of the water-solvated \emph{octamer} would be expected to reproduce the structural distributions even more closely.

\begin{figure}[tbhp]
  \centering
  \input{thesis_figures/benzene/water/fig_rdf.pgf}
  \mycaption{Water-solvated benzene dimer: distance distributions.}{\\ We compare the dimer GAP, the MSCG/FM pair potential, and the atomistic implicit solvent model with the all-atom reference model that explicitly includes water. In each case, a half-open harmonic restraint was applied to the COM distance starting at \SI{12}{\angstrom}, with force constant \smash{$\SI{50}{\kcal\per\mole\per\angstrom\squared} \approx \SI{84}{\kT\per\angstrom^2}$}. Only the many-body dimer GAP is able to capture the effective interactions and reproduce the all-atom distance distributions.  Note that the water molecules in the all-atom simulation mediate an effective interaction between the benzene molecules even beyond the all-atom force field cutoff of \SI{9}{\angstrom}.}
  \label{fig:waterbenzene:dimerdistance}
\end{figure}
\begin{figure}[tbp]
  \centering
  \input{thesis_figures/benzene/wat8/fig_rdf.pgf}
  \mycaption{Water-solvated benzene octamer: COM and site--site RDF.}{\\ By construction, the IBI pair potential reproduces the site--site RDF exactly. Nevertheless, it fails to capture the oscillatory structure in the COM RDF. In contrast, the dimer-descriptor GAP reproduces the RDFs to a reasonable accuracy, even though it was trained on water-solvated \emph{dimer} configurations.}
  \label{fig:waterbenzene:8rdf}
\end{figure}

\clearpage

\begin{figure}[H]
  \centering
  \input{thesis_figures/benzene/water/fig_orient_dot.pgf}
  \mycaption{Water-solvated benzene dimer: orientational distributions (1/2).}{\\ Plane orientation measured by the absolute value of the dot product between the plane normal vectors, $\abs{n_1 \cdot n_2}$; for more explanation see \cref{fig:benzene:comparison:orient1}. This shows that the dimer GAP provides a much better approximation to the all-atom model with explicit water than MSCG/FM pair potential or atomistic implicit solvent model.}
  \label{fig:waterbenzene:dimerorient1}
\end{figure}
\begin{figure}[H]
  \centering
  \input{thesis_figures/benzene/water/fig_orient_d2.pgf}
  \mycaption{Water-solvated benzene dimer: orientational distributions (2/2).}{\\ Plane orientation measured by the absolute value of the dot product between the plane normal vector $n_\text{plane}$ of a benzene molecule and the direction $n_\text{COM}$ between the benzene COMs. The all-atom reference distribution is accurately reproduced by the dimer GAP, whereas MSCG/FM pair potential and atomistic implicit solvent model show significant discrepancies.}
  \label{fig:waterbenzene:dimerorient2}
\end{figure}
\begin{figure}[H]
  \centering
  \input{thesis_figures/benzene/wat8/fig_orient_dot.pgf}
  \mycaption{Water-solvated benzene octamer: orientational distributions (1/2).}{\\ Dot product between plane normal vectors against the COM separation for all pairs of benzene molecules.}
  \label{fig:waterbenzene:8orient1}
\end{figure}
\begin{figure}[H]
  \centering
  \input{thesis_figures/benzene/wat8/fig_orient_d2.pgf}
  \mycaption{Water-solvated benzene octamer: orientational distributions (2/2).}{\\ Dot product between plane normal vector and normal vector between COMs against the COM separation for all pairs of benzene molecules.}
  \label{fig:waterbenzene:8orient2}
\end{figure}
\begin{figure}[p]
  \centering
  \input{thesis_figures/benzene/wat8/fig_adf.pgf}
  \mycaption{Water-solvated benzene octamer: Angular Distribution Function}{of the benzene COMs, given by the 2D distribution of the angle $\theta_\text{c12}$ between the centres of mass of a central molecule and two neighbours against the distance $r_\text{c2}$ to the second neighbour. Only triplets in which the distance $r_\text{c1}$ to the first neighbour is less than \SI{7}{\angstrom} were included. Note that the difference between dimer GAP (trained on dimer MF) and all-atom reference ADF is in line with the difference between the all-atom RDFs for water-solvated benzene dimer and octamer, \cf \cref{fig:benzene:comparison:comdist} on \mypageref{fig:waterbenzene:comparisonrdf}.}
  \label{fig:waterbenzene:8comadf}
\end{figure}

\FloatBarrier
\section{Recovering a Pair PMF from the CG Potential}
\label{sec:waterbenzene:compmf}

As a further demonstration of the power of the GAP approach, we compare the GAP-predicted PMF with the COM PMF obtained from Constrained MD with constraints on the centres of mass of two benzene molecules in water (as shown in \cref{fig:benzene:comparison:pmf}).

GAP predicts an approximation to the $N$-body PMF $A(\vec{R}^N)$ for all CG coordinates after averaging over the ``internal'' atomistic degrees of freedom. To compare this with the one-dimensional COM PMF $A_\text{COM}(r)$ shown in \cref{fig:benzene:comparison:pmf}, we need to further average over the CG degrees of freedom orthogonal to the COM distance $r$ between the two molecules, corresponding to \cref{eq:md:pmf,eq:cg:CGpotential} on a higher level:
\[
  A_\text{COM}(r) = \intdd{\vec{R}^N} \exp\big(-\beta A(\vec{R}^N) \big) \delta\big(r - r(\vec{R}^N)\big) .
\]
For the water-solvated benzene dimer, this procedure corresponds to a Boltzmann average of the PMF over rotations and deformations of the monomers.
Fortunately, within the GAP approach, contributions to the PMF have already been separated into monomer and dimer contributions. The monomer descriptor captures the contributions that are independent of the presence of other molecules and, hence, the distance between the two molecules. Therefore, we can effectively obtain the Boltzmann average over all monomer configurations by simply discarding the monomer prediction and only considering the contribution to the PMF predicted by the dimer descriptor (assuming negligible deformations of the monomers due to the dimer interaction).

We carry out the Boltzmann average over rotations by generating 1000 random, uniformly distributed orientations for the benzene monomers and predicting the dimer contribution for these orientations for a range of centre-of-mass separations.
The results are shown in \cref{fig:waterbenzene:gappmf} on the following page, where it can be seen that the dimer GAP prediction closely follows the COM PMF obtained in all-atom MD, whereas the pair potentials fail to describe its oscillatory structure.
\begin{figure}[H]
  \centering
  \input{thesis_figures/benzene/water/fig_compmf_watdim.pgf}
  \mycaption{COM Potential of Mean Force.}{We compare the PMF obtained through all-atom Constrained MD simulations with the COM PMF predictions of the CG interaction potentials. The COM PMF predictions were obtained from calculating the Boltzmann-weighted average over monomer orientations of the dimer terms in the potentials. This supports our claim that the GAP approach can accurately capture the many-body Potential of Mean Force and separate it into monomer and higher-order contributions.}
  \label{fig:waterbenzene:gappmf}
\end{figure}

\section{Simulation Speed}
\label{sec:waterbenzene:speed}

The overall speed of a molecular dynamics simulation for a given model or interaction potential is defined in terms of how much model time we can simulate in a certain amount of wall-clock time, typically measured in \si{\ns\per\day}. This depends on the system size, the number of computing cores, and the time step $\Delta t$.
The run time of molecular dynamics simulations generally scales linearly\footnote{The Ewald summation that deals with electrostatic interactions in periodic systems actually scales as $\order{N \log N}$, but the difference to linear scaling is negligible in practice.} with the number of particles $N$.
Ideally, doubling the number of cores would halve the time needed to run a simulation.
However, due to parallel overheads, the run time will not necessarily scale inversely with the number of cores.
In \cref{tab:waterbenzene:speed} we compare the speed of all-atom simulations with that of CG pair potentials and dimer GAP.
Parallel scaling behaviour of GAP simulations compared to all-atom MD is shown in \cref{fig:waterbenzene:speed}.

\begin{table}[htbp]
  \centering
  \begin{tabular}{llS[table-format=1.1]S[table-format=4]S[table-format=2]S[table-format=2.1]S[table-format=2.1]S[table-format=1.1]}
    \toprule
    system & model & {$\Delta t$ / \si{\fs}} & {$N_\text{site}$} & {cores} & {\si{\ms\per step}} & {\si{\ns\per\day}} & {$\tfrac{\si{\ns\per\day}}{\si{core}}$} \\
    \midrule
    dimer   & all-atom           & 0.5  & 2646 & 16  &  2.4 &   18   & 1.1 \\
    octamer & all-atom           &      & 2700 & 16  &  2.4 &   18   & 1.1 \\
            &                    &      &      &  8  &  3.9 &   11   & 1.4 \\
            &                    &      &      &  4  &  7.2 &    6   & 1.5 \\
            &                    &      &      &  2  & 14   &    3   & 1.5 \\
            &                    &      &      &  1  & 28   &    1.5 & 1.5 \\
    dimer   & dimer GAP          & 4    &    6 &  1  & 44   &    7.9 & 7.9 \\
    octamer & dimer GAP          &      &   24 &  4  & 42   &    8.2 & 2.0 \\
    octamer & IBI pair potential &      &   24 &  4  & \num{0.14} & {\num{2400}} & {\num{600}} \\
    \bottomrule
  \end{tabular}
  \caption{Simulation speeds for water-solvated benzene dimer and octamer.}
  \label{tab:waterbenzene:speed}
\end{table}
\begin{figure}[tbhp]
  \centering
  \input{thesis_figures/benzene/water/gaptiming.pgf}
  \mycaption{Parallel scaling.}{Simulation speeds for all-atom and dimer GAP MD, where $S$ indicates the number of sparse points. Ideal scaling would correspond to a linear relationship between simulation speed and number of computing cores. These results show that the parallel scaling improves for larger system sizes. The dashed line shows the result for $S=5000$ sparse points scaled by the factor \num{2.5}; this demonstrates the linear scaling with $S$, and also underlines that larger calculations scale better to a higher number of cores.}
  \label{fig:waterbenzene:speed}
\end{figure}

\FloatBarrier
\section{Summary}

Here we have demonstrated a GAP-CG model that is significantly faster than the underlying all-atom simulation, while providing a good reproduction of the all-atom distributions and correlations. In contrast, implicit solvent simulations using pair potentials, whether describing the solute with atomic detail or using CG sites, fail to reproduce all-atom distributions and correlations.

We note that the water-solvated benzene is expected to be at the lower end of speedups that are attainable in solvent-free CG models using GAP. Here, the time step in the CG simulations is limited by the comparatively stiff monomer term. When considering a more drastic coarse-graining of larger molecules, the resulting monomer interactions are likely to be softer, thereby allowing a larger time step in the CG simulation. Moreover, larger molecules would require a larger box, resulting in an even higher speedup compared to all-atom simulations. In cases where real dynamics are not relevant, the CG simulations are further accelerated by the smoother CG interactions, which lead to reduced friction in the CG model. 

We acknowledge that the GAP for the water dimer did not reproduce all of the all-atom distributions perfectly. There is also a visible discrepancy between GAP-predicted and all-atom mean forces along parts of the cutlines. However, there are no intrinsic reasons that would prevent the \descdimer GAP from accurately reproducing the water dimer PMF. Investigating this discrepancy will be a subject of future research.

In further work, this GAP-CG model could, for example, be used in a random search for minima. This would not be feasible with the all-atom potential due to the numerous local minima of the solvent configuration in which any minimiser would get stuck.

\chapter{Conclusions}
\label{ch:conclusion}

In this thesis I showed that the Gaussian Approximation Potential (GAP), previously used for interatomic potentials, can recover the many-body potential of mean force of a coarse-grained system to a high accuracy. I demonstrated for the examples of methanol clusters, bulk methanol and benzene, and water-solvated benzene clusters that molecular dynamics simulations using this GAP-CG approach are significantly better at reproducing distributions and correlations of the underlying fine-grained model than currently predominating approaches. Whereas other CG approaches generally describe effective interactions in the coarse-grained model using pairwise radial interactions between CG sites, GAP is based on molecular terms that assign contributions to the free energy to the internal configurations of monomers and to dimer and trimer configurations. Using Gaussian process regression, GAP infers these free energy contributions from observations of mean or instantaneous collective forces on the CG sites.

Coarse-grained interaction potentials that are based on matching forces were generally thought not to be as good at recovering \emph{structural} distributions as approaches that directly focus on reproducing the structure, for example, by iteratively refining potentials to match Radial Distribution Functions (RDFs). In contrast, such structure-based approaches generally fail to reproduce thermodynamic properties of the underlying model. In the present work I demonstrate that this lack of representability is actually due to the mistaken assumption that pairwise interactions between individual sites can provide a good approximation to the true CG interactions. The flexibility of the monomer, dimer, and trimer descriptors in the GAP approach can represent general many-body interactions (for example, a molecular dimer descriptor for a three-site description of a CG molecule corresponds to 6-body interactions in terms of CG sites). This allows us to recover the structural distributions of the underlying system \emph{better} than structure-based approaches that only optimise the interactions based on a subset of distributions (usually the RDFs).
GAP provides a flexible and powerful approach for representing not just interatomic potential energy surfaces, but also, as this thesis has shown, for coarse-grained free energy surfaces.
Compared to ``atomistic GAP'' for interpolating the potential energy surfaces of all-atom systems, GAP-CG faces additional challenges. Whereas both function values and gradient observations are available for the potential energy surface, in the coarse-graining approach we only have access to the gradients (mean or instantaneous collective forces) and from these need to reconstruct the free energy surface. This makes it harder both to train an accurate GAP and to validate its accuracy. Without function value observations, the Gaussian process regression requires more training data to reconstruct the unknown energy surface to comparable accuracy. Moreover, when the GAP is trained from instantaneous collective forces, the training set contains a significant amount of noise.
When mean forces are available, the GAP force predictions can be compared with mean forces for a separate test set of configurations that were not used in the training. Based on comparison with the test set mean forces, it is possible to tune the hyperparameters and thereby optimise the GAP predictions. When accurate mean forces are not available as a reference, we need to rely on intuition for selecting the values of the hyperparameters. However, in the present work I developed working values for the hyperparameters and showed that the GAP prediction accuracy is fairly insensitive to changes of the hyperparameters, so this is not as serious an issue as might have been thought initially.

\myvspace
Many-body interactions are computationally expensive.
The evaluation of a GAP interaction is significantly more expensive than the evaluation of a simple tabulated spline (by two to three orders of magnitude). %
Hence, in simulations of bulk systems where many interactions need to be calculated, GAP-CG is significantly slower than even the all-atom simulations. In bulk systems, packing effects are responsible for a significant part of the overall structure and, for this reason, CG pair potentials already provide reasonable accuracy.

GAP-CG will be useful in cases where it is faster than all-atom MD whilst being significantly more accurate than CG models based on pair potentials.
This can be the case in systems with a large volume of solvent, for which most of the computing time in typical all-atom MD simulations is spent on calculating non-bonded interactions between solvent molecules. If we are interested in the effective behaviour of the solutes, we can remove the solvent degrees of freedom. A solvent-free GAP-CG simulation can significantly reduce the computational cost compared to the all-atom simulation. Even though simpler methods based on pair potentials would be faster still, this would be at the cost of a significantly reduced accuracy with respect to the reproduction of properties of the underlying system.

\myvspace
GAP-CG does not supersede previously introduced approaches, but provides another tool whose usefulness depends on the application. GAP may be especially useful when combined with other approaches. %
In the following, I discuss scaling to larger system sizes (\cref{sec:conclusions:scaling}), limitations of the GAP approach to coarse-graining (\cref{sec:conclusions:limitations}), and an outlook to the future of this approach (\cref{sec:conclusions:future}).

\section{Scaling}
\label{sec:conclusions:scaling}

GAP-CG will only provide a speed advantage for CG models that contain a significantly smaller number of CG interactions than the all-atom system. The precise number of interactions depends on the specific configuration and can change in each time step, as atoms and molecules move in and out of the interaction cutoffs. For this reason, it is common practice to simply compare the number of interaction \emph{sites}, and the ratio between the total number of atoms and the number of CG sites provides a proxy for measuring potential speedups due to coarse-graining.

In systems of comparable size to the cutoff distances, the number of dimer interactions increases quadratically with the number of sites.
In large systems, because of the cutoffs used in both all-atom and CG models,
the number of interactions that have to be computed scales linearly with the system size.

In \cref{ch:waterbenzene} I discussed the coarse-graining of water-solvated benzene and showed that GAP-CG MD is faster than the corresponding all-atom simulation (though this depends on the number of sparse points used in the Gaussian process regression). There, the ratio between the number of atoms and the number of CG sites is on the order of one hundred. For larger biomolecular simulations, coarse-graining can lead to significantly larger computational gains, as I will discuss in the following.

As an example, let us consider DNA, the carrier of genetic information in all living organisms.
In order to provide a large strand of DNA with the freedom to arbitrarily bend and flex, a sufficiently large simulation box and hence increasingly large volumes of solvating water molecules are needed. By excluding the solvent from the coarse-grained description, we vastly reduce the number of particles and thus the number of interactions that need to be evaluated during the simulation. I have quantified this saving by explicitly solvating DNA systems of different sizes in water and counting the total number of atoms. The results are presented in \cref{tab:gap:cg-wins} on the following page. For comparison, numbers for the water-solvated benzene systems and for the HIV-1 capsid model that was mentioned in the Introduction are also included.

Hence, for larger models, GAP-CG may be faster than all-atom MD by an additional order of magnitude. For models in which internal fluctuations of the CG monomer occur on longer time scales than for the comparatively stiff (high-frequency) methanol or benzene molecules discussed in this thesis, it may also be possible to further increase the time step used in the GAP-CG MD simulation.

\newcommand*{\bp}[1]{\SI{#1}{bp}}
\begin{table}[bh]
  \centering
  \begin{tabular}{lS[table-format=1e1]S[table-format=4]S[table-format=1e1]}
    \toprule
    \multicolumn{1}{l}{solute} & \textcell{atoms (solvated)} & \textcell{CG sites} & \textcell{ratio} \\
    \midrule
    benzene (\num{2} molecules, $c=\SI{0.13}{\M}$)	&   3e3 &    6 & 5e2 \\ %
    benzene (\num{8} molecules, $c=\SI{0.5}{\M}$)	&   3e3 &   24 & 1e2 \\ %
    DNA (\num{1} base pair [\si{bp}])			&   3e3 &    6 & 5e2 \\
    DNA (linear, \bp{10})				&   1e4 &   60 & 2e2 \\
    DNA (half-circle, \bp{50})				&   2e5 &  300 & 7e2 \\
    DNA (half-circle, \bp{100})				&   1e6 &  600 & 2e3 \\
    HIV-1 capsid					&  64e6 & 6580 & 9e3 \\
    \bottomrule
  \end{tabular}
  \caption{Scaling of the number of atoms with system size, compared to the number of CG sites. The solvated benzene systems were discussed in \cref{ch:waterbenzene}. Solvated atom numbers for DNA are a conservative estimate based on a fixed \SI{8}{\angstrom} solvent margin around each \emph{initial} configuration of the DNA; in practice, the \num{50}~and \bp{100} systems would need an even larger simulation box for an accurate simulation. As for benzene, a three-site CG description is assumed for each DNA base. Numbers for the HIV-1 capsid are based on the data in \textcite{bigMD:largeAllAtom}, assuming a description using five CG sites per capsid protein.}
  \label{tab:gap:cg-wins}
\end{table}

\FloatBarrier

\section{Limitations of GAP-CG}
\label{sec:conclusions:limitations}

The main limitation of GAP is its computational cost compared to simpler CG approaches. As GAP consists of independent training and prediction stages, their respective computational costs will be discussed separately.

In the training stage, we need to construct and invert the covariance matrix between sparse points to obtain the sparse coefficients. This involves matrix multiplication with the matrix of covariances between sparse and training points. Overall, the GAP training scales as $N S^2$ in calculation time and as $N S$ in memory, where $N$ is the number of training points and $S$ is the number of sparse points. Due to the large number of points that are needed to accurately determine the interactions, this stage is primarily limited by the available memory. In many cases, the GAP training requires dedicated large memory nodes; however, this effort only needs to be expended once for a CG model.

In the prediction stage, the forces at a ``test point'' are obtained by simply calculating the dot product of the sparse coefficients with the vector of covariances between test point and sparse points. This operation is linear in $S$,\footnote{The computational cost of predicting the variance scales as $S^2$.} as demonstrated by \cref{fig:waterbenzene:speed}. The predictions are mainly compute-bound.

The performance of GAP-CG also depends on the dimensionality of the descriptors. Clearly, higher-dimensional descriptors will be more expensive. Increasing the descriptor dimensionality directly increases the computational cost of calculating covariances. More significant, however, is that this will also require a larger number of training and sparse points to accurately describe the local free energy contributions in a higher-dimensional space. For very high-dimensional descriptors, it will become difficult to achieve sufficient data coverage of the descriptor space.

In this work I used descriptors based on the list of \emph{all} pairwise distances between the sites involved in monomers, dimers and trimers. As discussed in \cref{sec:gap:descriptors}, this generally provides an \emph{overcomplete} description. We may obtain better performance by constructing cleverer, lower-dimensional descriptors for the systems we want to investigate, for example, by only considering a certain subset of the distances.

When comparing the speed of GAP with other approaches, it is important to remember that GAP is essentially a development code and has not yet benefited from as many years of performance optimisation as common MD packages for established potentials such as \AMBER or \LAMMPS.
In the course of this work, I already significantly improved the parallel performance of GAP MD.

As we saw in \cref{fig:waterbenzene:speed}, the parallel scaling with the number of cores is better for larger systems. Descriptors are all independent of each other and only depend on the positions of CG sites within a localised region. There is no reason in principle that would prevent GAP from parallelising to much larger systems.
It is worth noting that MD packages are typically designed with all-atom simulations in mind. Solvent-free CG MD will require new algorithms that take advantage of the sparse nature of the CG system; further discussion of this issue can be found in \textcite{voth:parallel}.

\section{Outlook and Future Work}
\label{sec:conclusions:future}

In this work I demonstrated that GAP-CG can significantly improve the simultaneous representation of both thermodynamic and structural properties of the fine-grained system by CG simulations.
Building on this achievement, there are a range of different directions that would be worth pursuing. I will first discuss the use of GAP to describe the monomer term in highly coarse-grained models. I will then discuss the use of GAP for a better description of coarse-grained interactions in adaptive resolution simulations. Finally, I will present an outlook towards improving not just representability, but also transferability.

\paragraph{Highly coarse-grained models.}
GAP is very promising for highly coarse-grained models in which CG sites do not represent just a few atoms, but whole side chains or even whole residues; for example, this can be a lipid of 144 atoms that is represented by only three to five sites\cite{voth:lipidhcg}. This results in significantly fewer CG interactions, and hence the overhead of GAP might not be too much of a disadvantage. At the same time, potentials of mean force for highly coarse-grained models are likely to be more complex, containing more correlations than in less drastic coarse-graining. For example, CG descriptions of conjugated polymers require complicated directional interactions. Thus, in highly coarse-grained models, separating the potential of mean force into pairwise site--site interactions will be a worse approximation, and we expect this to be a good area for directly applying GAP-CG.

In this thesis I focused on modelling the \emph{non-bonded} interactions between two and more molecules. However, on a highly coarse-grained scale it turns out that even monomer interactions cannot be separated into the commonly used bond, angle, and dihedral terms. In fact, we have already planned a cooperation with the Voth group in which we will explore the use of GAP for describing the monomer term for individual lipid molecules, combined with their force-matched non-bonded interactions between different molecules.

\paragraph{Adaptive resolution simulations.}
GAP may prove to be useful in adaptive resolution simulations where part of the system is described at a higher resolution than the rest. An example would be a water-solvated protein with an active binding site and different competing solutes. The binding site and nearby water and solute molecules might require a fully atomistic description, whereas the rest of the protein and solutes that are far away from the binding site do not need such a detailed representation.
A significant problem in such simulations is the difference in chemical potential between the CG and the all-atom description of a molecule, due to an incorrect CG description of the thermodynamics.
This leads to an artificial drift across the border between CG and all-atom description\cite{AdaptiveResolution}.
So far, simulations need to be corrected manually for this drift, which is a difficult and \textit{ad-hoc} procedure. With the improved representation of both structure and forces in GAP-CG, these manual corrections might no longer be necessary.

\paragraph{Transferability.}
Another major open question in CG research concerns the \emph{transferability} of CG potentials between different state points. The averaging over microscopic degrees of freedom leads to an entropic component in the CG interactions that inherently depends on the state point, that is, temperature, pressure/density, and so forth. CG potentials that include density or temperature as extended variables have been investigated within the MSCG approach\cite{MSCG:transferringTemperature,MSCG:densityDependent,MSCG:densityDependent2,CGtransferabilityExtendedEnsemble}. So far, we have not discussed this dependence in detail but, in principle, external parameters such as the temperature could be explicitly added as extended variables to the Gaussian process regression within the GAP approach. This would allow ``reuse'' of contributions to the potential of mean force that only change slowly with temperature.

\appendix
\chapter{Appendix}

\section{GAP hyperparameters}

\begin{table}[htbp]
  \centering
  \begin{tabular}{lc}
    \toprule
    hyperparameter & dimer \\
    \midrule
    core potential: \\
    \multicolumn{2}{l}{\qquad\descmonomer GAP from grid MF} \\
    \multicolumn{2}{l}{\qquad{}and non-bonded short-range core} \\[0.5ex]
    covariance cutoff & \SI{9}{\angstrom} \\
    \qquad transition width & \SI{3}{\angstrom} \\
    energy scale $\delta$ & \SI{1}{\eV} \\
    length scale $\theta$ & \\
    \qquad bonded & \SI{3}{\angstrom} \\
    \qquad non-bonded & \SI{2}{\angstrom} \\[0.5ex]
    \multicolumn{2}{l}{transfer function [\cref{eq:gap:transferfunction}]} \\
    \qquad scaling factor $\alpha$ & \num{4} \\
    \qquad centre $r_0$ & \SI{4}{\angstrom} \\
    \qquad transition width $w$ & \SI{2}{\angstrom} \\[0.5ex]
    sparse points $n_X$ & 5000 \\
    sparse point selection & uniform \\[0.5ex]
    noise level (forces) $\sigma_f$ & \SI{0.5}{\meV\per\angstrom} \\
    \bottomrule
  \end{tabular}
  \caption{Hyperparameters for the methanol \descdimer GAP.}
  \label{tab:methanol:gap:dimer}
\end{table}

\begin{table}[htbp]
  \centering
  \begin{tabular}{lc}
    \toprule
    hyperparameter & trimer \\
    \midrule
    core potential: \\
    \multicolumn{2}{l}{\qquad\descdimer GAP (\cref{tab:methanol:gap:dimer})} \\[0.5ex]
    covariance cutoff & \SI{6}{\angstrom} \\
    \qquad transition width & \SI{3}{\angstrom} \\
    energy scale $\delta$ & \SI{1}{\eV} \\
    length scale $\theta$ & \\
    \qquad bonded & \SI{3}{\angstrom} \\
    \qquad non-bonded & \SI{2}{\angstrom} \\[0.5ex]
    sparse points $n_X$ & 5000 \\
    sparsification method & random \\[0.5ex]
    noise level (forces) $\sigma_f$ & \SI{0.5}{\meV\per\angstrom} \\
    \bottomrule
  \end{tabular}
  \caption{Hyperparameters for the methanol \desctrimer GAP.}
  \label{tab:methanol:gap:trimer}
\end{table}

\begin{table}[htbp]
  \centering
  \begin{tabular}{lcc}
    \toprule
    hyperparameter & monomer & dimer \\
    \midrule
    core potential: \\
    \multicolumn{3}{l}{\qquad{}non-bonded short-range core, \cf \cref{fig:benzene:bulk:pairpot}} \\[0.5ex]
    covariance cutoff                  & ---                 & \SI{8}{\angstrom}   \\
    \qquad transition width            &                     & \SI{4.5}{\angstrom} \\[0.5ex]
    energy scale $\delta$              & \SI{10}{\eV}        & \SI{1}{\eV}         \\
    length scale $\theta$ & \\
    \qquad bonded                      & \SI{0.3}{\angstrom} & \SI{3}{\angstrom}   \\
    \qquad non-bonded                  & {---}               & \SI{1.5}{\angstrom} \\[0.5ex]
    \multicolumn{3}{l}{transfer function [\cref{eq:gap:transferfunction}]} \\
    \qquad scaling factor $\alpha$     & {---}               & \num{4}             \\
    \qquad centre $r_0$                & {---}               & \SI{4}{\angstrom}   \\
    \qquad transition width $w$        & {---}               & \SI{2}{\angstrom}   \\[0.5ex]
    sparse points $n_X$                & 100                 & 2000                \\
    sparsification method              & uniform             & uniform             \\[0.5ex]
    noise level (forces) $\sigma_f$ & \multicolumn{2}{c}{\SI{10}{\meV\per\angstrom}} \\
    \bottomrule
  \end{tabular}
  \caption{Hyperparameters for the bulk benzene \descdimer GAP.}
  \label{tab:benzene:bulkgap:dimer}
\end{table}

\clearpage
\section{Pair potentials}
\begin{figure}[H]
  \centering
  \input{thesis_figures/methanol/trimer/pairpotentials.pgf}
  \mycaption{Methanol trimer pair potentials.}{From top to bottom: \pairCC, \pairCO and \pairOO interactions. We show MSCG/FM splines obtained from trimer MF (and, for comparison, the corresponding potentials from the dimer system) and the IBI pair potentials that match the trimer site--site distance distributions.}
  \label{fig:methanol:trimer:pairpotentials}
\end{figure}
\begin{figure}[p]
  \centering
  \input{thesis_figures/methanol/tetramer/pairpotentials.pgf}
  \mycaption{Methanol tetramer pair potentials.}{From top to bottom: \pairCC, \pairCO and \pairOO interactions. We show MSCG/FM splines obtained from tetramer MF and the IBI pair potentials that match the tetramer site--site distance distributions.}
  \label{fig:methanol:tetramer:pairpotentials}
\end{figure}
\begin{figure}[p]
  \centering
  \input{thesis_figures/methanol/bulk/pairpotentials.pgf}
  \mycaption{Bulk methanol pair potentials}{generated by force-matching (MSCG/FM) and structure-matching (IBI), as well as the two ``core-only'' short-range potentials.}
  \label{fig:methanol:bulk:pairpot}
\end{figure}
\begin{figure}[p]
  \centering
  \input{thesis_figures/benzene/bulk/pairpotentials.pgf}
  \mycaption{Bulk benzene pair potentials}{generated by force-matching (MSCG/FM) and structure-matching (IBI), as well as the short-ranged repulsive core potential.}
  \label{fig:benzene:bulk:pairpot}
\end{figure}
\begin{figure}[p]
  \centering
  \input{thesis_figures/benzene/water/pairpotentials.pgf}
  \mycaption{Pair potentials for water-solvated benzene.}{The MSCG/FM potential is based on matching the mean forces of the water-solvated dimer and the IBI potential is based on reproducing the site--site RDF of the water-solvated octamer.}
  \label{fig:waterbenzene:pairpot}
\end{figure}

\clearpage
\section{Length scale hyperparameter and transfer function}

\begin{figure}[H]
  \centering
  \input{thesis_figures/benzene/bulk/fig_hyper_thetatransfer.pgf}
  \mycaption{Influence of length scale hyperparameter and transfer function on GAP force prediction error.}{Force RMSE in \si{\kT\per\angstrom} for different short- and long-range length scales and for different values of the transition width $w$ and centre $r_0$ of the transfer function between the two regimes. (All other hyperparameters as given in \cref{tab:benzene:bulkgap:dimer}.) If no transfer function is used, there is only a single length scale. The RMSE for different values of this length scale is as follows:}
  \vspace*{1ex}
  \begin{tabular}{S[table-format=1.1]S[table-format=1.2]}
    \toprule
    {$\theta / \si{\angstrom}$} & {RMSE / \si{\kT\per\angstrom}} \\
    \midrule
    0.5 & 3.57 \\
    1.0 & 2.48 \\
    1.5 & 2.07 \\
    2.0 & 2.32 \\
    \bottomrule
  \end{tabular}
  \label{fig:gapcg:hyper:thetatransfer}
\end{figure}

\clearpage
\section{Mean forces for benzene dimer in different environments}
\label{appendix:benzenemeanforces}

\begin{figure}[htbp]
  \centering
  \includegraphics[width=0.6\textwidth]{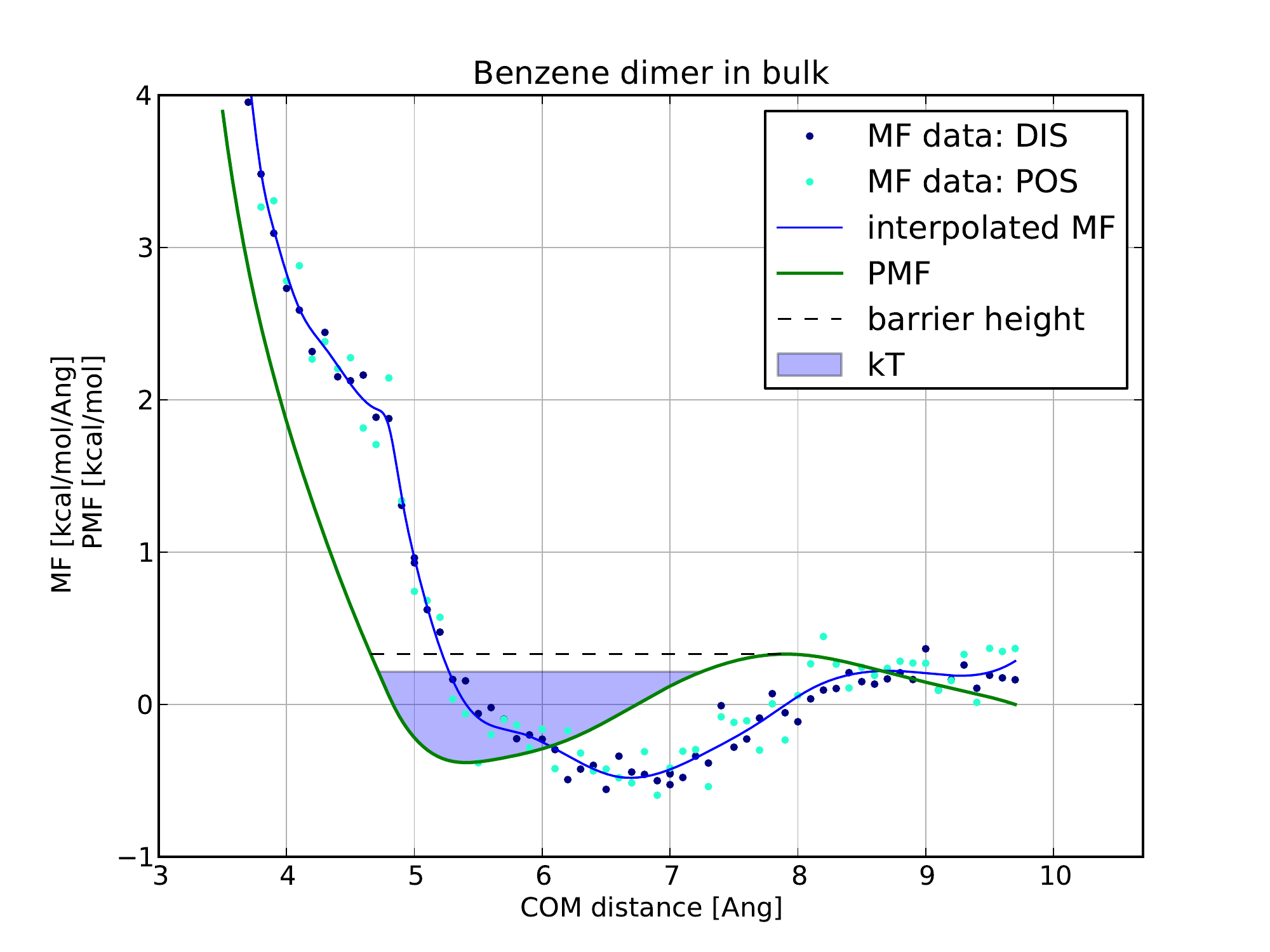}
  \includegraphics[width=0.6\textwidth]{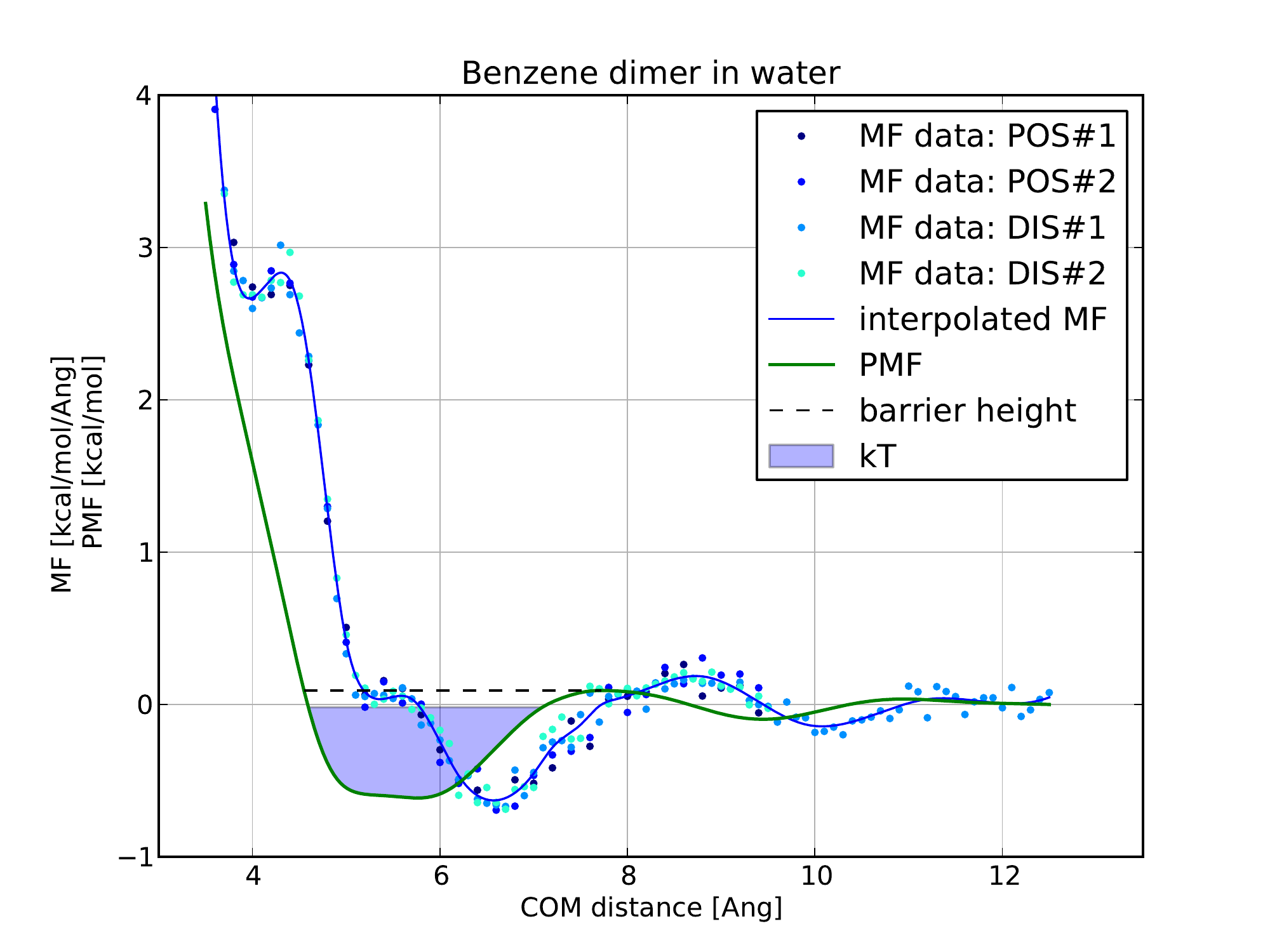}
  \includegraphics[width=0.6\textwidth]{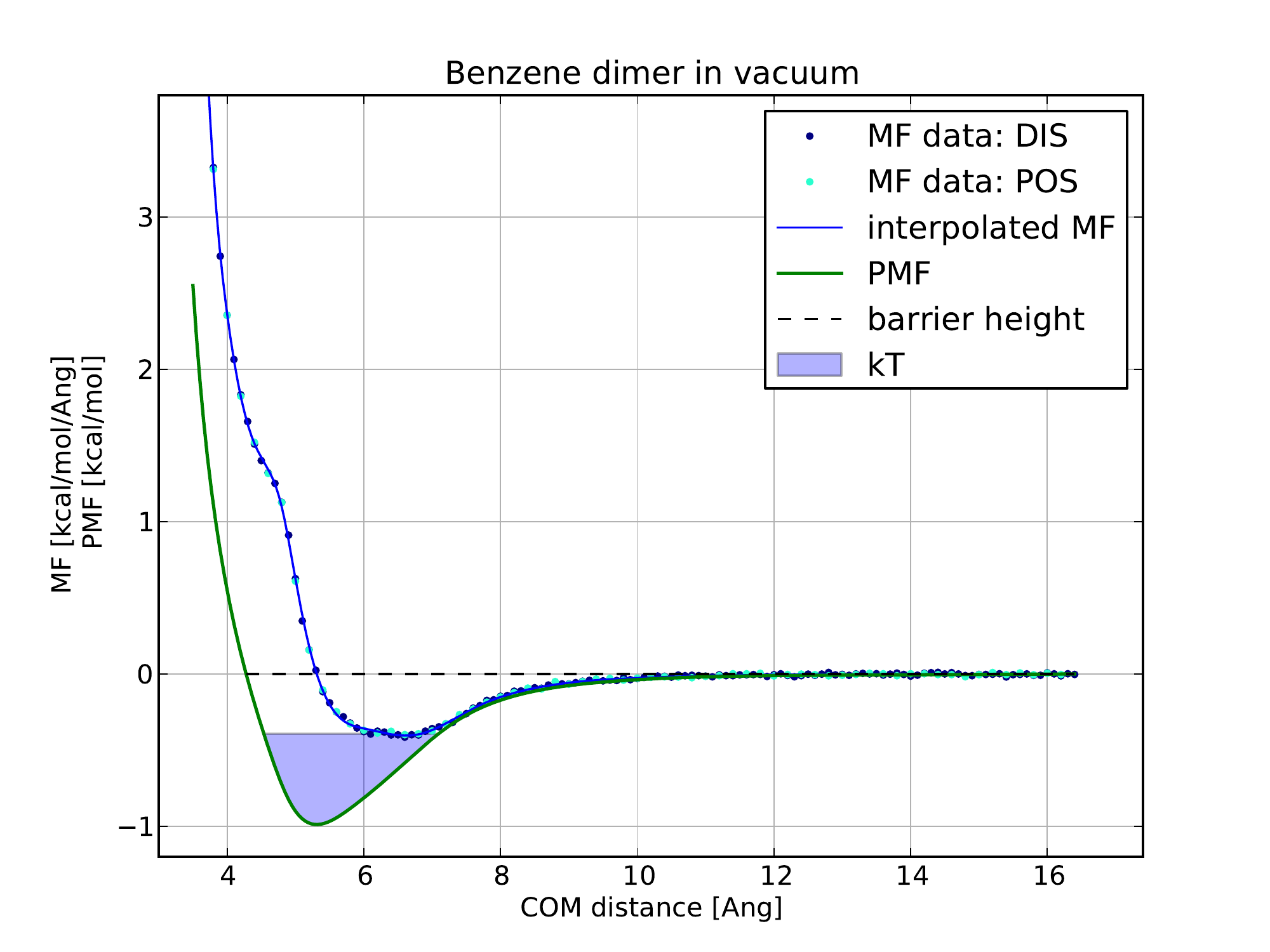}
  \caption{Benzene in different environments: COM mean forces and PMF.}
  \label{fig:appendix:benzenecommf}
\end{figure}

In \cref{ref:benzene:comparison} we compare the Potential of Mean Force (PMF) for the COM distance between two benzene molecules in different environments: in gas phase, in bulk benzene, and solvated by water. We obtained the mean forces (MF) by Constrained MD using \PMFlib.
It is possible to use either a single constraint on the distance between the COMs or position constraints on the coordinates of both COMs.
The mean force obtained for a distance constraint, $F^\text{MF}_\text{dis}$, has an additional entropic component due to the rotation of the constraint in space. It is connected to the mean force between the two COMs for position constraints, $F^\text{MF}_\text{pos}$, by
\[
  F^\text{MF}_\text{pos}(r) = F^\text{MF}_\text{dis}(r) - 2 \frac{\kT}{r} ,
\]
where $r$ is the distance between the COMs. We applied this correction to all mean forces obtained for distance constraints.
The resulting mean forces were interpolated by splines which could then be integrated to obtain the PMF.
The individual mean force observations, the spline interpolation, and the final PMF are shown in \cref{fig:appendix:benzenecommf} on the following page.

\mycleardouble

\begin{spacing}{1.0}
\bibliography{references}
\bibliographystyle{myunsrt2} %
\end{spacing}

\end{document}